\documentclass{ws-procs9x6}

\def\Dslash{{\rm D}\!\!\!\!/\,}

\newcommand\str{{\rm str}}
\newcommand\tr{{\rm tr}}
\newcommand\sdet{{\rm sdet}}
\newcommand\Tr{{\rm Tr}}

\def\CL{\mathcal{L}}
\def\CM{\mathcal{M}}
\def\CO{\mathcal{O}}
\def\bar{\overline}
\def\eff{{\rm eff}}

\def\GSM{{\rm GSM}}
\def\LQCD{\Lambda_{\rm QCD}}
\def\Lchi{\Lambda_{\chi}}
\def\chpt{\raise0.4ex\hbox{$\chi$}PT}
\def\spose#1{\hbox to 0pt{#1\hss}}
\def\ltapprox{\mathrel{\spose{\lower 3pt\hbox{$\mathchar"218$}}
 \raise 2.0pt\hbox{$\mathchar"13C$}}}
\def\gtapprox{\mathrel{\spose{\lower 3pt\hbox{$\mathchar"218$}}
 \raise 2.0pt\hbox{$\mathchar"13E$}}}
\def\inapprox{\mathrel{\spose{\lower 3pt\hbox{$\mathchar"218$}}
 \raise 2.0pt\hbox{$\mathchar"232$}}}

\begin{document}

\title{Applications of Chiral Perturbation theory to 
lattice QCD\footnote{\uppercase{L}ectures given at 
\uppercase{ILFTN} \uppercase{W}orkshop on
``\uppercase{P}erspectives in \uppercase{L}attice \uppercase{QCD}'',
\uppercase{N}ara, \uppercase{J}apan, \uppercase{O}ct 31-\uppercase{N}ov 11, 
2005.}}

\author{Stephen.~R. Sharpe} 

\address{Physics Department,\\ University of Washington, \\
Seattle, WA 98195-1560, USA \\
E-mail: sharpe@phys.washington.edu }

\maketitle

\abstracts{These lectures describe the use of
effective field theories to
extrapolate results from the parameter region where
numerical simulations of lattice QCD are possible
to the physical parameters (physical quark masses, infinite volume,
vanishing lattice spacing, etc.).
After a brief introduction and overview, I discuss
three topics: 1) Chiral perturbation theory (\chpt) in the continuum;
2) The inclusion of discretization effects into \chpt, focusing on the
application to Wilson and twisted-mass lattice fermions;
3) Extending \chpt\ to describe partially quenched QCD.
}

\section{Overview and Aims}
\label{sec:overview}

More than 30 years after Wilson introduced lattice QCD\cite{Wilson74},
and more than 25 years after Creutz's pioneering numerical studies of
non-abelian gauge theories\cite{Creutz80}, we can
now simulate lattice QCD, including quarks, with parameters that
approach their physical values. This is the result not only
of advances in computer power 
but also of improvements in algorithms and actions.\footnote{%
Some of these improvements are reviewed in Tony Kennedy's lectures.}
In particular, we can simulate QCD with pion masses of $250\;$MeV
or lower, with the minimum value depending on the choice of fermion action.
Such masses should allow a controlled extrapolation to
the physical pion masses, one that can give errors
at the few percent level\cite{MILCfpi}.
One of the aims of the field is to provide results with this accuracy
for many hadronic quantities, allowing both
tests of the method and predictions for unmeasured quantities.
This goal has begun to be attained\cite{Kronfeld05}.

Despite the successes just outlined, it is important to keep
in mind the limitations of LQCD (lattice QCD). Simulations
are, and will remain for the foreseeable future, 
limited in scope---one or two particle states
in a box unlikely to exceed $L=3-5\;$fm, with lattice spacings
unlikely to be smaller than $a\approx 0.05\;$fm, and pion
masses unlikely to drop below $200\;$MeV.\footnote{%
There are important exceptions, such as the very small lattice spacings
used to match QCD with a $b$ quark onto heavy quark effective 
theory\cite{NPHQET}, which are possible because $L$ can also be reduced.}
In order to connect these results to those for physical
quark masses in the continuum and infinite volume limits one needs
a quantitative theoretical understanding of how to 
extrapolate.\footnote{%
For percent accuracy one must also account
for the effects of electromagnetism.}
Such an understanding can be provided by chiral perturbation theory
generalized to include discretization errors, and is the topic
of these lectures.

One way to think of this situation is that
LQCD is a powerful tool with several adjustable parameters (``knobs'').
While we are able to turn these knobs independently
(unlike in the physical world where they are fixed),
we cannot turn them to their physical values.
Thus we are stuck simulating theories 
with unphysical values of the parameters, and
we need additional theoretical input.

\begin{figure}[ht]
\centerline{\epsfxsize=3in\epsfbox{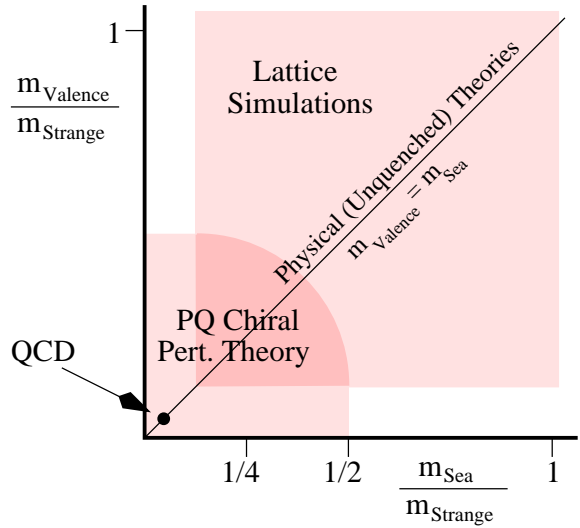}}   
\caption{PQQCD: using $m_{\rm val}\ne m_{\rm sea}$ and PQ\chpt\
to better extrapolate (from the overlap of the shaded regions)
to the physical theory.}
\label{fig:PQQCD}
\end{figure}

In fact, there are several other knobs (beyond quark masses, $a$, $L$
and $\alpha_{\rm EM}$) that we can adjust independently.
We can use different sea and valence quark masses---giving partially
quenched (PQ) theories---or we can go further and use different actions for
valence and sea quarks---``mixed action'' simulations. 
An interesting example of the latter is to
use valence fermions with good chiral symmetry (Domain-Wall or Overlap)
and cheaper sea quarks (staggered or Wilson-like).
Both PQ and mixed action theories are ``really'' unphysical: they not only have
unphysical values of the parameters but they are also not unitary.
Nevertheless, they are well-defined
Euclidean statistical systems, with long-distance correlations,
and it is plausible that they can be described by an effective chiral
theory. Furthermore, in both cases there are points in parameter
space for which the theories are physical, which ``anchor'' the effective
theories. I will discuss this in detail for PQQCD in sec.~\ref{sec:PQPQChPT}, 
and for now only illustrate the situation with Fig.~\ref{fig:PQQCD}.
The aim is to use the freedom provided by having
extra knobs, which are relatively cheap to turn, in order to
improve the accuracy of the extrapolation to the physical point:
``physical results from unphysical simulations''\cite{ShSh}.
This is an essential feature of the MILC collaboration's work
on decay constants and quark masses\cite{MILCfpi}.

A different example of unphysical theories is the use of
``rooted'' staggered fermions. Each staggered flavor leads to
four degenerate ``tastes'' in the continuum limit, 
and the standard approach to obtain a single
continuum fermion per flavor is to take the
fourth root of the fermion determinant.
As the taste symmetry is broken
for $a\ne0$, this rooting leads, however, to a {\em non-local} single-flavor
fermion action on the lattice\cite{BGSh}.
The implications of a non-locality that formally vanishes as
$a\to0$ are controversial---does one remain in the universality
class of QCD? 
I will not discuss this issue here, but only note that the theory
at $a\ne 0$ is undoubtedly unphysical, and the chiral-continuum
extrapolation can only be done if one has an effective theory
which describes the unphysical features. This is provided by
``(rooted) staggered \chpt''\cite{LeeSh,AB,rSChPT}. Whatever the
outcome of the rooting controversy, this is another example
of using effective field theory to obtain physical results from
unphysical simulations.

Due to limitations of time and space, I will discuss only
a subset of applications of \chpt\ to LQCD in these lectures.
I begin with a brief introduction to \chpt\ in the continuum,
with an emphasis on lessons for the lattice. I follow that
with the example of incorporating discretization errors into \chpt\ for
twisted-mass fermions, which includes Wilson and improved Wilson
fermions as a subset. In this case the theory is physical, but
gives a nice example of the power of adding an extra knob (the twist
angle) and of the utility of \chpt.
Finally, I discuss PQ\chpt, i.e. chiral perturbation theory for PQQCD.

I will mainly focus on the theoretical set-up and 
on general issues of the applicability of \chpt. I will not provide
a review of the status and accuracy of the state-of-the-art extrapolations.
My hope is that this introduction to the tools will allow the reader to
critically assess current work.
 
\section{Review of $\chi$PT in the continuum}
\label{sec:chptcont}

In this section I describe the construction of the
chiral Lagrangian in the continuum. There are many good books
and lectures on this topic. I have found those
by Donoghue, Golowich \& Holstein\cite{Donoghue}, Ecker\cite{Ecker},
Georgi\cite{Georgi}, Kaplan\cite{Kaplan}, Kronfeld\cite{Kronfeld},
Manohar\cite{Manohar} and Pich\cite{Pich} very useful.

\subsection{Effective Field Theories in general}

In these lectures I consider two examples of effective field theories (EFTs):
\chpt\ as an EFT for QCD,
and Symanzik's effective continuum theory for lattice QCD
(the latter to be discussed in sec.~\ref{sec:symanzik}). 
Thus it is useful to begin with
a discussion of EFTs in general. If you are unfamiliar with the subject
then some of this section may be hard to follow in detail, but
my aim is to begin with a broad-brush sketch, which will be filled in later.

The generic situation is that we have an underlying theory in which there is
a separation of scales. In the theories of interest we have:
\begin{eqnarray}
\textrm{\chpt:}&& p_\pi \sim m_\pi \ll m_\rho, m_N\,; 
\\
\textrm{Symanzik:}&&
p_{\rm quark}, p_{\rm gluon}\sim \Lambda_{\rm QCD} \ll \pi/a
\,.
\end{eqnarray}
Note that in the former case there is a separation of masses, with
the ``pions'' (by which I mean the light pseudoscalars: $\pi$,
$K$ and $\eta$) being lighter than all other hadrons,
while in Symanzik's theory we choose to consider momenta much
smaller than the lattice cut-off. In both cases there is a good reason
to split off the low-scale physics. For \chpt\ it is because
the pion sector changes most rapidly as we approach the chiral limit
(as we will see in detail). For Symanzik's theory we want to understand
the impact of lattice spacing errors on the quarks and gluons which
dominate the non-perturbative contributions to hadronic quantities.

\begin{figure}[ht] 
\centerline{\epsfxsize=4.1in\epsfbox{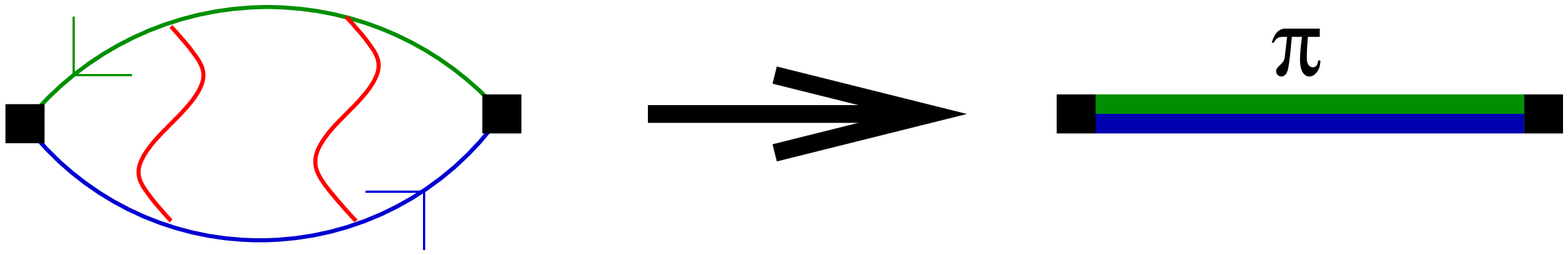}}   
\vskip 0.5cm
\centerline{\epsfxsize=4.1in\epsfbox{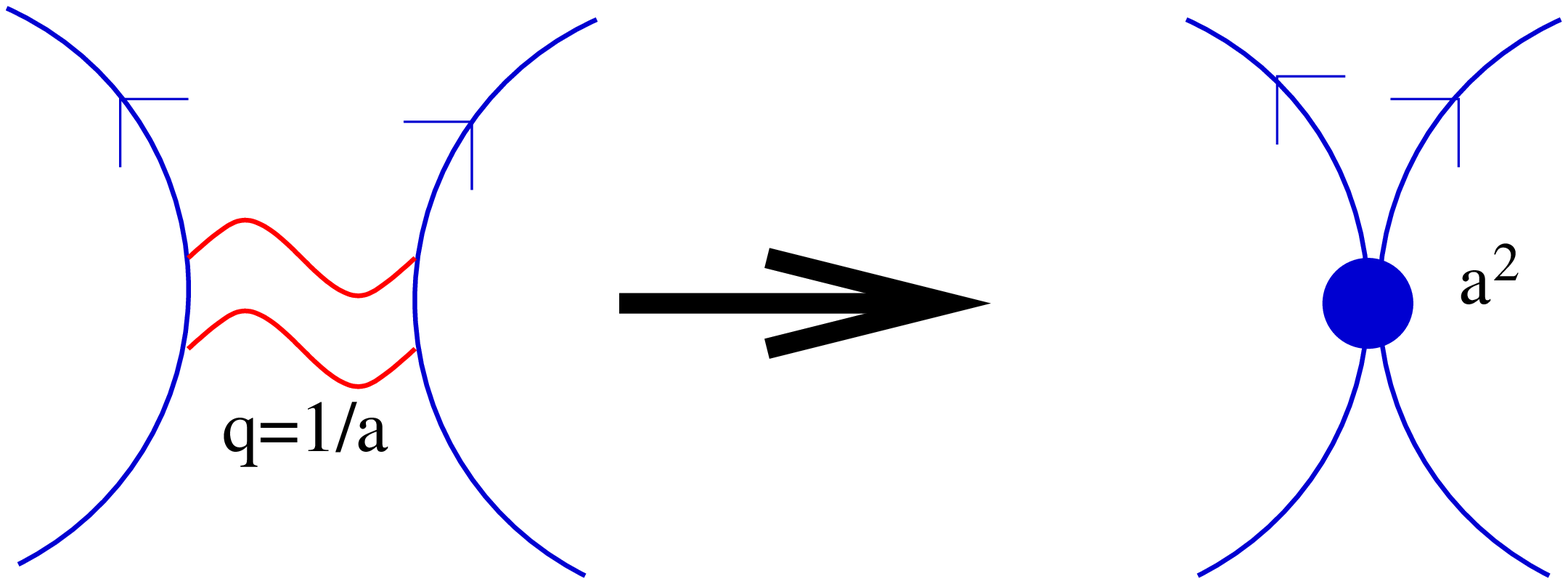}}  
\caption{Generation of effective field theories. }
\label{fig:EFT}
\end{figure}

Crudely speaking,
we now introduce a momentum cut-off $\Lambda$ lying between the two scales,
and ``integrate out'' the high-momentum degrees of freedom.
This process is illustrated schematically in Fig.~\ref{fig:EFT}.
It leaves only pions with low momenta in \chpt, and continuum-like quarks
and gluons in the Symanzik theory.
These degrees of freedom interact via
vertices that are quasi-local, with a physical size $\Lambda^{-1}$.
This quasi-locality follows because we consider only
external momenta satisfying $p \ll \Lambda$, so the high-momentum degrees
of freedom are always highly virtual. 
The vertices are then expanded
in powers of $p/\Lambda$, yielding local operators with increasing numbers
of derivatives.
The low-momentum modes themselves can become nearly on shell
(or exactly on shell if we continue to Minkowski space),
but the resulting analytic structure of correlation functions
(leading to cuts in Minkowski space) is maintained in the effective theory.

This description is impractical to implement in most cases. In particular, we do
not know how to integrate out quarks and gluons from QCD analytically
to yield a theory of pions, since confinement is a non-perturbative phenomenon.
Even in the Symanzik theory, where one might have expected that
quarks and gluons with $p\sim \pi/a$ would have
been perturbative since $1/a\gtapprox 2\;$GeV ($a\ltapprox 0.1\;$fm),
it turns out that accurate results
mostly require non-perturbative calculations.\footnote{%
This is found when implementing the 
improvement program for Wilson fermions \cite{ALPHA}.}
The beauty of the EFT method, however, 
is that we do not actually need to do the
integrations. Instead, following Weinberg\cite{Weinberg} 
we can rely on the general properties of EFTs. 
If the underlying theory is physical, 
its S-matrix will be unitary, Lorentz-invariant, satisfy cluster decomposition,
and transform appropriately under the internal symmetries.
These properties must be maintained by the EFT, which is, after all,
designed to reproduce the S-matrix of the low-momentum
degrees of freedom. The only known way to do this is with
a local, Lorentz-invariant Lagrangian.
It should be constructed solely from the low-energy degrees of freedom,
and satisfy the same internal symmetries as the underlying theory.
All possible terms consistent with these symmetries must be included---this
precludes the need to explicitly integrate-out degrees of freedom,
at the price of introducing unknown constants.

One notable feature of the resulting $\CL_\eff$ is that it is not
renormalizable, and thus valid only over a limited energy range.
This is an intrinsic part of the construction: we know that 
the EFT breaks down when
$p\gtapprox \Lambda$. Non-renormalizability does not, however, imply a lack
of calculability. As we will see, one can expand quantities in
powers of $p/\Lambda$, with a finite number of unknown coefficients
at each order. The limitations of the method are then
(a) the need to introduce unknown coefficients and (b) an
unavoidable truncation error. This error, however, decreases as
the separation in scales increases (i.e. as $m_\pi\to 0$ or $a\to 0$).

As just described, the justification of EFT is based on properties of
the S-matrix, and thus rooted in Minkowski space. While this is fine
for the development of continuum \chpt\ (my first topic),
the natural objects in lattice simulations (my second topic) are
Euclidean (finite-volume) correlation functions of local operators.
In particular, the discretization errors are constrained by the
symmetries of a Euclidean lattice.
Thus an alternative approach to developing and justifying
an EFT is needed. This has been provided 
for the case at hand by Symanik\cite{Symanzik},
using an extension of renormalization theory.
The result (established to all orders in perturbation theory)
is that the recipe given above still applies: keep all local
terms consistent with the symmetries of the underlying
theory (in this case the discrete symmetries of a Euclidean lattice).
My third topic, PQQCD, is also strictly limited to Euclidean space,
but here neither of the previous justifications apply, and one must
make further assumptions.

Because my second and third topics involve Euclidean theories, I have
chosen to couch the discussion of the first (\chpt) also in Euclidean space.
This allows later sections to build on the earlier notation.
In fact, it is perfectly legitimate, having determined the
Minkowski-invariant local effective Lagrangian, to rotate this
to Euclidean space. The result will be the most general
Euclidean-invariant local Lagrangian (consistent with the other symmetries,
which are unaffected by the rotation). This Lagrangian will reside
in the functional integral which generates the Euclidean correlation
functions of the theory.

\subsection{Chiral symmetry in QCD and its breaking}
\label{sec:brokenchiral}

Without further ado, let me turn to the first concrete example, \chpt.
The fermionic part of Euclidean Lagrangian for QCD is given by
\begin{eqnarray}
{\CL}_{QCD} =\overline Q_L \Dslash Q_L+\overline Q_R \Dslash Q_R
+\overline Q_L M Q_R+\overline Q_R M^\dagger Q_L
\,.
\label{eq:LQCD}
\end{eqnarray}
where I have included only the $N=3$ light quarks,
$Q^{tr} = (u,d,s)$, $ \overline Q= (\bar u, \bar d,\bar s)$.
I will also consider the $N=2$ theory without the strange quark.
Left- and right-handed fields are defined with projectors
$P_\pm=(1 \pm \gamma_5)/2$, and are $Q_{L,R} = P_\mp Q_{L,R}$
and  $\overline Q_{L,R} = \overline Q_{L,R} P_\pm$.
There is no problem with $\overline Q_L$ and $Q_L$ being
defined with the different projectors, since
$\overline{Q}$ and $Q$ are independent fields.
The key fact is that, 
in the massless limit, left- and right-handed quarks can
be rotated independently, so the Lagrangian has a
${G}=SU(3)_L\times SU(3)_R$ chiral symmetry under which
\begin{equation}
Q_{L,R} \to U_{L,R} Q_{L,R}\,,\quad
\overline Q_{L,R} \to \overline Q_{L,R} U_{L,R}^\dagger\,:\quad
U_{L,R} \in SU(3)_{L,R}\,.
\end{equation}
There is also the overall vector $U(1)$ symmetry, which counts quark
number, while the apparent
axial $U(1)$ symmetry is broken by the anomaly.

Quark masses enter through the mass matrix $M$, which is conventionally
taken to be $M=\textrm{diag}(m_u,m_d,m_s)=M^\dagger$.
These masses break the chiral symmetry: any non-zero values violate
the axial symmetries (those with $U_L = U_R^\dagger$),
while the vector symmetries ($U_L=U_R$)
are broken unless the masses are degenerate.
We can, however, formally retain the chiral symmetry of $\CL_{QCD}$
by treating $M$ as a complex ``spurion field'' transforming as
$M \to U_L M U_R^\dagger$ and $M^\dagger \to U_R M^\dagger U_L$.
This is a convenient trick for keeping track of the symmetry-breaking
caused by the mass term.

Since chiral symmetry is key to all that follows,
and quark masses break this symmetry, we must require that $M$ be small.
What does small mean? One criterion is that
$M$ should be small compared to the QCD scale,
$m_q \ll \Lambda_{QCD}\sim 300\;$MeV.
A more precise criterion will arise from $\chi$PT:
$m_{\pi,K,\eta} \ll \Lambda_\chi\equiv 4\pi f_\pi\approx 1200\;$MeV.
It follows that in physical QCD, with $(m_u\!+\!m_d)/2\approx 4\;$MeV, 
$SU(2)_L\times SU(2)_R$ is a very good approximate symmetry, while
$SU(3)_L\times SU(3)_R$ is more badly broken since
$m_s\approx 100\;$MeV and $m_{K,\eta} \approx \Lambda_\chi/2$.
This brings up an important question for lattice applications of \chpt:
can approximate chiral symmetry
can be used to determine the strange quark mass dependence
when $m_s^{\textrm{lat}}\approx m_s$?
If not, then we can only use chiral symmetry to
guide extrapolations in $m_u$ and $m_d$.

Chiral perturbation theory is an expansion about the chiral limit,
$M=0$, so I first discuss massless QCD.
It is expected in this theory that the
exact chiral symmetry is spontaneously broken by the vacuum. 
This is based on an accumulation of evidence about QCD itself: 
the lightness of the $\pi$, $K$ and $\eta$ are
consistent with their being {\em pseudo}-Nambu-Goldstone bosons (PGBs)
of the broken {\em approximate} chiral symmetry of real QCD;
the absence of (approximate) parity-doubling in the hadron spectrum, 
e.g. $m_N(P=+)\ne m_N(P=-)$, as would be required by
an unbroken (approximate) chiral symmetry;
and the accumulated successes of \chpt, especially in the 
$SU(2)$ sector.
It is also expected that the order parameter for chiral symmetry
breaking is the condensate:\footnote{%
There has been some controversy over whether
the condensate is the dominant order parameter\cite{stern}, but
this standard picture is now strongly favored\cite{ColangeloLeutwyler},
and is supported by lattice calculations of $\langle \bar q q\rangle$,
so I will accept it here.}
\begin{eqnarray}
\langle \bar q q\rangle = 
\langle (\bar q_L q_R + \bar q_R q_L)\rangle
\sim \Lambda_{\rm QCD}^3 \ne 0\,,\quad q=u,d,s\,.
\label{eq:cond}
\end{eqnarray}
The vector symmetry is not expected to be spontaneously broken,
based on the presence of approximate $SU(3)_V$ multiplets
in the hadron spectrum, and on theoretical 
considerations\cite{VafaWitten}.
This implies that the condensates are equal in the massless theory, 
$\langle \bar u u\rangle=\langle \bar dd\rangle= \langle \bar ss\rangle$.

The form of the condensate in eq.~(\ref{eq:cond}) is 
a convention.
There is in fact a manifold of equivalent vacua
related by chiral transformations, and parameterized by the 
orientation of the condensate in flavor space:
\begin{equation}
\Omega_{ij} = \langle Q_{L,i,\alpha,c} \overline Q_{R,j,\alpha,c} \rangle
\ {\lower0.6ex\hbox{$\stackrel{G}{\longrightarrow}$}}\ 
U_L\, \Omega \,U_R^\dagger\,.
\label{eq:Omega}
\end{equation}
Here I have shown the summed Dirac and color indices explicitly
($\alpha$ and $c$, respectively), since $Q$ and $\bar Q$ are
in the opposite order from usual. This order makes the flavor matrix
structure (indices $i$ and $j$) transparent, and is particularly useful
when discussing PQ\chpt\ below.
The assumption of unbroken vector symmetry implies that
$\Omega_{ij} = \omega \,\delta_{ij}$ must lie in the manifold,
and so the general point is $\omega U_L U_R^\dagger$. 
In this language, chiral symmetry breaking is 
equivalent to $\omega$ being non-zero, for then
the vacuum is left invariant only by a subgroup $H$ of $G$:
\begin{equation}
\underbrace{SU(3)_L \times SU(3)_R}_{{ G}}
\longrightarrow
\underbrace{SU(3)}_{{ H}}
\,,
\end{equation}
The nature of $H$ is simplest using
the conventional vacuum orientation, 
$\Omega_{ij} = \omega \,\delta_{ij}$, 
with $\omega = -\langle \bar q q\rangle$.
Then $H=SU(3)_V$ (with $U_L=U_R$), while the
axial transformations with $U_L =U_R^\dagger$ are broken
($\Omega = \omega \to \omega U_L^2$). 

Goldstone's theorem then implies that there are 8 
massless Nambu-Goldstone bosons (NGB---labeled $\pi^b$, and
corresponding to the $\pi$, $K$, and $\eta$), 
each coupled to one of the eight broken axial generators.
For the conventional vacuum orientation, one has
\begin{equation}
\langle \pi^b(p) | \overline Q\gamma_\mu\gamma_5 T^a Q(0)|0\rangle
= -i f_\pi p_\mu \delta^{ba}
\,,
\end{equation}
with $T^a$ being $SU(3)$ generators.
The masslessness of the NGB follows from the fact that
rotations of the condensate with four-momentum $p$ cost zero energy
as $p\to 0$, in which limit they become global rotations.

We now have the ingredients with which to construct an EFT:
a separation of scales ($m_{\rm GB}=0$ compared to the scale
of other hadron masses---$m_\rho$, $m_\eta'$, $m_{\rm proton}$, etc., 
with $m_{\rm had}\approx 1\;$GeV),
and a knowledge of the symmetries of the underlying theory.
The EFT will contain only NGB as dynamical degrees of freedom,
and should be valid as long as $p_{\rm GB} \ll m_{\rm had}$.
The spurion field $M$ should also be added to include the effects of
quark masses.\footnote{%
The EFT can also contain {\em static} sources, representing heavy particles with
$m\gtapprox m_{\rm had}$, off which the NGBs can scatter.
These can represent the interesting cases of
vector mesons, baryons, or heavy-light hadrons.}

Representing NGB fields is probably the most conceptually non-trivial step 
of the construction, because (a) the underlying theory is written in terms
of quarks rather than mesons (one is not just ``thinning'' degrees of freedom,
one is also changing basis) and (b) the choice is not 
unique\cite{Coleman,Callan}. Because of point (b), the strategy is simply to find
any representation which works---it turns out that
by field redefinitions one can then switch to other choices as desired. 

Since there are no precise rules to follow, it is useful to proceed by
analogy. To this end, I recall the
canonical example of spontaneous symmetry breaking:
a complex scalar field with a ``Mexican hat'' potential
\begin{equation}
V = -\mu^2 \phi^\dagger \phi + {\lambda} (\phi^\dagger \phi)^2/2
\,.
\end{equation}
There is a $G=U(1)$ phase symmetry, $\phi\to e^{i\alpha}\phi$,
which is spontaneously broken if  $\mu^2>0$, 
for then there is a non-zero vacuum expectation value 
(VEV) $|\langle\phi\rangle|=v=\sqrt{\mu^2/\lambda}$.
The vacuum manifold consists of the phase of $\langle\phi\rangle$, and
is thus $U(1)$.
The symmetry breaking is ${ G} \longrightarrow { H}=1$, implying
a single NGB corresponding to phase rotations.
The EFT for this massless mode is analogous to that we wish to
write down for the pions in QCD.

The advantage of this theory, compared to QCD, is that we can
directly construct the EFT by integrating out heavy fields,
as long as $\lambda$ is small enough that we can use perturbation theory.
Having done so, we can see how the EFT could be obtained using the
symmetries alone, and use this to guide the construction for QCD.

I decompose the field in a way which differs from that used,
say, in studying the Higgs:
$\phi(x) = v e^{\rho(x)} e^{i\theta(x)}$ rather than
$\phi(x) = v + h(x)$.
This choice picks out the NGB degree of freedom, $\theta$, explicitly,
and allows the phase symmetry to act linearly:
$e^{i\theta[x]} \to e^{i\alpha} e^{i\theta[x]}$.
If we integrate out the heavy radial degree of freedom, $\rho$,
we obtain an effective Lagrangian
in terms of $e^{i\theta}$. But we do not need to do any work to
determine the general form of $\CL_{\rm eff}$---we need only
require locality, reality, Euclidean and $U(1)$ invariance. 
The result is
\begin{equation}
 \CL_{\rm eff} =
c_2 \partial_\mu(e^{i\theta}) \partial_\mu (e^{-i\theta})
+
c_4 \partial_\mu (e^{i\theta}) \partial_\mu (e^{-i\theta})
\partial_\nu (e^{i\theta}) \partial_\nu (e^{-i\theta})
+ \dots \,,
\end{equation}
where $c_i$ are unknown constants.\footnote{%
Since $U(1)$ is abelian, $\CL_{\rm eff}$ can be
simplified to $c_2 (\partial_\mu\theta)^2
+c_4 [(\partial_\mu\theta)^2]^2$. I do not pursue this
as similar manipulations fail for the non-abelian chiral groups
relevant for QCD.}
Terms without derivatives
on every factor of $e^{\pm i\theta}$ can be brought
into the form shown (up to total derivatives) using
$e^{i\theta} e^{-i\theta}=1$ and the abelian nature of the group.
The result is a massless NGB having interactions proportional to $p^4$.
It is an interesting exercise to check the latter result
in perturbation theory using the conventional
expansion in terms of $h(x)$---the $p^4$ arises
from cancellations between non-derivative interactions.

We learn two things from the $U(1)$ example. First,
to use the exponential of the ``pion'' fields, since it transforms
linearly under $G$, and simplifies the implementation
of the symmetries. Second, that a ``fixed radius''
form ($|e^{i\theta}|=1$ above) automatically includes only
the NGB, excludes heavy degrees of freedom, and enforces
the spontaneous breakdown of the symmetry (i.e. $U(1)$ is
broken for any value of $\langle\theta\rangle$).

The QCD analog of $\langle\phi\rangle$ is the condensate
$\Omega_{ij}$ of eq.~(\ref{eq:Omega})---both map out
the corresponding vacuum manifolds. 
The analog of fixed length angular fluctuations 
is obtained by promoting $\Omega/\omega$ to a dynamical field
$\Sigma_{ij}(x)$, corresponding roughly to fluctuations in the 
condensate. Just as $\Omega/\omega$ is in $SU(3)$, so is $\Sigma$,
and the chiral transformation properties are the same:
\begin{equation}
\Sigma(x) \in SU(3):\qquad
\Sigma(x) 
\ {\lower0.6ex\hbox{$\stackrel{G}{\longrightarrow}$}}\ 
U_L \Sigma(x) U_R^\dagger 
\,.
\end{equation}
Any VEV of $\Sigma$ breaks $ { G}$ to $ { H}=SU(3)$,
leading to the desired number of NGBs. The fixed radius of
$\Sigma$ (i.e. $\Sigma \Sigma^\dagger=\Sigma^\dagger\Sigma=1$)
implies that the {\em only} degrees of freedom in $\Sigma$ are the NGBs.
For example, if $\langle\Sigma\rangle=1$ we can expand as
\begin{equation}
\Sigma(x) = \exp\left( 2i\Pi(x)/f\right)
= \exp\left( 2 i \pi^a(x) T^a/f\right)\,,\qquad
a=1,8,
\label{eq:expandSigma}
\end{equation}
in terms of the eight ``pion'' fields and a 
constant $f$ to balance dimensions.
Note that although $\Sigma$ transforms linearly, this
is not the case for the pion fields (e.g. $\delta \pi_a$
contains terms with any odd powers of $\pi_b$).
Thus constructing $\CL_{\rm eff}$ in terms of the pion field
directly would be very difficult.

\subsection{Constructing the pionic effective Lagrangian\protect\cite{GL}}
\label{sec:construct}

\subsubsection{Building blocks for $\CL_{\rm eff}$}
\label{sec:blocks}

We are now in business. The ingredients are $\Sigma$ and 
$\Sigma^\dagger$, as well as the spurions $M$ and $M^\dagger$.
I recall their transformation properties
under the chiral symmetry group $G=SU(3)_L\times SU(3)_R$:
\begin{equation}
\Sigma \to U_L \Sigma U_R^\dagger\,,\ \ 
\Sigma^\dagger \to U_R \Sigma^\dagger U_L^\dagger\,,\ \
M \to U_L M U_R^\dagger\,,\ \ 
M^\dagger \to U_R M^\dagger U_R^\dagger\,.
\end{equation}
It is useful to construct objects which transform solely under
the left-handed (LH) or right-handed (RH) sub-groups
(and which I call respectively LH and RH building blocks),
since they simplify enumeration of operators:\footnote{%
I use the convention throughout that
derivatives only act on the objects immediately to their right. The arrows
implicitly denote transformation under $G$.}
\begin{itemize}
\item[{ LH:}]
$
{L_\mu} =\Sigma \partial_\mu \Sigma^\dagger
=-\partial_\mu \Sigma \Sigma^\dagger = { - L_\mu^\dagger}
\longrightarrow U_L { L_\mu} U_L^\dagger
$
\item[{LH:}]
$
M \Sigma^\dagger \longrightarrow U_L { (M \Sigma^\dagger)}
 U_L^\dagger\,,\qquad
{ \Sigma M^\dagger} \longrightarrow U_L { (\Sigma M^\dagger)} U_L^\dagger
$
\item[{ RH:}]
$
{ R_\mu} =\Sigma^\dagger \partial_\mu \Sigma
=-\partial_\mu \Sigma^\dagger \Sigma = { - R_\mu^\dagger}
\longrightarrow U_R { R_\mu} U_R^\dagger
$
\item[{ RH:}]
$
{M^\dagger \Sigma} 
\longrightarrow U_R { (M^\dagger \Sigma)} U_R^\dagger\,,\qquad
{ \Sigma^\dagger M} \longrightarrow 
U_R { (\Sigma^\dagger M)} U_R^\dagger
$ .
\end{itemize}
where I have repeatedly used the unitarity of $\Sigma$.
From the fact that $\textrm{det}(\Sigma)=1$ one learns
that $L_\mu$ and $R_\mu$ are traceless, e.g.
\begin{equation}
0 = \partial_\mu (\det\Sigma) =\partial_\mu (\exp\tr \ln\Sigma)
= \det\Sigma\ \tr(\partial_\mu \Sigma \Sigma^{-1})
= - \tr({ L_\mu})
\,.
\end{equation}
Thus $L_\mu$ and $R_\mu$ (``Weyl derivatives'')
are elements of the Lie algebra, $su(3)$.

Another symmetry of QCD 
is parity. Since $\Sigma \sim q_L \bar q_R$ and 
$\Sigma^\dagger \sim q_R \bar q_L$, the transformations 
in the EFT are
[with $x_P=(-\vec x,x_4)$]
\begin{equation}
\Sigma(x)\leftrightarrow\Sigma^\dagger(x_P),
\ \
M \leftrightarrow M^\dagger,
\ \
L_i(x) \leftrightarrow -R_i(x_P),
\ \
L_4(x) \leftrightarrow R_4(x_P).
\end{equation}
If one expands about $\langle\Sigma\rangle=1$, as in eq.~(\ref{eq:expandSigma}),
then the pion field transforms as $\Pi(x) \to -\Pi(x_P)$.
There are also $C$ and $T$ symmetries, which I do not show explicitly.

I now enumerate terms which are local, real and satisfy the symmetries of QCD.
These are just products of the building blocks above, with LH and
RH blocks combined separately into traces to 
make them invariant under $G$. In fact, for the terms I display,
one need only use LH building blocks as the results
equal their ``parity conjugates'' (p.c.).
Since in the end we expand in powers of momenta,
it is useful to classify terms according to the number of derivatives.
Similarly, as $M$ is treated as small, one should classify
according to the number of spurion insertions.
We will see that one should usually count
two derivatives for each spurion.


There are no non-trivial terms without derivatives or spurions: these
would be constructed from $\tr[(\Sigma\Sigma^\dagger)^n]$
or powers or $\det\Sigma$, but both are constants.
Euclidean invariance rules out a single derivative.
The only independent term with two derivatives is 
\begin{enumerate}
\item[1.]
$\tr(L_\mu L_\mu) = -\tr(\partial_\mu \Sigma \partial_\mu\Sigma^\dagger)
=\tr(R_\mu R_\mu)$,
\end{enumerate}
while the only
term with no derivatives and one mass insertion is 
\begin{enumerate}
\item[2.]
$\tr(M \Sigma^\dagger) + \tr(\Sigma M^\dagger)$\,.
\end{enumerate}
There are  five terms with four derivatives:
\begin{enumerate}
\item[3.]
$\left[\tr(L_\mu L_\mu)\right]^2$
\item[4.]
$\tr(L_\mu L_\nu)\tr(L_\mu L_\nu)$
\item[5.]
$\tr(L_\mu L_\mu L_\nu L_\nu)$ 
{ [not independent for two light flavors]]}
\item[6.]
$\tr(L_\mu L_\nu L_\mu L_\nu)$ 
{[not independent for 2 or 3 light flavors]]}
\item[7.]
The Wess-Zumino-Witten (WZW) term involving 
$\epsilon_{\mu\nu\rho\sigma}$\cite{WZW}\,;
\end{enumerate}
two terms with 
two derivatives and one mass insertion:
\begin{enumerate}
\item[8.]
$\tr(L_\mu L_\mu)\ \tr(M\Sigma^\dagger+\Sigma M^\dagger)$
\item[9.]
$\tr(L_\mu L_\mu [M\Sigma^\dagger+\Sigma M^\dagger])$\,;
\end{enumerate}
and three terms with two mass insertions:
\begin{enumerate}
\item[10.]
$\left[\tr(M\Sigma^\dagger+\Sigma M^\dagger)\right]^2$
\item[11.]
$\left[\tr(M\Sigma^\dagger-\Sigma M^\dagger)\right]^2$
\item[12.]
$\tr(M\Sigma^\dagger M\Sigma^\dagger+ M^\dagger\Sigma M^\dagger \Sigma)$ .
\end{enumerate}
Each of these terms appears in $L_{\rm eff}$ with an independent
unknown coefficient, except for terms 5 and 6 which, as noted,
are not independent for certain chiral groups.
I will not discuss the interesting structure of the WZW term,
as it is complicated, and
does not contribute to the simple processes considered here.
I will also not continue the enumeration beyond this point. 
This has been done as part of
next-to-next-to-leading order (NNLO) calculations\cite{twoloopchpt}, 
but is beyond the scope of this introduction.

With the enumeration of operators in hand, 
I turn to the predictions. I will assume a power counting
with $\partial^2\sim M$ 
and justify this {\em a posteriori}.
At leading order (LO) we have (the superscript on $\CL$ counting derivatives):
\begin{equation}
\CL^{(2)}_\chi = 
\frac{f^2}{4} \tr\left(\partial_\mu\Sigma \partial_\mu \Sigma^\dagger\right)
-
\frac{f^2 B_0}{2} \tr(M \Sigma^\dagger + \Sigma M^\dagger )
\,,\label{eq:L2first}
\end{equation}
where the unknown ``low energy constants'' (LECs) have been given their
standard names $f$ and $B_0$. Since we are expanding about massless QCD, the only
scale that can appear is $\LQCD$, so we expect $f\sim B_0\sim \LQCD$.
Up to this stage, {$M$} is a complex spurion field.
To include the effects of quark masses we set
it to its physical value: 
$M\to M_0=\textrm{diag}(m_u,m_d,m_s)=M_0^\dagger$.
This makes the potential---the second term in eq.~(\ref{eq:L2first})---depend
on the direction of $\Sigma$, and we must determine its VEV by
minimizing this potential.
In other words, the quark masses, which break the chiral symmetry,
pick out a preferred direction in the vacuum manifold.
If all quark masses can be chosen positive (as is apparently the case in
reality), then one finds
that the VEV is $\langle\Sigma\rangle=1$.
I stress that this result is convention dependent: we could choose
$M = U_L M_0 U_R^\dagger$, in which case $\langle\Sigma\rangle= U_L U_R^\dagger$.

\subsubsection{Brief aside on vacuum structure}

\begin{figure}[bt]
\centerline{\epsfxsize=3.6in\epsfbox{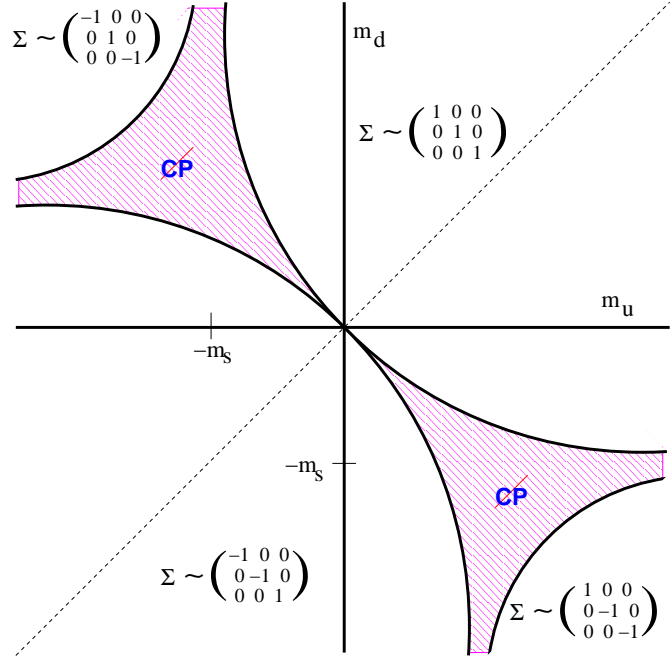}}   
\caption{Phase structure at LO with 3 flavors and fixed $m_s$
(from Creutz\protect\cite{CreutzCP}). }
\label{fig:phaseCP}
\end{figure} 

It is instructive
to consider the vacuum structure of {\em two flavor} theory in a little more detail.
Then we can write $\Sigma=\exp(i\theta \vec n\cdot\vec\tau)$,
implying $\Sigma+\Sigma^\dagger= 2 \cos\theta \times 1$, and 
thus ${ V}^{(2)} \propto - \tr(M) \cos\theta$.
This is minimized by
{$\langle\Sigma\rangle=1$} if {$\tr M > 0$}, 
and by {$\langle\Sigma\rangle=-1$} if {$\tr M < 0$}.
Note that at LO all that matters is the average quark mass $\tr M/2$; the
difference $m_u-m_d$ does not enter. 
There is a first order phase transition when $\tr M$ changes
sign at which the condensate flips sign but maintains its magnitude.
Note, however, that the physical theory is the same for both signs of $\tr M$:
one can go between them with a chiral rotation satisfying $U_L U_R^\dagger=-1$.
I will discuss
how discretization errors effect this transition in 
sec.~\ref{sec:Aokiregime} below.

With three flavors (or any odd number), the situation is more
complicated because $\Sigma=-1$ is not an element of $SU(3)$.
Changing the sign of $M$ leads to a different theory
(one with the original $M$ plus a $\theta-$term with $\theta=\pi$).
Without going into details, I show in Fig.~\ref{fig:phaseCP}
the phase structure
if $m_s$ is fixed and positive while the other two masses change.
The shaded region is where CP is spontaneously broken.
In the real world we are very likely in the right-hand upper quadrant,
but it is striking that such interesting physics lurks not far away
and is contained in the LO potential.

\subsubsection{Properties of pseudo-Nambu-Goldstone bosons at leading order}

Assuming positive quark masses so that $\langle\Sigma\rangle=1$,
we can study pion properties by inserting eq.~(\ref{eq:expandSigma})
into $\CL^{(2)}$ and expanding:
\begin{eqnarray}
\CL^{(2)}_\chi
&=& 
\tr(\partial_\mu \Pi \partial_\mu \Pi) 
+ \frac{1}{3f^2} \tr([\Pi,\partial_\mu \Pi] [\Pi,\partial_\mu \Pi])
+ O([\partial\Pi]^2\Pi^4)
\nonumber\\
&&\ \  
+ 2 B_0\; \tr(M \Pi^2) 
       - \frac{2B_0}{3f^2} \tr(M\Pi^4) 
       + O(M\Pi^6) .
\label{eq:L2exp}
\end{eqnarray}
We can now understand the choice of factors multiplying the kinetic
term in (\ref{eq:L2first}): they are chosen so that
the pion kinetic term is correctly normalized
[using $\tr(T^a T^b) = \delta^{ab}/2$].
If $M=0$ [the first line of (\ref{eq:L2exp})], 
the pions are massless as required by Goldstone's theorem,
and their interactions all involve derivatives,
as exemplified by $(\partial\Pi)^2\Pi^2$ term shown. 
Note that the non-abelian nature of the group
allows there to be interaction terms with two 
(as opposed to four) derivatives,
in contrast to the $U(1)$ example described above.
This interaction term is non-renormalizable, as are
subsequent terms involving more pion fields, with
the dimensions balanced by factors of $f$.

Including $M$, the pions become massive pseudo-Goldstone bosons
(PGB). \chpt\ predicts at LO that the pion mass
{\em squared} is proportional to the quark mass.
The explicit form is simplest for degenerate quarks: {$m_\pi^2 = 2B_0 m_q$}.
This answers the following potential puzzle: how can physical quantities
like $m_\pi^2$ be related to scheme- and scale-dependent 
quantities like $m_q$? The answer is that $B_0$ cancels the scheme dependence
in $m_q$. This works for all the terms in the second line of 
(\ref{eq:L2exp}),
as they contain the common factor $B_0 M$. For this reason it is useful
to give the combination a name, specifically  $\chi_q=2B_0 m_q$.

The spurion terms also give rise to
higher-order, non-renormalizable interactions among pions.
Note that all vertices in $\CL^{(2)}$ contain an even 
number of pions,
because $\CL^{(2)}$ is invariant under $\Pi(x) \leftrightarrow - \Pi(x)$.
This is an accidental symmetry, which does not correspond to a symmetry
of QCD (note that it differs from parity), and is broken by NLO terms
in \chpt. More precisely, it is broken by the WZW term, which allows,
for example, interactions between five PGB.

 LO \chpt\ makes a number of predictions. First, it is clear that
once $f$ and $\chi$ have been determined (e.g. from PGB masses and
scattering amplitudes), all higher order vertices are predicted.
But there are also predictions from the structure of the
quadratic and quartic interactions alone.
The former give relations between PGB masses, that latter between
pion scattering in different channels (e.g. $I=0,1,2$ in the two-flavor
theory). I will discuss the mass relations here.


First we need to place the physical particles in the pion matrix $\Pi$.
This can be done using the vector symmetry 
corresponding to diagonal phase rotations, 
$U(1)_{V,u}\times U(1)_{V,d}\times U(1)_{V,s} \in U(3)_V$,
which is unbroken by quark masses, and under which the pion
field transforms linearly: $\Pi \to U_V \Pi U_V^\dagger$.
The $\pi^\pm$, $K^\pm$, $K^0$ and $\overline{K^0}$ are charged 
under these symmetries\footnote{%
In the following, I refer to all such particles as ``charged'',
having in mind this generalized definition rather than electric charge.}
(e.g. the $\pi^+$ has u-ness $+1$ and d-ness $-1$),
and so live in definite off-diagonal positions in $\Pi$.
Isospin then determines the position of the $\pi_0$, and orthogonality
that of the $\eta$. Including normalizations, the result is
 \begin{equation}
\Pi = \left(\begin{array}{ccc}
\frac{\pi^0}{2}+\frac{\eta}{\sqrt{12}} & 
                     \frac{\pi^+}{\sqrt2} & \frac{K^+}{\sqrt2}\\
\frac{\pi^-}{\sqrt2} & -\frac{\pi^0}{\sqrt2}+\frac{\eta}{\sqrt{12}} & 
                                               \frac{K^0}{\sqrt2}\\
\frac{K^-}{\sqrt2} &\frac{\overline{K^0}}{\sqrt2} &
-\frac{2\eta}{\sqrt{12}}
	    \end{array}\right)  .
 \end{equation}
Inserting this into {$-2B_0 \tr(M\pi^2)$}, we find that
charged particle masses are proportional to the average
mass of the quarks they contain: 
{ $m_{q_i q_j}^2 = B_0(m_i+m_j)$, $i\ne j$}.
While there are no predictions (masses of three pairs of CPT conjugate mesons
are given in terms of three quark masses),
one can determine quark mass ratios from the experimental PGB masses, e.g.
\begin{equation}
\frac{m_{K^+}^2+m_{K^0}^2}{2 m_{\pi^+}^2} = 
\frac{m_\ell+m_s}{2 m_\ell} 
\approx 13 
\qquad \left(m_\ell = \frac{m_u+m_d}{2}\right)\,.
\end{equation}
This implies $m_s/m_\ell\approx 25$, up to NLO \chpt\ 
and electromagnetic (EM) corrections. This is how we know
that the strange quark is so much heavier than the up and down quarks.
The corresponding determination of $(m_u-m_d)/m_\ell$
(or equivalently $m_u/m_d$)
from $(m_{K^+}^2-m_{K^0}^2)/m_{\pi^+}^2$ fails, however,
since the NLO corrections 
($\Delta(m_u/m_d) \propto m_s/\Lambda_{\rm QCD}$\cite{KaplanManohar})
and EM contributions are potentially as large as the LO 
term in \chpt. 
A determination of $m_u/m_d$ can be achieved
by a direct lattice calculation of the
meson masses using non-degenerate quarks together
with an estimate of the EM contributions.
The most accurate results at present\cite{MILCfpi}
are $m_u/m_d=0.43(8)$ and $m_s/m_\ell=27.4(4)$.

The first predictions of \chpt\ occur in the neutral sector.
The $\pi^0$ and $\eta$ mix, but with an angle
 {$\theta\sim(m_u-m_d)/m_s$} that (despite the uncertainty
in $m_u/m_d$) we know to be very small. Thus
\begin{eqnarray}
m_{\pi^0}^2 &=& m_{\pi^+}^2 + O(\theta^2 m_K^2) + \dots \,,
\\
\underbrace{m_\eta^2}_{ (548\ \textrm{MeV})^2} &=& 
\underbrace{(2[m_{K^+}^2+m_{K^0}^2]-m_{\pi^+}^2)/3}_{ (566\ \textrm{MeV})^2} 
+ O(\theta^2 m_K^2) +\dots\,.
\end{eqnarray}
Both predictions are well satisfied. The first,
that of approximate isospin symmetry for the pions, 
holds not because $m_u/m_d \approx 1$ (which is not true),
but because $(m_u-m_d)/\LQCD\ll 1$.
The second is the famous Gell-Mann--Okubo (GMO) relation.

\subsubsection{Lessons for lattice simulations}

(I) Leading order \chpt\ works to $\sim 10\%$ in GMO relation,
despite the fact that this is a three-flavor relation involving
the strange quark.
This gives hope that three-flavor \chpt\ can be used to extrapolate
from the lattice simulations that are presently being undertaken,
which have
{$m_s^{\rm phys}/2 \gtapprox m_{\ell}^{\rm lat} \gtapprox 2 m_\ell^{\rm phys}$}
and $m_s^{\rm lat}\approx m_s^{\rm phys}$.
The issue is whether the NLO corrections are generically this small,
and I return to this below.

\noindent
(II) Assuming the validity
of \chpt, the ratio $m_{\pi^+}^2/m_q$ 
determines the {\em physical} $B_0$ (in whatever scheme the quark mass
is defined in) {\em even when the quarks are degenerate
and have  masses differing from (usually larger than) their physical values.}
Such a determination is an example of obtaining physical results
from simulations with unphysical parameters. It works as long
as there are $N=3$ dynamical quarks: $B_0$
(like all LECs) depends on $N$, and so one must
simulate with the same number as in QCD. Of course, one also needs
NLO corrections in \chpt\ to be small.

\noindent
(III) The isospin limit,
$m_u=m_d$, is close to physical QCD. Working in this limit
simplifies simulations, e.g. by reducing the
number of adjustable parameters, and by canceling disconnected
contributions to neutral correlators such as in the $\pi^0$
propagator: \\
\vspace{-.5truecm}
\begin{figure}[h]
\begin{center}
\includegraphics[width=8cm]{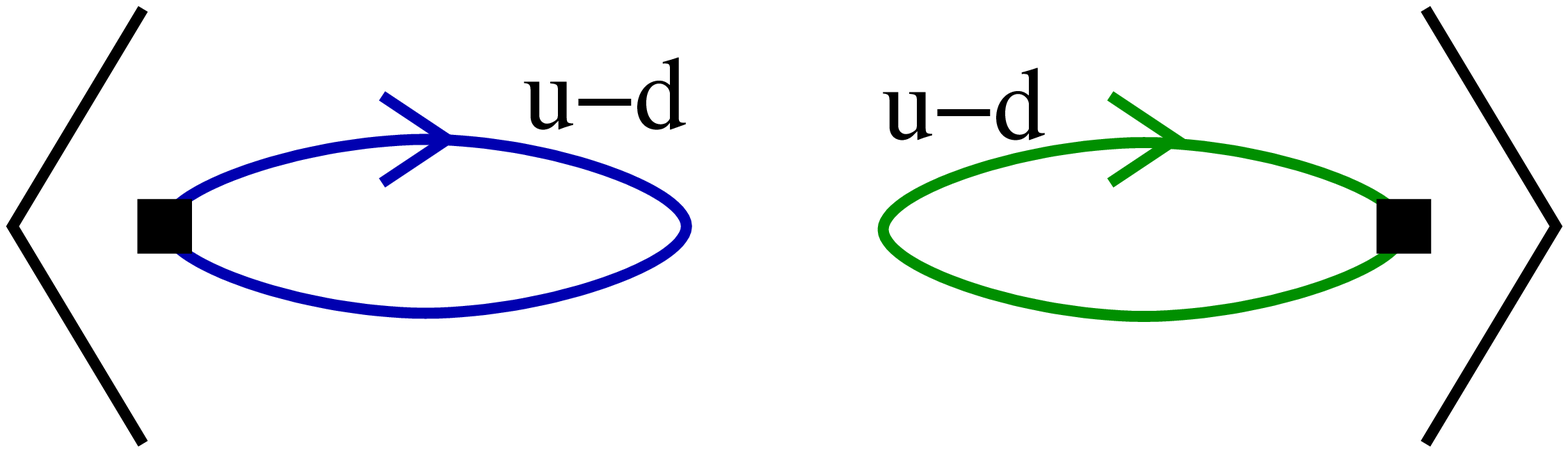}
\end{center}
\end{figure}
\vspace{-.5truecm}

\noindent
The error one makes in hadron masses by setting $m_u=m_d$ is generically
$\sim (m_u-m_d)/\LQCD\sim 1\%$,  comparable 
to those from EM contributions.
Of course, once one can attain 1\% accuracy in the isospin
limit, further improvement requires the
calculation of disconnected and EM contributions

\subsubsection{Power counting in \chpt\ ($M=0$)}\label{sec:powercont}

I now turn to the questions of power counting
and predictivity of non-renormalizable theories: how are contributions
ordered, and by what factor are higher order terms suppressed?
As already noted, the ordering turns out to be in powers of momenta-squared
and mass insertions, so the NLO effective Lagrangian,
$\CL^{(4)}$, contains the terms proportional to
$\partial^4$, $\partial^2 M$ and $M^2$
enumerated above. 
Setting $M=0$ to simplify discussion, and expanding, one finds, schematically:
\begin{eqnarray}
\CL^{(2)} &\sim& f^2 \tr(L_\mu L_\mu) 
\sim (\partial\Pi)^2 + \frac{\Pi^2(\partial\Pi)^2}{f^2} + \dots
\\
\CL^{(4)} &\sim& L_{GL}\tr(L_\mu L_\mu)^2 + \dots
\sim L_{GL}\left[\frac{(\partial\Pi)^4}{f^4} + 
                 \frac{\Pi^2(\partial\Pi)^4}{f^6}\right] + \dots
\,,
\end{eqnarray}
where $L_{GL}$ are unknown dimensionless LECs,
first enumerated by Gasser and  Leutwyler\cite{GL}.
Taking $\pi\pi$ scattering as an example
(with, say, dimensional regularization to avoid power divergences),
the contributions up to quartic in momenta are:
\vskip 1.0cm
\begin{minipage}{1cm}
$\CL^{(2)}_{\rm tree}$:
\end{minipage}
\begin{minipage}{2cm}
\begin{center}
\includegraphics[width=2cm]{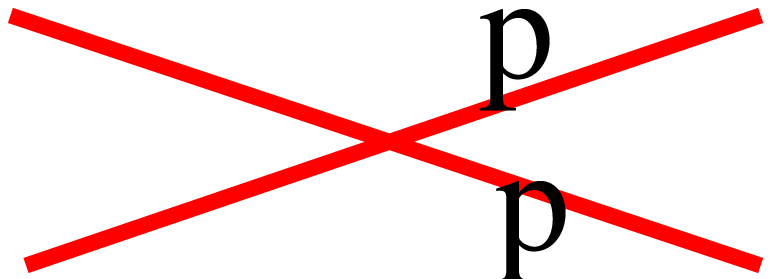}
\end{center}
\end{minipage}
\begin{minipage}{1cm}
{\small $\sim \frac{p^2}{f^2}$}
\end{minipage}
\hfill
\begin{minipage}{1cm}
$\CL^{(4)}_{\rm tree}$:
\end{minipage}
\begin{minipage}{2cm}
\begin{center}
\includegraphics[width=2cm]{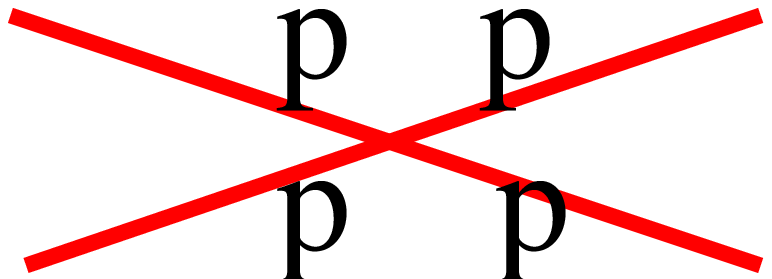}
\end{center}
\end{minipage}
\begin{minipage}{2cm}
{\small $\sim L_{GL} \left(\frac{p^2}{f^2}\right)^2$}
\end{minipage}
\vskip 1.0cm
\begin{minipage}{1.5cm}
$\CL^{(2)}_{\rm 1-loop}$:
\end{minipage}
\begin{minipage}{5cm}
\begin{center}
\includegraphics[width=4.5cm]{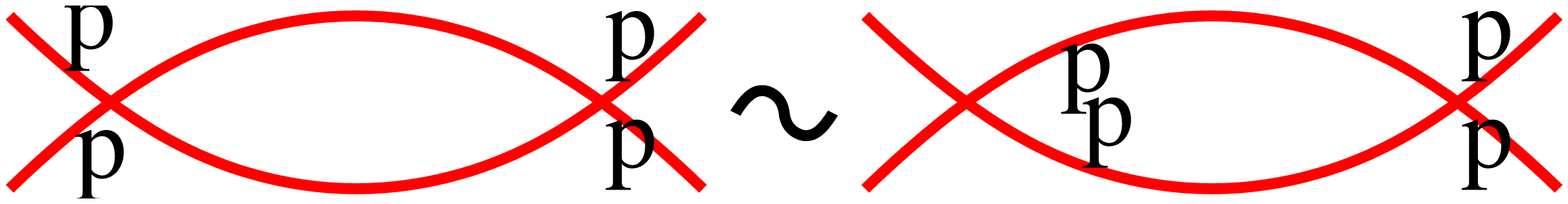}
\end{center}
\end{minipage}
\begin{minipage}{3.cm}
{$\sim \left(\frac{p^2}{f^2}\right)^2 \frac{\ln(p^2/\mu^2)}{(4\pi)^2}$}
\end{minipage}
\vskip 1.0cm

\noindent
where I have shown only a representative loop diagram, and
the positions of the ``p's'' in the diagrams indicate whether
they refer to external or loop momenta.
Contributions from $\CL^{(4)}$ at tree-level,
and $\CL^{(2)}$ at one-loop,
are proportional to $p^4$ (up to logs), where $p$ is a generic
external momentum. These terms are suppressed relative to the tree-level
contribution from $\CL^{(2)}$ by $p^2/f^2$ (up to logs).

It is straightforward to generalize from this example to a power-counting
scheme, using the fact that
each pion field brings with it a factor of $1/f$.
One finds that, for all processes involving PGB, the expansion parameters
are (up to logs) $p^2/f^2$ and
(if we reintroduce $M$) $\chi/f^2 \sim m_\pi^2/f^2$.
At LO contributions come from $\CL^{(2)}_{\rm tree}$,
at NLO from $\CL^{(4)}_{\rm tree}$ and $\CL^{(2)}_{\rm 1-loop}$,
and at NNLO from 
$\CL^{(6)}_{\rm tree}$, 
$(\CL^{(2)}+L^{(4)})_{\rm 1-loop}$
and $\CL^{(2)}_{\rm 2-loop}$, etc..
These are, respectively, ``trivial'', ``easy'' and ``hard''
to calculate (though the latter is done), 
with NNNLO being ``very hard''.

We can now understand the nature of predictions from \chpt.
At each order there are a finite number of LECs (2 at LO, 10 at NLO,
90 at NNLO). We pick an order to work at, say NLO.
We determine the LECs from the appropriate number of physical quantities,
and then make predictions for all other quantities. These predictions
will be accurate up to truncation errors (of NNLO size in our example).
In the continuum, these errors can only be estimated. On the lattice,
we can attempt to fit them as part of the extrapolations.

The 
example discussed above also illustrates
the general relationship between terms analytic and non-analytic in $p^2$
and $M$. 
Non-analytic terms (often called ``chiral logs'') come from pion loops
and involve lower order vertices
(e.g. LO vertices in the NLO calculation of the example).
In particular, they do not involve the LECs of the order being worked at,
and in this sense are predictions. 
They do, however, depend on the renormalization scale $\mu$,
as exemplified by the $p^4\ln\mu$ term in the example.
A physical result cannot depend on $\mu$, and indeed
the dependence can be canceled by renormalizing
the LECs: {$L_{GL} \to L_{GL}(\mu)$}.
Renormalization theory shows that all divergences can be removed in this
way as long as all terms consistent with the symmetries
are included in $\CL_\eff$, which was, of course, our starting point.
Thus the LECs can be thought of as counterterms which
contain our ignorance about the short distance parts of
loops, crudely speaking the parts with $|p|>\mu$. It is then
natural to set $\mu\approx m_{\rm had}$, since $m_{\rm had}\approx 1\;$GeV
is the scale above which we know the EFT to be ineffective.

One can argue for a natural value for this ``cut-off'' scale from
within \chpt\ itself, and at the same time estimate the LECs at this scale. 
The above example should make plausible the
result that the LECs satisfy renormalization group equations of
the form {$d L_{GL}/d\ln(\mu) = O(1) \times (4\pi)^{-2}$}.
The $(4\pi)^{-2}$ comes from the four-dimensional loop integral.
Now we do not know the value of the $L_{GL}$ at the natural
matching scale $\mu\sim m_{\rm had}$, nor do we know what this matching
scale is within a factor of two or so.
But we do know how the LECs change with $\mu$,
{$|L_{GL}(2\mu) - L_{GL}(\mu) |\approx 1/(4\pi)^2$},
so that even if $L_{GL}\approx 0$ at a possible matching scale,
it would be of order $1/(4\pi^2)$ at another.
This motivates assuming that the natural size is $|L_{GL}(\mu)| \approx 1/(4\pi)^2$.
This can be generalized into a self-consistent scheme to all orders.
Of course, this argument does not rule out LECs with larger magnitudes,
but it turns out that those LECs that have been determined from experiment
have the expected magnitude.

If we make this assumption, then the NLO contributions are
suppressed relative to the LO by $L_i p^2/f^2 \sim p^2/\Lambda_\chi^2$
(and $m_\pi^2/\Lambda_\chi^2$ when we include $M$), with
$\Lambda_\chi=4\pi f$. We will shortly see that $f\approx f_\pi$, 
in which case $\Lambda_\chi = 1.2\;$GeV. This value
is completely consistent with our original
expectation that \chpt\ would break down for $|p|\gtapprox m_{\rm had}$.

\subsubsection{Lessons for lattice simulations (continued)}
(IV)
We can use \chpt\ to extend the reach of lattice calculations to multiparticle
processes. Simulations can only access properties of single or
two particle states, with the latter being a challenge away from
threshold. However we can use lattice calculations to determine the LECs,
and then use \chpt\ to calculate inaccessible processes.
For example, one can determine {${\mathcal A}(K\to\pi\pi)$} using
unphysical, but more accessible,  matrix elements. 

\noindent
(V)
The downside of relying on \chpt\ is the inevitable presence of
a truncation error. If we want to estimate this error, we need
fits including the next higher order terms. For example, to 
reliably determine the 
$L_{GL}$ discussed above, one must do fits including, at least 
approximately, NNLO terms. In fact, the full NNLO expression 
in PQ\chpt\ is available for PGB masses and decay 
constants\cite{twoloopPQ}.

\subsubsection{Technical aside: adding sources}
\label{sec:sources}

Electroweak currents can be used to probe PGBs.
For example, the weak leptonic
decay $\pi\to\ell \bar\nu$  is proportional to the matrix element
of the axial current,
$\langle 0 | A_\mu | \pi\rangle$, itself proportional to the pion decay
constant $f_\pi$. It is thus important to incorporate vector and
axial currents into the \chpt\ framework. Other operators are of interest too,
and I discuss here the scalar and pseudoscalar densities.
The aim is to map such operators at the quark level into operators
in the EFT in such away that matrix elements agree, up to truncation
errors in \chpt. This mapping should be based only on the symmetries of
the operators and the action.

A convenient method for doing the mapping is to start with the QCD action 
including position-dependent sources for left and right-handed currents ($l_\mu$ and
$r_\mu$, respectively), and for the densities ($s$ and $p$):
\begin{equation}
\CL_{\rm QCD} = \overline Q_L (\Dslash -i \gamma_\mu{l_\mu})Q_L
+ \overline Q_R (\Dslash -i \gamma_\mu{r_\mu}) Q_R
+ \overline Q_L({s}+i{p}) Q_R
+ \overline Q_R({s}-i{p}) Q_L
\,.
\label{eq:LQCDsources}
\end{equation}
Functional derivatives of the partition function 
$Z_{\rm QCD}({l_\mu,r_\mu,s,p})$ with respect to
the sources (which are hermitian matrices in flavor space)
bring down the corresponding currents and densities.
Note that $s$ and $p$
are just a (position-dependent) rewriting of the mass spurion, $M\to s+i p$.
After derivatives with respect to the sources have been taken, they are
to be set to zero, except for $s+ip$, which is set to $M$.\footnote{%
For the standard diagonal mass term, $s$ is set to $M$ and $p$ to zero.}

One nice feature of $Z_{\rm QCD}({l_\mu,r_\mu,s,p})$ is that it represents
what one can actually calculate in QCD---correlation functions of
gauge-invariant operators. It allows one to access PGB physics,
because the axial currents (and pseudoscalar densities) couple to pions.
For example, a $4$-point correlator of axial currents, analytically
continued to Minkowski momenta, can be amputated and then evaluated
at the pole positions so as to obtain the pion scattering amplitude 
using the LSZ formula. Similarly one can obtain
matrix elements of currents and densities between PGBs (or other states).

Thus it is natural to phrase the matching of QCD with the EFT in terms of
this partition function\cite{GL}
\begin{equation}
{Z_{\chi}({l_\mu,r_\mu,s,p})
=  Z_{\rm QCD}({ l_\mu,r_\mu,s,p})} +\textrm{truncation errors}
\quad ({ p_\pi,m_{\pi}\ll\Lchi})
\,.
\label{eq:Zmatching}
\end{equation}
This simply says that the correlation functions
of the two theories must match for momenta such that PGBs
give the dominant contributions, up to errors due to the truncation
of \chpt. The matching  ensures that masses,
scattering amplitudes and matrix elements agree in the two theories.

This is nice packaging, but to give it content we need to learn how
to add sources to $Z_\chi$.
This can be done by generalizing the spurion method to enforce
a {\em local} {$SU(3)_L\times SU(3)_R$} symmetry.
{$\CL_{\rm QCD}$} in (\ref{eq:LQCDsources})
is invariant if {$l,r_\mu$} transform as gauge fields,
\begin{equation}
{l_\mu} \to U_L {l_\mu} U_L^\dagger 
+ i U_L \partial_\mu U_L^\dagger\,,\ \ 
{r_\mu} \to U_R {r_\mu} U_R^\dagger 
+ i U_R \partial_\mu U_R^\dagger\
\end{equation}
(with $U_{L,R}=U_{L,R}(x)$),
while $s$ and $p$ transform as before,
$\chi = 2B_0(s + i p) \to U_{L} \chi U_R^\dagger$.
Here I have introduced the convenient matrix $\chi$.
Note that this invariance only works because the left and right-handed
currents are Noether currents for the corresponding symmetries.
One would like to argue that, in order to satisfy
eq.~(\ref{eq:Zmatching}), $\CL_\chi$ must also be invariant
under local chiral transformations.
This turns out to be correct, 
but the argument, given by Leutwyler\cite{Leutwyler}, is subtle.
First one must deal with anomalies:
$Z_{\rm QCD}$ is not invariant under local chiral transformations,
but has a variation which is a known functional of the sources alone.
This must be matched by the variation in $Z_\chi$,
which in turn requires
the presence of the WZW term in $\CL_\chi$.
This allows the remainder of $\CL_\chi$ to be invariant,
but this is not automatic. Leutwyler shows that it
is possible to bring it to an invariant form
if one uses the freedom of
adding total derivatives (which do not change the action)
and changing variables.

The invariance of {$\CL_\chi$} under local transformations
can be accomplished in the usual way: by replacing normal derivatives
with covariant derivatives which transform homogeneously, e.g.
\begin{equation}
{D_\mu \Sigma = \partial_\mu\Sigma - i{ l_\mu}\Sigma + i\Sigma{r_\mu}
\to U_L (D_\mu \Sigma) U_L^\dagger}\,.
\label{eq:ChPTcovder}
\end{equation}
Note that this fixes the normalization of the {$l,r_\mu$} terms,
implying that the normalization of the currents in the EFT is 
known.\footnote{%
One could also obtain the correct normalization in the absence of
symmetry breaking by matching the Noether currents in the two theories.}
This is possible because the currents generate a non-abelian group,
whose algebra is schematically $[L,L]\sim L$, in which normalizations are fixed.
Terms without derivatives, e.g. mass terms, are automatically invariant
under the local symmetry, so do not need to be changed.
The inclusion of sources also allows new types of terms in $\CL_\chi$,
as will be seen shortly.

\subsubsection{Final form of chiral Lagrangian}
\label{sec:final}

We now have the ingredients to construct the EFT through NLO
including sources. The LO result 
differs from the earlier form (\ref{eq:L2first}) 
only by $\partial_\mu\to D_\mu$  and $2 B_0 M\to \chi$:
\begin{equation}
\CL^{(2)}_\chi = 
\frac{f^2}{4} \tr\left(D_\mu\Sigma D_\mu \Sigma^\dagger\right)
-
\frac{f^2}{4} \tr(\chi \Sigma^\dagger + \Sigma \chi^\dagger ) \,,
\label{eq:L2}
\end{equation}
The addition of the sources allows us to
match currents with QCD, e.g. equating
{$\delta Z/\delta { l_\mu(x)}|_{ l=r=p=0, s=M}$} 
in QCD and \chpt\ implies
\begin{equation}
\overline Q_L \gamma_\mu T^a Q_L \simeq
(i f^2/2) \tr(T^a \Sigma \partial_\mu \Sigma^\dagger)
=-(f/2)\partial_\mu \pi^a + \dots\,.
\end{equation}
This result can also be obtained using the Noether procedure.
It allows us to determine $f$: from the vacuum to pion matrix
element one finds ${ f} = f_\pi\approx 93\;$MeV.
Similarly, we can relate $B_0$ to quark-level quantities.
Taking {$\delta Z/\delta { s(x)}|_{{ l=r=p=0, s}=M}$} gives
\begin{equation}\overline Q Q \simeq
-(f^2 B_0/2) \tr(\Sigma + \Sigma^\dagger)
=-N f^2 B_0 + O(\pi^2)\,, 
\end{equation}
the VEV of which gives $\langle \overline q q\rangle = -{ f^2 B_0}$
(with $q=u$, $d$, or $s$).
This is a place where the lattice can contribute, since
the condensate is not experimentally measurable but
can be directly calculated on the lattice.
Alternatively, since the PGB masses give the combination $B_0 m$, 
as seen above, a lattice determination of the quark masses 
gives another result for $B_0$. Comparing the two determinations
tests the accuracy of \chpt\ at LO.

At NLO the addition of sources leads to new terms. For three flavors,
the enumeration in sec.~\ref{sec:blocks} gave
8 independent terms plus the WZW term. Sources allow
two additional terms involving the field-strength tensors
$L,R_{\mu\nu}$ (constructed from the source gauge fields $l,r_{\mu}$
in the usual way), as well two terms involving sources alone.
The latter are multiplied by so-called 
high-energy coefficients (HECs). In sum, one has
\begin{eqnarray}
\CL^{(4)} &=& 
- { L_1}\, \big[\tr(D_\mu \Sigma D_\mu \Sigma^\dagger)\big]^2
- { L_2}\, \tr(D_\mu \Sigma D_\nu \Sigma^\dagger)
      \tr(D_\mu \Sigma D_\nu \Sigma^\dagger) 
\nonumber\\ &&
+ { L_3}\, \tr(D_\mu \Sigma D_\mu \Sigma^\dagger
      D_\nu \Sigma D_\nu \Sigma^\dagger)
\nonumber\\ &&
+ { L_4}\, \tr(D_\mu \Sigma^\dagger D_\mu \Sigma)
         \tr(\chi^{\dagger} \Sigma +  \Sigma^\dagger\chi)
+ { L_5}\, \tr(D_\mu \Sigma^\dagger D_\mu \Sigma
         [\chi^{\dagger} \Sigma +  \Sigma^\dagger\chi])
\nonumber\\ &&
- { L_6} 
\big[\tr(\chi^{\dagger} \Sigma \!+\! \Sigma^\dagger\chi)\big]^2
-\! { L_7} \big[\tr(\chi^{\dagger} \Sigma \!-\! \Sigma^\dagger\chi)\big]^2
- { L_8}\, \tr(\chi^\dagger\Sigma\chi^\dagger\Sigma \!+\! { p.c.})
\nonumber\\ &&
+ { L_{9}}\, i\tr(L_{\mu\nu} D_\mu \Sigma D_\nu \Sigma^\dagger + p.c.)
+ { L_{10}}\, \tr(L_{\mu\nu} \Sigma R_{\mu\nu} \Sigma^\dagger) 
\nonumber\\ &&
+ { H_1}\, \tr(L_{\mu\nu} L_{\mu\nu} + p.c.)
- { H_2}\, \tr(\chi^\dagger\chi) + \CL_{WZW}
\,,
\label{eq:L4}
\end{eqnarray}
There is also a term {$\propto \tr(D_\mu \chi^\dagger D_\mu \Sigma)$},
which contributes only to the mass dependence of the currents,
but it can be removed by a change of variables for $\Sigma$,
and is thus redundant. 
The dimensionless {$L_i$} are the well-known ``Gasser-Leutwyler coefficients''.
A subset of them can be determined experimentally to good accuracy and 
a different subset is straightforward to determine on the lattice, as
illustrated below.

What about the HECs, $H_{1,2}$? Since these multiply terms which do not involve
$\Sigma$ they give contact terms in correlation
functions (e.g. $H_2$ gives a contribution to 
$C_P(x)=\langle \bar Q \gamma_5 T^a Q (x) \bar Q\gamma_5 T^a Q(0)\rangle$
proportional to $H_2 \delta(x)$,
and also contributes to the mass dependence of the condensate).
Thus they do not contribute to physical scattering amplitudes or
matrix elements. Why does one need them? They are required
if one wants to describe the mass dependence of the condensate
(which can be calculated on the lattice) within \chpt,
and in order that quark level Ward identities are satisfied
(e.g. $m \int d^4x C_P(x) \propto \langle \bar q q\rangle$).\footnote{%
One might be concerned that, because the short-distance behavior
in QCD and \chpt\ are different (operators having different dimensions
in the two theories), the whole notion of matching the partition
functions with sources is flawed. How can correlators be matched when
the operator product expansions differ? 
The answer is that the matching is done only for $|p|<\Lambda_\chi$,
so one avoids the short-distance regime in the underlying theory.}
For the matching to work, the HECs must depend on the regulator used
for QCD. For example, if QCD is regulated using a lattice, then
$H_2 \propto 1/a^2$ because it must reproduce
$\langle\bar q q\rangle\propto m/a^2$. 
This is in stark distinction to the LECs which do not depend
on the underlying regulator (leading to their different name),
and are physical parameters.

\subsection{Examples of NLO results}

With $\CL^{(2,4)}$ in hand, it is straightforward to calculate NLO
results for physical quantities. For example, the pole
in the two-point function of the left-handed current gives the PGB mass,
while the residue is proportional to $f_{\rm PGB}^2$ (just as in lattice
simulations). Recall that the LO result is 
$m_{\rm PGB,0}^2= (\chi_{q1} + \chi_{q2})/2 = 2 B_0 (m_{q1}+m_{q2})/2$.
At NLO, there is a tree-level contribution
\begin{center}
\begin{minipage}{1.5cm}
 $\delta m_{\rm PGB}^2 \sim$
\end{minipage}
\begin{minipage}{2cm}
\begin{center}
\includegraphics[height=0.6cm]{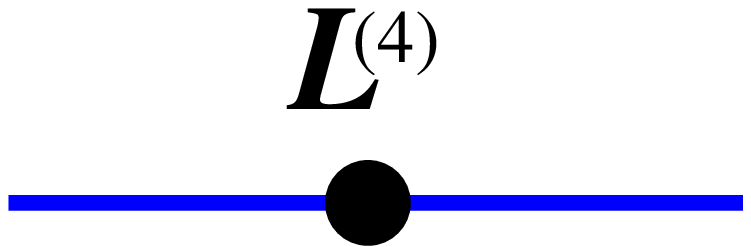}
\end{center}
\end{minipage}
\begin{minipage}{5cm}
$\sim \chi\, {L}\, \frac{\chi}{f^2} \sim 
\chi { (16 \pi^2 L)} \frac{m_{\rm PGB,0}^2}{\Lchi^2}$, 
\end{minipage}
\end{center}
and a 1-loop contribution 
\begin{center}
\begin{minipage}{1.5cm}
 $\delta m_{NG}^2 \sim$
\end{minipage}
\begin{minipage}{1.5cm}
\begin{center}
\includegraphics[width=1.5cm]{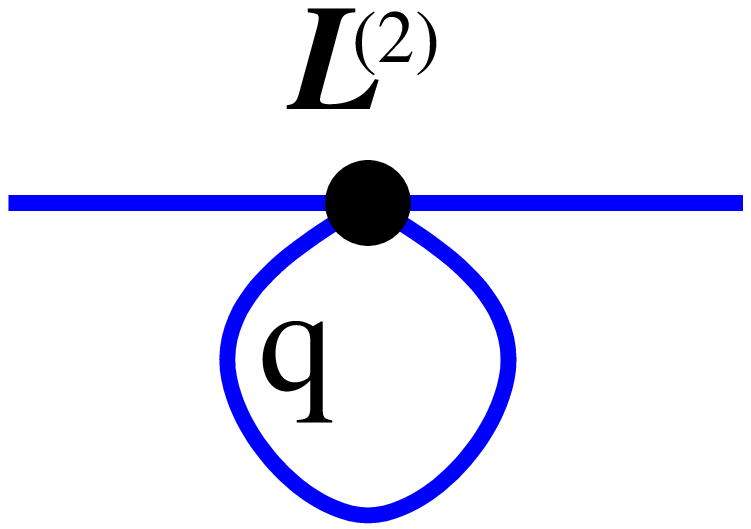}
\end{center}
\end{minipage}
\begin{minipage}{7cm}
$ \sim \frac{\chi}{f^2} \int_q \frac1{q^2+m_{\rm PGB,0}^2}
 \sim \chi \frac{m_{\rm PGB,0}^2}{\Lchi^2}
\ln\left(\frac{m_{\rm PGB,0}^2}{\mu^2}\right)$ .
\end{minipage}
\end{center}

I quote the final result for the $\pi^\pm$ as an example\cite{GL}:
\begin{equation}
m_{\pi^\pm}^2 = \chi_\ell \left\{ 1 +
\frac{8}{f^2} [ 
\underbrace{{ (2 L_8\!-\!L_5)} \chi_\ell}_{\textrm{ valence}} 
+ 
\underbrace{{ (2 L_6\!-\!L_4)}(2 \chi_\ell+\chi_s)}_{\textrm{ sea}}]
+ \underbrace{\frac{3 L_\pi\! -\! L_\eta}{6}}_{\textrm{ logs}} \right\} 
\,,
\label{eq:mpipm}
\end{equation}
where $\chi_l=(\chi_u+\chi_d)/2$, and the chiral logs are 
(using dim. reg.)
\begin{equation}
 L_\pi = \frac{m_\pi^2}{\Lchi^2} \ln\left(\frac{m_\pi^2}{\mu^2}\right) \,,
\qquad
L_\eta = \frac{m_\eta^2}{\Lchi^2} \ln\left(\frac{m_\eta^2}{\mu^2}\right) 
\,.
\label{eq:chirallog}
\end{equation}
The $\mu$ dependence is absorbed by the $L_i$, as discussed above.
Looking ahead to the discussion of PQQCD,
I have separated the analytic terms into those that arise
from the masses of the quarks which compose the pion (``valence'')
and those of the quarks in loops (``sea''). How this is done
will be explained in sec.~\ref{sec:PQPQChPT}. Within QCD itself
this separation is not useful, as corresponding 
valence and sea quarks have the same masses.

Another example is the ratio of decay constants, which is
\begin{equation}
\frac{f_K}{f_\pi} = 1 +
\frac{2}{f^2}  
\underbrace{{ (L_5)} (\chi_s-\chi_\ell)}_{\textrm{ valence}} 
+ \underbrace{\frac58 L_\pi -\frac14 L_K - \frac38L_\eta}_{\textrm{ logs}}
\,.
\label{eq:fKbyfpi}
\end{equation}
This result allows $L_5$ to be determined from experiment. The result
depends on the choice of $\mu$, a conventional value being $\mu=m_\rho$.

These results illustrate the general structure at NLO: there are 
corrections analytic in $\chi$ and dependent on the masses of valence 
and sea quarks, and chiral logs that are non-analytic in $\chi$. 
The expansion parameter $m_{\rm PGB}^2/\Lambda_\chi^2$ is clear in the logs,
but is obscured in the analytic terms by the convention for the $L_i$
(which are numerically of size $\ltapprox 1/(16\pi^2)$.)

\subsubsection{Lessons for lattice simulations (continued)}

\begin{figure}[t]
\centerline{\epsfxsize=3.5in\epsfbox{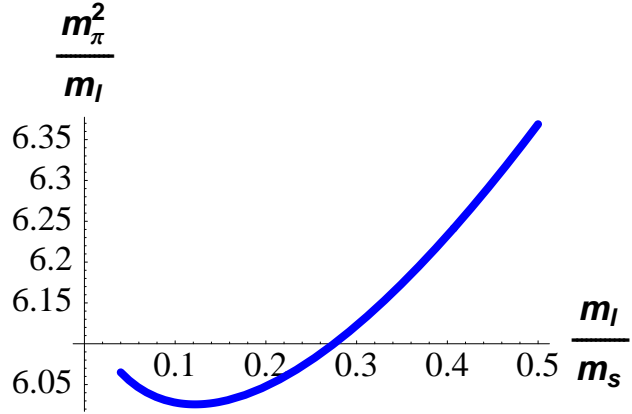}}   
\caption{Typical behavior of $m_\pi^2/m_l$ as a function of $m_l/m_s$ at NLO.}
\label{fig:log}
\end{figure}

\begin{figure}[ht]
\centerline{\epsfxsize=3.5in\epsfbox{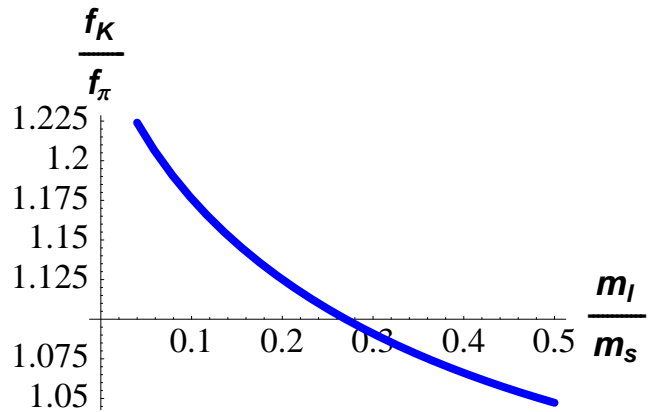}}   
\caption{Typical behavior of $f_K/f_\pi$ as a function of $m_l/m_s$ at NLO.}
\label{fig:fkbyfpi}
\end{figure}

\noindent (VI)
Non-analytic terms become important at small masses.
To illustrate this, I plot $m_{\pi^\pm}^2/m_l$ versus $m_l/m_s$
in Fig.~\ref{fig:log}, with $m_l$ the average light-quark mass.
I hold the strange mass fixed at $m_s=0.08\;$GeV, 
and use values for the LECs that are representative of those from 
\chpt\ analyses: $f=0.093\;$GeV, $L_5=1.45\times 10^{-3}$,
$L_8=10^{-3}$, $L_4=L_6=0$ (with $\mu=m_\rho$ here and below). 
The curve's lower end is approximately the physical point.
At LO the \chpt\ prediction is a constant. The analytic
NLO corrections lead to linear dependence on $m_l$, and the logs to
curvature. Clearly, to obtain 1\% accuracy one must simulate
down to $m_l/m_s\approx 0.1$ in order to see 
and fit to the predicted curvature.\footnote{%
Extrapolation can be simplified 
in some cases by considering ``golden (silver) ratios''
in which the chiral logs completely (partially) cancel\cite{Becirevic}.
By contrast, for some quantities the chiral logs
are enhanced, e.g
$\langle r^2\rangle_\pi \sim \ln(m_\pi^2/\mu^2)$ rather
than $m_\pi^2 \ln(m_\pi^2)$.
}
This has been achieved in the MILC simulations.
In my opinion, seeing curvature consistent with \chpt\ predictions
is a necessary check on lattice techniques.
Similar comments hold for $f_K/f_\pi$, plotted in fig.~\ref{fig:fkbyfpi}
using the same parameters except now $f=0.085\;$GeV (chosen
so as to better match the experimental value at the lower end of the
curve). Here the linear terms are larger, but 
an extrapolation accurate to a few percent still requires inclusion of
the curvature. 

\noindent (VII) 
The results further illustrate the utility of the lattice for
obtaining LECs.
In nature one can vary the valence content by considering
different PGBs, while the sea content is fixed.
Thus $L_4$ and $L_6$ are not accessible using the results above.
On the lattice, however, one can change the sea quark masses
and thus determine these LECs more easily.
Using PQ simulations (varying valence and sea masses independently)
further simplifies the determinations, as will be discussed in
sec.~\ref{sec:PQLs}.

\subsubsection{Volume dependence from \chpt}
\begin{figure}[ht]
\centerline{\epsfxsize=3.5in\epsfbox{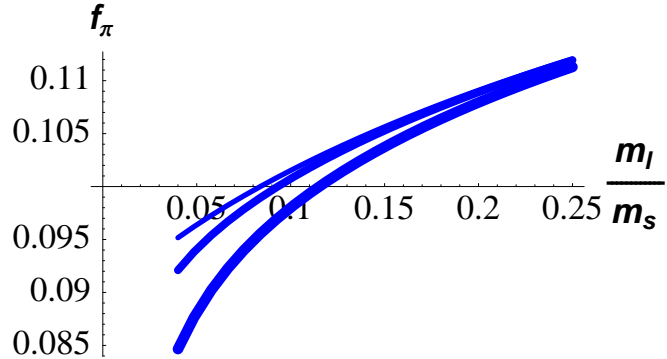}}   
\caption{{$f_\pi$} vs. $m_l/m_s$ at $a=0.1\;$fm, with $L=2.4\;$fm (thick), 
$3.2\;$fm (medium) and $\infty$ (thin), using
$m_s=0.08\;$GeV, $f=0.08\;$GeV, $L_5=1.45\times 10^{-3}$,
$L_4=0$.}
\label{fig:fpifinitevol}
\end{figure}

For single particle matrix elements, PGB loops
loops give the leading correction due to the use
of a finite spatial volume. 
To obtain the leading volume dependence, one simply
replaces the momentum integrals with sums, e.g.
\begin{itemize}
\item[]
\begin{minipage}{1.5cm}
\begin{center}
\includegraphics[width=1.5cm]{mpi_loop.eps}
\end{center}
\end{minipage}
\begin{minipage}{7cm}
$\to \int_q \left(\frac1{q^2+m_{NG}^2}\right) 
\to \int_{q_4}\sum_{\vec q=2\pi \vec n/L}
\left(\frac1{q^2+m_{NG}^2}\right)$\,.
\end{minipage}
\end{itemize}
This is now done routinely when fitting lattice data. 
Figure~\ref{fig:fpifinitevol} illustrates the rapid growth in
finite volume shifts as the quark mass is reduced.

Unfortunately, recent work suggests that an accurate estimate of volume
corrections requires the inclusion of at least the dominant part of the
two-loop contributions\cite{Colangelo}. 
The requisite calculation has only been done
for a few quantities, so for others
one may be forced to work in volumes large enough 
that finite volume effects are negligible. For determining such volumes
one-loop results should be a reliable guide if used conservatively.

If the quark mass is reduced at fixed box size 
satisfying $L \gg 1/\LQCD$,
one eventually enters the so-called ``$\epsilon-$regime''
where $m_{NG} L\ltapprox 1$. Here the pion propagator
is completely distorted by finite volume effects. It turns
out, however, that one can still use \chpt\
to predict the form of correlation functions\cite{epsilonregime}.
I will not discuss this regime further, but note that there
is an ongoing effort to determine the LECs of QCD (including
electroweak interactions) by comparing the results of simulations
in the $\epsilon-$regime to the predictions of \chpt\cite{pilarh}.

\subsubsection{Convergence of \chpt}

I have only scratched the surface of calculations in \chpt, which
have been extended to include the electroweak Hamiltonian
in the PGB sector and to NNLO 
(as reviewed, e.g., by Bijnens\cite{Bijnens}).
Many quantities are relevant for lattice simulations---I give
a list below in sec.~\ref{sec:statusPQChPT} when discussing PQ\chpt.
Here I only discuss what has been learned about the important
question of convergence of \chpt.

I quote one example\cite{Amoros}, obtained from a fit of the $N=3$ NNLO formulae
to a number of experimental inputs, but with
the NNLO LECs estimated approximately
using ``single resonance saturation''.
One of the input quantities is $f_K/f_\pi$,
and this turns out to have a chiral expansion\cite{Bijnens}
\begin{equation}
f_K/f_\pi = 
\underbrace{1}_{\textrm{LO}}
+ \underbrace{0.169}_{\textrm{NLO}}
+ \underbrace{0.051}_{\textrm{NNLO}} \,,
\label{eq:fKfpiexpansion}
\end{equation}
showing reasonable convergence.
The convergence is less good, however, for the PGB masses.

The naive conclusion from this and similar results is
that NNLO terms in \chpt\ are needed for good accuracy.
This is, I think, correct if one does a global fit 
to several quantities using $SU(3)$ \chpt. One might,
in practice, be able to get away with including only
the {\em analytic} terms at NNLO (whose form is easy to determine)
rather than the full two-loop expression. This is the approach
used in the MILC analysis\cite{MILCfpi}.
This amounts to mocking up the two-loop contributions by
changing the NLO and NNLO LECs. While this makes the
results for these LECs
less reliable, I expect it to impact the extrapolated
results for physical quantities only at the level of NNNLO 
corrections.\footnote{%
This is based on the following argument. The dominant NNLO
terms are those involving $m_s^2$, either explicitly
or through factors of $m_K$ or $m_\eta$.
These are of size $(m_K/\Lchi)^4\approx 0.03$ relative to LO terms,
consistent with the result in eq.~(\ref{eq:fKfpiexpansion}). 
(This is to be compared to purely light quark 
NNLO contributions---$(m_\pi/\Lchi)^4\approx 0.0002$---and
mixed light-strange contributions---$m_\pi^2 m_K^2/\Lchi^4 \approx 0.003$.)
The $m_s^2$ terms can involve logarithms of $m_K$ or $m_\eta$, but
not $m_\pi$, since they cannot be singular when $m_u=m_d\to 0$.
It follows that the dominant NNLO logarithms are 
being evaluated far from the meson masses
where they are non-analytic ($m_K,m_\eta=0$), and thus can
be well represented by analytic terms. This will be especially true
if NNNLO LECs are included, as in some MILC fits. The subleading NNLO
logarithms involving $m_\pi$ will be much less well represented
by analytic terms, but these
are numerically smaller than the NNNLO contributions proportional
to $m_s^3$.}
Clearly, though, a full NNLO fit would be preferable.

Another approach which reduces the impact of NNLO terms
is to use $SU(2)$ \chpt\ alone, treating $m_s$ as heavy. 
After all, the actual extrapolation being done in
present simulations is for the light quarks alone, with
$m_s$ fixed near its physical value.
In this approach the kaon and eta are treated as heavy particles,
and one makes no assumption about the convergence
of the expansion in $m_s$. The idea is that this removes
the dominant contribution to the corrections
in (\ref{eq:fKfpiexpansion}) and sums them to all orders.
In practice, this approach has been used primarily in the
baryon sector.

\subsubsection{Extension to ``heavy'' particles}

I will not describe \chpt\ technology for including
heavy particles here, but I do want to mention the
form of the results. ``Heavy'' means $m_{\rm had}\gtapprox \Lambda_\chi$,
and the approach is to expand in $1/m_{\rm had}$ so that at LO the hadron
is a static source for PGBs. In this way one can include the dominant
long-distance physics which gives rise to curvature at small light-quark mass.

The form of the resulting chiral expansion depends on the quantity
considered.
For heavy-light meson decay constants 
it is similar to that for PGB properties,
e.g.
\begin{equation}
F_{B} \sim F_{B,0} (1 +  
\underbrace{m_\pi^2}_{\textrm{analytic}} 
+
\underbrace{m_\pi^2\ln(m_\pi)}_{\textrm{ chiral log}} 
+ \dots)
\end{equation} 
One new feature is that the non-analytic terms are not predicted in terms of the LO
LECs, but involve an additional coefficient, {$g_{\pi B B^*}$}.

For baryons and vector meson masses the expansion differs further,
involving odd powers of $m_{PGB}$.\footnote{%
There is a similar ``$m_\pi^3$'' contribution to heavy-light meson masses but
there the leading term is $M_{\rm heavy-light}\gg\LQCD$,
so the correction is less important.}
\begin{equation}
M_H \sim M_0 + \underbrace{m_\pi^2}_{\textrm{ analytic}} 
+ \underbrace{g_{\pi HH'}\ m_\pi^3}_{\textrm{ leading loop}} 
+ \underbrace{m_\pi^4\ln(m_\pi)}_{\textrm{ subleading loop}}
+ \ m_\pi^4 + \dots 
\end{equation}
This means that
the expansion is in powers of $m_\pi/\Lchi$ 
(c.f. $(m_\pi/\Lchi)^2$ for PGBs and heavy-light mesons),
so that the convergence is generically poorer.
Thus it is even more important to use light quark masses when studying
baryon properties.\footnote{%
For vector mesons, and unstable baryons, the chiral expansion is yet more
complicated because of the opening of the decay channel as the quark mass
is reduced.}

\section{Incorporating discretization errors into \chpt}\label{sec:disc}

In this lecture I describe how for Wilson and twisted-mass fermions 
one can incorporate discretization errors into \chpt,
and what one learns by doing so. The method is general, and
has been applied also to staggered fermions\cite{LeeSh,AB}, 
and to mixed-action theories\cite{BRSMA}.
See also the review by B\"ar\cite{Barlat04}.

\subsection{Why incorporate discretization errors?}\label{sec:whydisc}

At first sight, it may seem strange to incorporate the effects of the 
{\em ultraviolet} (UV) cut-off of
the underlying theory into the EFT describing its {\em infrared} (IR)  behavior.
The key point is that the UV effects break the chiral symmetry which determines
the IR behavior. One way of saying this is that discretization errors
lead to a non-trivial potential in the vacuum manifold which is otherwise flat due to
the symmetry. As we will see, symmetry breaking
due to quark masses and discretization errors
have comparable effects on the potential
if $m/\Lambda\approx (a\Lambda)^2$, with $\Lambda$ a scale of $O(\LQCD)$. 
The appropriate value of $\Lambda$ depends on the action, and it is quite
possible that this condition is satisfied even for relatively fine
lattices and light quarks. For example, if $\Lambda=500\;$MeV 
and $a^{-1}=2\;$GeV, then it is satisfied when $m=30\;$MeV.
Thus it is imperative to study the impact of discretization errors.\footnote{%
As an aside, I note that if one uses lattice fermions with an exact
on-shell chiral symmetry (overlap, perfect or Domain-wall fermions with
$N_5\to\infty$)
then the considerations of this lecture become almost trivial. These fermions are
described by {\em continuum} \chpt, but with LECs that depend on $a^2$ and
must be extrapolated to the continuum limit.
The only exception is that there are additional terms induced by
the breaking of Euclidean symmetry, but these are of very high order
in the meson sector, as discussed below.}
\begin{center}
\begin{minipage}{2.2in}
One question that often arises in the present context is whether
one should first extrapolate $a\to0$ and then use continuum \chpt,
or do combined extrapolation in $a\to0$ and $m\to m_{\rm phys}$.
The possibilities are illustrated to the right.
\end{minipage}
\hfill
\begin{minipage}{2.in}
\centerline{\epsfxsize=2.in\epsfbox{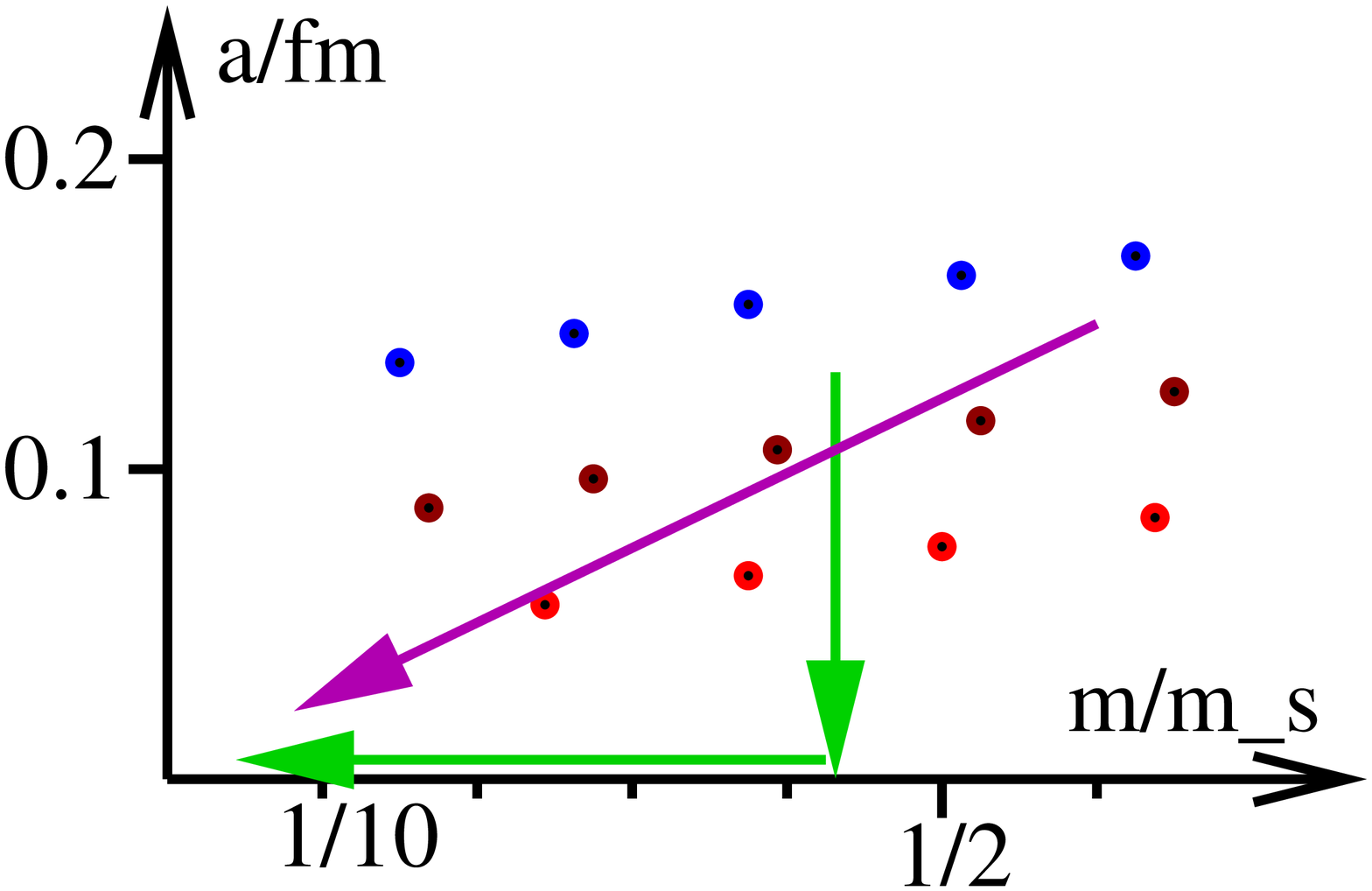}}   
\end{minipage}
\end{center}
An apparent advantage of the first approach is that one does not have to rely
on the validity of \chpt\ to do the continuum extrapolation; 
one simply uses a standard polynomial {\em ansatz}.
There are, however, several reasons to use the second, ``combined'', approach
if \chpt\ formulae are available: 
\begin{itemize}
\item
It incorporates relations between discretization
errors in different quantities that follow from the specific way in which
chiral symmetry is broken.
\item
It accounts for non-analyticities in $a$ which arise because of pion loops
(e.g. for staggered fermions one has, schematically,
$m_\pi^2 \sim m_q [1 + (m_q + a^2) \ln(m_q + a^2) + \dots]$).
These might well be missed in a simple polynomial continuum extrapolation.
The ``$a^2$'' in the chiral  logs reduce the curvature, as clearly
observed in the MILC results\cite{MILCfpi}.
It should be kept in mind, however, that 
``$a^2$'' always means ``up to logs'', so not all non-analyticities are included.
\item
It accounts for changes in orientation of the condensate, which can be rapid
with twisted-mass fermions if $m\sim a^2$, as discussed below.
\end{itemize}

\subsection{General strategy}
One proceeds in two steps\cite{ShSi}. First, following 
Symanzik\cite{Symanzik},
determine the continuum EFT describing the interactions of quarks and gluons
with $|p|\ll 1/a$. Discretization errors enter with explicit factors of $a$,
and are controlled by the symmetries (or lack thereof) of the underlying
lattice theory. Second, use standard techniques to develop \chpt\
for the Symanzik EFT. Since the latter is a continuum theory, this is no
different conceptually from determining the effect of beyond-the-standard-model
physics on the IR properties of QCD. The two steps are illustrated in 
Fig.~\ref{fig:EFT} above.

\subsection{Application to Wilson \& twisted mass fermions}
Twisted mass lattice QCD (tmLQCD)\cite{tm} has received a lot of recent
attention because of its improved algorithmic properties and because,
at maximal twist, physical quantities 
(including matrix elements) are automatically $O(a)$ improved\cite{FR}.
Here it also serves as an excellent example, particularly as it contains
Wilson fermions as a subset.\footnote{%
For more extensive discussion of tmLQCD 
and its applications see the lectures by Sint, and recent
reviews\cite{Shindler,FR05review}.}
%

Twisting the mass in continuum QCD simply means doing
an $SU(3)_L\times SU(3)_R$ rotation. The standard diagonal
mass $M_0$ (which is hermitian assuming real $m_q$) is
rotated into $M=U_L M_0 U_R^\dagger$ which is not hermitian.
The example I consider in detail has two degenerate flavors and 
\begin{equation}
M =  m_q e^{i\tau_3\omega}
= m_q (\cos\omega + i \sin\omega \tau_3)
\equiv m + i\mu \tau_3\,,
\label{eq:standardtm}
\end{equation}
resulting in a Lagrangian mass term containing a $\gamma_5\tau_3$ part:
\begin{equation}
\overline Q_L M Q_R +
\overline Q_R M^\dagger Q_L = \overline Q (m + i \mu\gamma_5\tau_3) Q
\,.
\end{equation}
Note that $m_q$ is the physical quark mass, with $m$ and $\mu$ respectively
the untwisted and twisted components.
\begin{center}
\begin{minipage}{2.5in}
The ``geometry'' of the parameters is shown to the right.
Although it naively appears that parity and flavor are broken,
we know this is not the case since
physics is unchanged by the chiral rotation.
Thus $\omega$ is a redundant parameter.
Usually, we keep the symmetries manifest by working at $\omega=0$,
but it is important to know that we do not need to do so. 
The continuum 
\end{minipage}
\hfill
\begin{minipage}{1.7in}
\centerline{\epsfxsize=1.5in\epsfbox{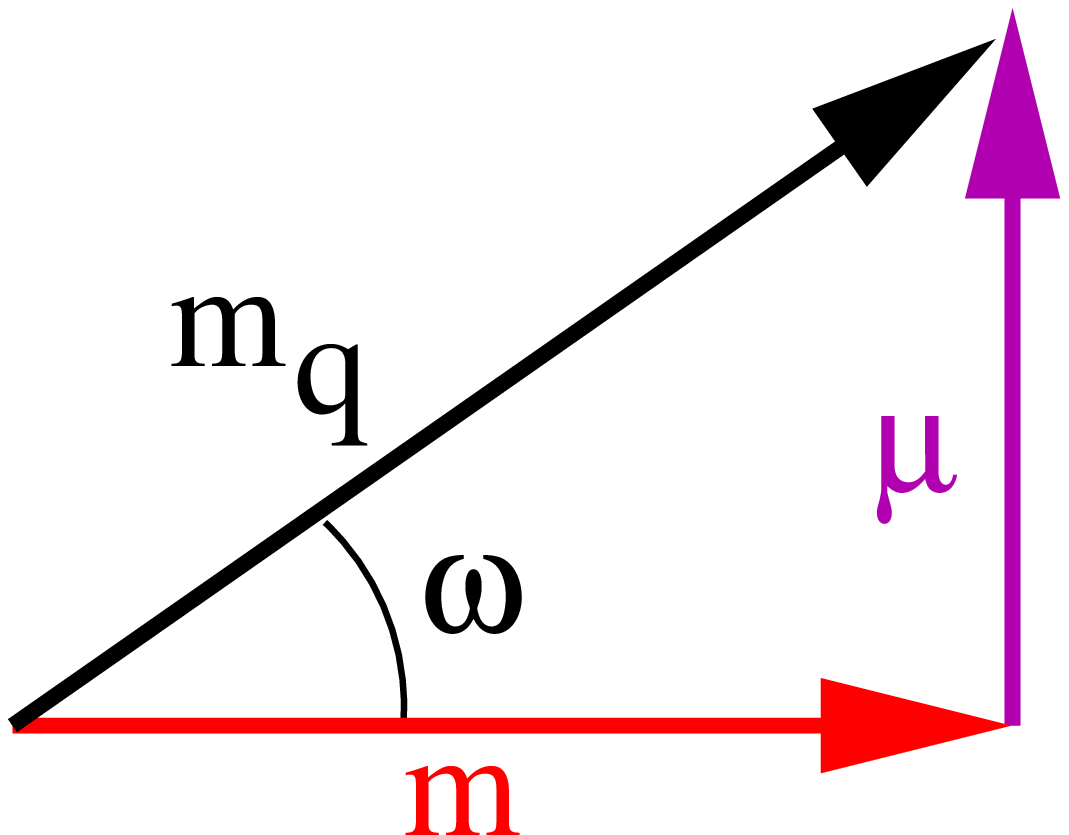}}   
\end{minipage}
\end{center}
\vspace{-.05cm}
$\chi$PT analysis described above goes through for any $\omega$,
as long as we expand about the rotated vacuum.

The situation is quite different after discretization.
The lattice action is 
\begin{equation}
 S_{l,\rm glue} + a^4 \sum_x
\overline \psi_l D_W \psi_l
+
\overline \psi_{l,L} M \psi_{l,R}
+
\overline \psi_{l,R} M^\dagger \psi_{l,L} 
\,,
\end{equation}
with $M$ the twisted mass, and
the subscript ``$l$'' indicating ``lattice''.
Here $D_W$ is Wilson's doubler-free derivative,
\begin{equation}
\Dslash \longrightarrow D_W=
\frac12 \sum_\mu \gamma_\mu (\nabla_\mu^\ast+\nabla_\mu) 
- \frac{1}{2} \sum_\mu (\nabla_\mu^\ast\nabla_\mu)
\end{equation}
($\nabla$ and $\nabla^\ast$ are forward and backward derivatives,
respectively).
Since the second term in $D_W$ 
(the ``Wilson term'', in which I have set the Wilson parameter $r=1$)
breaks chiral symmetry, one cannot
rotate away the twist in the mass. The theories with 
mass term
{$M_0$} and {$U_L M_0 U_R^\dagger$} are different
on the lattice.
In fact, the full fermion matrix {$D_W + M P_R + M^\dagger P_L$} has
positive determinant (and is thus useful in practice) 
only for special {$M$}. 
One such choice is two flavors with the twisted mass (\ref{eq:standardtm}),
for which the lattice action is
\begin{equation}
{S}_{\rm tmLQCD} = S_{l,\rm glue} + a^4 \sum_x
\overline \psi_l\left( D_W + m_0 + i\gamma_5 \tau_3 \mu_0 \right)\psi_l 
\,.
\label{eq:StmLQCD}
\end{equation}
Here $m_0$ and $\mu_0$ are, respectively,
the  bare untwisted and twisted mass (in lattice units).
For the remainder of this lecture I will focus entirely on this theory.

\subsection{Determining the local effective Lagrangian}\label{sec:symanzik}

Symanzik\cite{Symanzik} showed how to study the approach of the lattice theory to
its continuum limit. The first step is to understand this limit itself.
The lattice provides a legitimate regularization of QCD (one that is
awkward from a perturbative point of view, but has the great advantage of
being non-perturbative), and so one obtains continuum QCD (in this case
tmQCD) as the cut-off $1/a$ is sent to infinity. This has been
established to all orders in perturbation theory\cite{Reisz} and it 
assumed to hold non-perturbatively. One must appropriately tune the 
(``relevant'') lattice
bare parameters to reach the continuum limit. In particular, since the
Wilson term mixes with the identity operator, $m_0$ is additively renormalized,
and must be tuned to $m_c(a) < 0$, while the twisted mass has nothing to mix
with and is multiplicatively renormalized.\footnote{%
One must also tune the bare
coupling $g_0\to 0$ in the usual way.}
The resulting continuum theory is
\begin{equation}
\CL_{\rm tmQCD} = \CL_{\rm glue} + 
\bar{\psi} (\Dslash  + m + i \gamma_5 \tau_3 \mu) \psi\,,
\label{eq:LtmQCD}
\end{equation}
with $\CL_{\rm glue}$ the usual continuum gluon action,
and with the continuum field and physical masses being
\begin{equation}
\psi = a^{-3/2} Z\ \psi_l\,,\quad
m = Z_S^{-1} (m_0 - m_c)/a\,, \quad {\rm and} \quad
\mu = Z_P^{-1} \mu_0/a\,.
\label{eq:mdefs}
\end{equation}
Here $Z$, $Z_S$ and $Z_P$ are renormalization factors relating quantities
in the lattice regularization to those in the chosen continuum scheme. 
The corresponding geometry is illustrated  below.
\begin{figure}[ht]
\centerline{\epsfxsize=3.1in\epsfbox{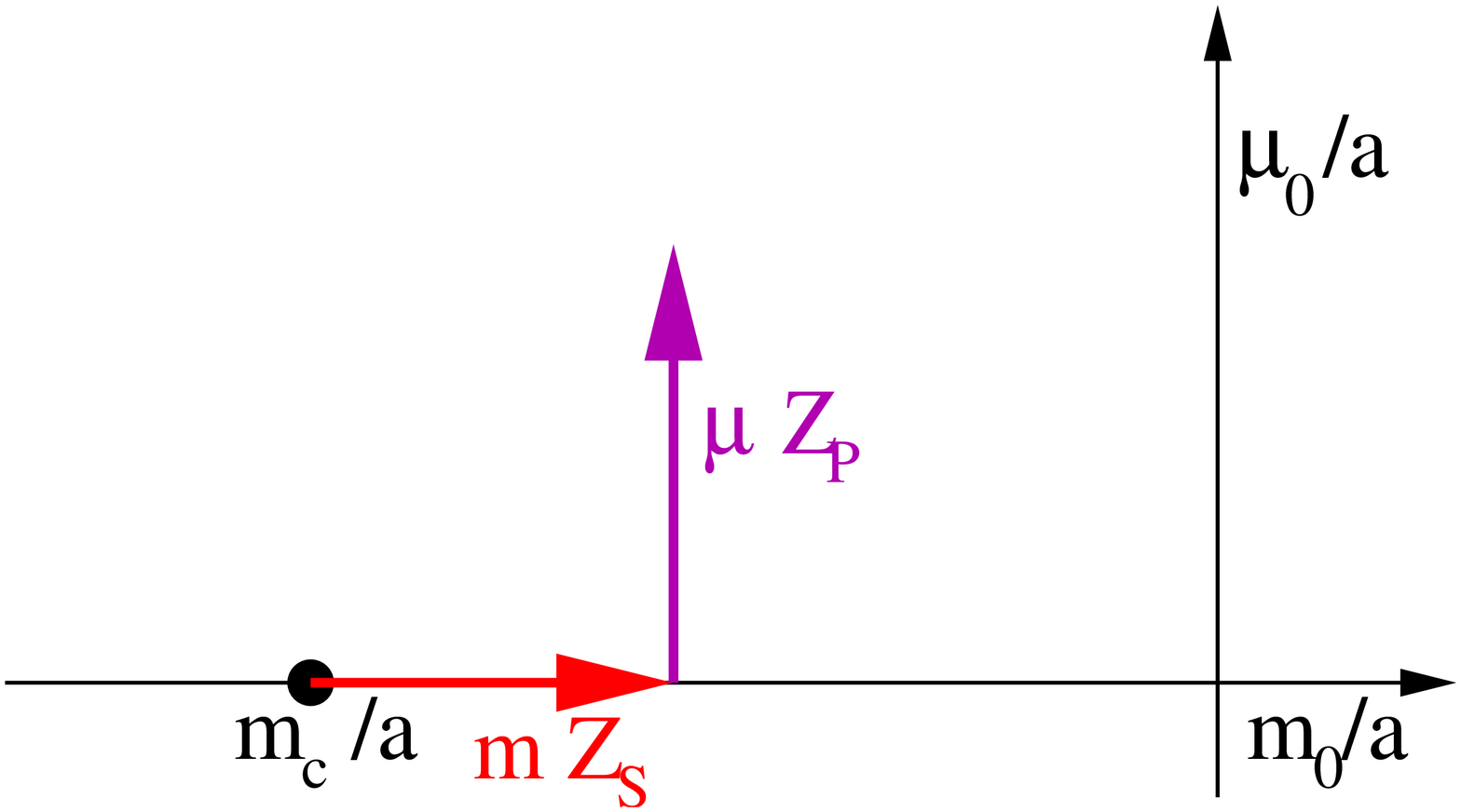} }
\end{figure}

Corrections to the continuum limit are suppressed by inverse powers
of the cut-off, i.e. by positive powers of $a$. Symanzik showed how
these can collected into a local effective Lagrangian 
\begin{equation}
\CL_{\rm Sym} = \CL_{\rm tmQCD} + a \CL^{(5)}
+ a^2 \CL^{(6)} + \dots,
\label{eq:LSymanzik}
\end{equation}
where $\CL_{\rm tmQCD}$ is the desired continuum Lagrangian
(note the absence of the ``L'' in the subscript), 
and $\CL^{(5)}$ and $\CL^{(6)}$ contain 
terms of dimension 5 and 6, respectively.\footnote{%
The number of effective Lagrangians in these lectures is approaching 
a confusing level. Note that I always use the subscript $\chi$
for the chiral Lagrangian, so that the six-derivative contribution
$\CL^{(6)}_\chi$ can be distinguished from $\CL^{(6)}$
in eq.~(\protect\ref{eq:LSymanzik}).}
The effective theory must be regularized, either
by a standard continuum regulator such as dimensional regularization,
or possibly with a finer lattice having {$a'\ll a$}. We will not
actually use $\CL_{\rm Sym}$ 
for concrete calculations so do not need to be more
specific. A key feature of $\CL_{\rm Sym}$ is that all factors of
$a$ are explicit---the effective theory does not ``know'' about the
lattice spacing in any other way. Note, however, that
``{$a$}'' includes logarithms,
of the form {$\sim a [1 + g(a)^2 \ln a + \dots]$}.
I return to this point below.

The content of eq.~(\ref{eq:LSymanzik}) is that all discretization
errors in all correlation functions can be reproduced by a set of
{\em local} insertions. This is established by a procedure akin
to renormalization (in which one determines divergent parts of graphs by
doing a Taylor expansion  in the external momenta,
or a lattice variant of this procedure\cite{Reisz},
 and then subtracts them), 
except that one subtracts more terms 
in the Taylor expansion (``over-subtraction''), 
thus including those proportional to powers of $a$.
As with renormalization, the consistency of this
 procedure requires that one include
{\em all terms in $\CL^{(n)}$ of the appropriate dimension which are
invariant under the symmetries of the theory}---here, of tm{\bf L}QCD.
These terms will have coefficients such that reflection positivity is
satisfied, since they arise from a theory in which it is satisfied.\footnote{%
This is true for the action used in the text.
Reflection positivity is violated with improved Wilson fermions or improved
gauge actions, but it is expected that this does not effect the long distance
physics which is being captured by $\CL_{\rm Sym}$.}
This procedure also works if one includes sources for external operators,
which should be treated using the spurion trick.
The procedure has been demonstrated to all orders in perturbation theory,
and is assumed to work non-perturbatively. 

The preceding discussion is nothing other than (a sketch of)
a derivation of an EFT. The usual EFT words---``separation of scales''---were
not mentioned but were implicit. $\CL_{\rm Sym}$ 
is only useful if $p \ll 1/a$,
for otherwise successive terms in the expansion, which give contributions
of relative size $a p$, are not suppressed.
Thus $\CL_{\rm Sym}$ is an EFT for quarks and gluons with energies far
below the cut-off scale. The set-up is the same as when considering
the impact of new short-distance physics on QCD, except that the
new physics here violates rotation and translation symmetries.
It indicates how  one can derive an EFT in a Euclidean context,
at least order-by-order in perturbation theory. One does not
need to rely on the S-matrix argument of Weinberg. This is important
because the underlying lattice theory is discretized in Euclidean
space. 

\begin{figure}[htb]
\centerline{\epsfxsize=4.1in\epsfbox{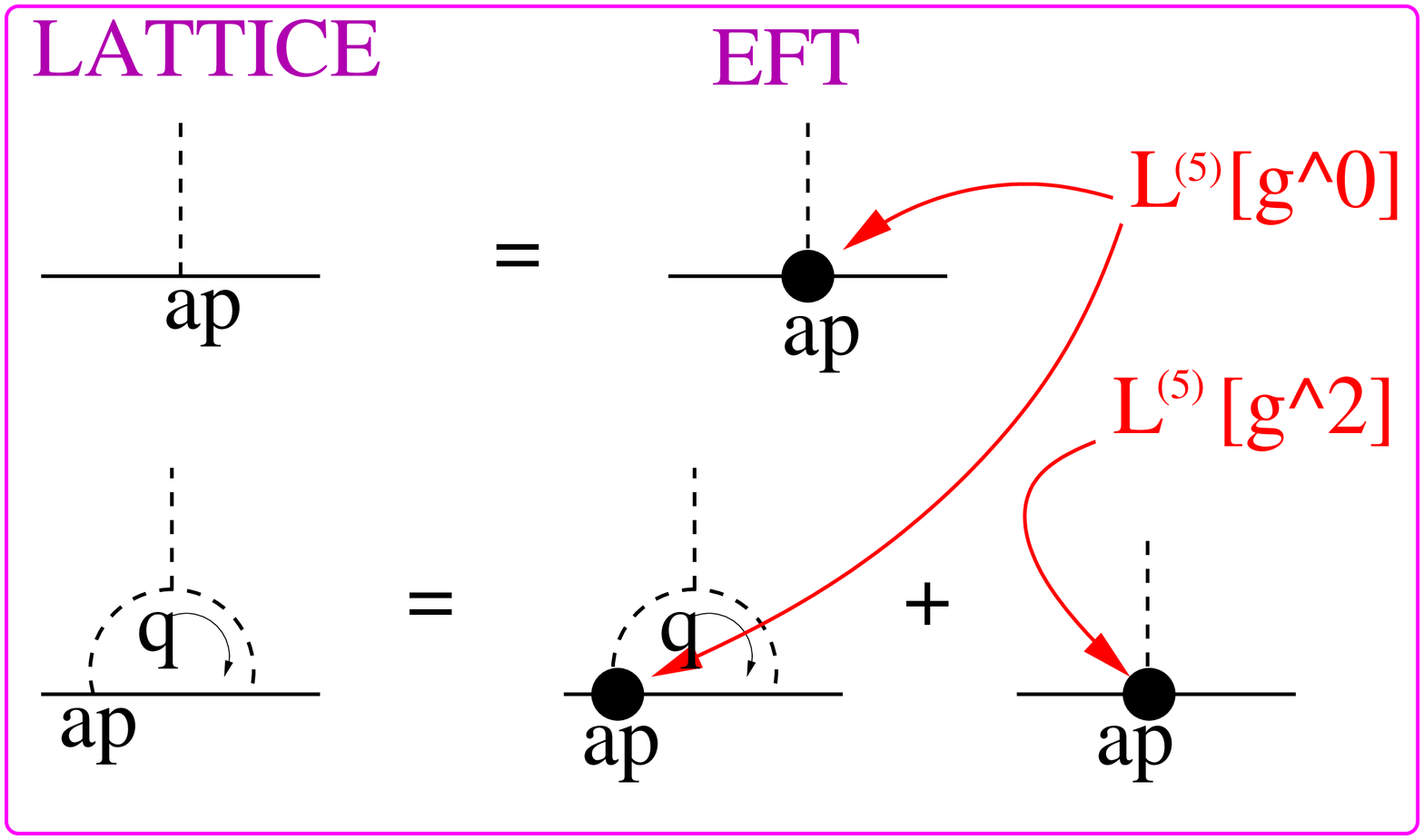} }
\caption{Matching lattice QCD with the Symanzik EFT. Dashed lines are
gluons, solid lines quarks, $p\ll 1/a$ an external momentum, 
$q$ a loop momentum.}
\label{fig:exofmatching}
\end{figure}

Let me illustrate these general words with a simple example.
Consider the quark-gluon vertex shown in Fig.~\ref{fig:exofmatching}.
In EFT language,
the counterterms {$\CL^{(5,6,\dots)}$} are to be determined by
matching correlation functions with those of the lattice theory.
At tree-level, the $O(a)$ terms in the lattice vertex can be matched
by adjusting the coefficients of operators in $\CL^{(5)}$ such as
$\bar \psi D^2 \psi$ as long as $|p|\ll 1/a$. 
At one-loop, for $|q|\ll 1/a$, the integrands
match by construction, giving a logarithmic divergence.
For larger $|q|$, however, the matching fails, leading to
a finite difference in the results (finite since both
theories are regulated). To match one must add an $O(g^2)$ contribution
to the coefficients of the operators in $\CL^{(5)}$.
One loop matching is schematically:
\begin{eqnarray}
a p g^2 (\ln[pa]+r_{\rm lat}) &=& a p g^2 (\ln[pa']+ r_{\rm EFT})
+ ap g^2 c^{(2)}_{\rm EFT}
\\
\Rightarrow\quad c^{(2)}_{\rm EFT} &=& r_{\rm lat} - r_{\rm EFT} + \ln[a/a'] \,,
\end{eqnarray}
where $r_{\rm lat,EFT}$ are the finite parts of the loop
diagram, and I have used a lattice regularization of the EFT with
spacing $a'$.
We can now see the generic form of the $a$ dependence of the coefficients
in $\CL^{(5)}$ (whose one-loop contribution is given here
by $g^2 c^{(2)}_{\rm EFT}$). There is explicit
logarithmic dependence, and an implicit logarithmic dependence
through $g$, which is evaluated in the lattice calculation at a scale $\sim a$.

As an aside, I note that, having determined the form of
$\CL^{(5)}$ in the EFT, one can add corresponding terms to
the lattice theory (i.e. terms having operators in $\CL^{(5)}$
as their classical continuum limit), and then adjust the
coefficients of the lattice terms to set those in
$\CL^{(5)}$ to zero. This is the ``improvement program'' at $O(a)$\cite{Symanzik},
and it has been implemented non-perturbatively\cite{ALPHA,Jansen}.
In the ``new physics'' context this would be called ``fine tuning'',
with negative connotations, but here we can again turn
all the knobs at our disposal to improve extrapolations.
The program can be extended, in principle, to any order,
but has in practice not been extended to $\CL^{(6)}$. Since
$\CL^{(6)}$ will play a key role in the following, this means that
the considerations below are essentially unaffected by improvement.

\subsection{Symanzik effective action for tmLQCD}\label{sec:SymtmLQCD}

We are now ready to determine the operators in 
$\CL^{(5)}$ and $\CL^{(6)}$. 
These must satisfy the symmetries of tmLQCD, eq.~(\ref{eq:StmLQCD}),
and be reflection positive.
The symmetries are
gauge invariance, lattice rotations and translations, 
charge conjugation and fermion number, but {\em not} flavor
$SU(2)$ nor parity (for generic $\omega$).
Only the $U(1)$ flavor subgroup generated by $\tau_3$, 
and combinations of parity with a discrete flavor rotations survive:
\begin{equation}
\mathcal{P}^{1,2}_F:\quad
\psi_l(x) \rightarrow \gamma_0 (i\tau_{1,2}) \psi_l(x_P)\,,\ \
\bar{\psi_l}(x) \rightarrow \bar{\psi_l}(x_P) (-i\tau_{1,2}) \gamma_0
\,, \label{eq:tmparity}
\end{equation}
Also useful is $\widetilde{\mathcal{P}}$: parity combined with
{$[\mu_0 \rightarrow - \mu_0]$}.

The continuum Lagrangian consistent with these symmetries is tmQCD,
eq.~(\ref{eq:LtmQCD}). To determine $\CL^{(5)}$ one simply enumerates
all allowed operators, generalizing the work done
for Wilson fermions\cite{ALPHA}. The result is\cite{ShWuI}
\begin{eqnarray}
\CL^{(5)} &=& 
b_1 \bar{\psi} i \sigma_{\mu\nu} F_{\mu\nu} \psi 
+ b_2 \bar{\psi} (\Dslash + m + i \gamma_5 \tau_3 \mu)^2
\psi \nonumber \\ 
&&+ b_3 m \bar{\psi} (\Dslash  + m + i \gamma_5 \tau_3 \mu)
\psi 
+ b_4 m {L}_{\rm glue} 
+ b_5  m^2 \bar{\psi} \psi  
\nonumber\\ 
&&+ b_6 \mu \bar{\psi} \big[ (\Dslash + m + i \gamma_5
\tau_3 \mu), i \gamma_5 \tau_3 \big]_+ \psi 
+ b_7  \mu^2 \bar{\psi} \psi 
\,,
\label{eq:L5full}
\end{eqnarray}
where I use continuum masses {$m,\mu$} rather than bare masses.
The coefficients {$b_i$} (the analog of the LECs of \chpt)
are real (from reflection positivity), 
and depend on {$g^2(a)$} and {$\ln a$}.
Among them  {$b_{6,7}$} are 
``new'' compared to Wilson case.
Many terms have been forbidden by lattice symmetries:
$\widetilde{P}$ forbids $m\mu\bar\psi\psi$ and
{$m^2 \bar\psi i\gamma_5\tau_3\psi$}, and it
requires {$\bar\psi\tau_3\gamma_5\psi$} to come with factor {$\mu_0\propto \mu$},
and the twisted Pauli term 
{$\bar\psi\sigma_{\mu\nu}F_{\mu\nu}\tau_3\psi$} to have factor of {$\mu$}
(so that it appears in {$\CL^{(6)}$});
flavor $U(1)$ forbids {$\bar\psi \tau_{1,2}\psi$};
and $\mathcal{P}^{1,2}_F$ forbids {$\bar\psi\gamma_5\psi$}, 
{$\widetilde F_{\mu\nu}F_{\mu\nu}$} and {$\bar\psi\tau_3\psi$}.

\setcounter{footnote}{0}
\renewcommand{\thefootnote}{\Roman{footnote}}

$\CL^{(5)}$ looks rather forbidding, with 7 unknown coefficients,
but this proliferation is misleading, for two reasons.
First, we will be doing a joint continuum-chiral expansion,
and working in  the ``generic small mass'' (GSM)
regime in which $m \sim \mu \sim a \LQCD^2$.
Thus each factor of $m$ or $\mu$ counts as an additional power of $a$.
We will work at NLO in this power counting.
Since $\bar\psi \Dslash \psi$ and $\CL_{\rm glue}$ map into LO
operators in \chpt, and $\CL^{(5)}$ comes with an overall factor
of $a$, any further factors of $m$ or $\mu$ make the operator
of next-to-next-to-leading order (NNLO). This allows us to
drop all except the $b_1$ term, and 
the $\bar\psi \Dslash^2 \psi$ parts of the $b_2$ and $b_3$ terms.\footnote{%
In the usual discussion of on-shell Symanzik improvement, one
drops terms vanishing by the LO equations of motion, which only
contribute to contact terms. This is not necessary here, but explains
the basis used in eq.~(\protect\ref{eq:L5full}).}
Second, we will be mapping $\CL_{\rm Sym}$ into \chpt, at which point
all that matters is the chiral transformation properties of the
operators. Now the Pauli ($b_1$) term and $\bar\psi \Dslash^2\psi$
transform the same way, so, since the coefficients in the mapping
to \chpt\ are unknown, we can drop the latter operator.
The outcome is 
\begin{equation}
\CL_{\rm NLO}^{(5)}= 
b_1 \bar{\psi} i \sigma_{\mu\nu} F_{\mu\nu} \psi\,,
\label{eq:L5NLO}
\end{equation}
which is unchanged from the result for (untwisted) Wilson fermions.

Moving onto $\CL^{(6)}$, there are now gluonic terms\cite{LW}
\begin{eqnarray}
\mathcal{L}_{\rm glue}^{(6)} &\sim& 
\Tr(D_\mu F_{\rho\sigma} D_\mu F_{\rho\sigma})
+ \Tr(D_\mu F_{\mu\sigma} D_\rho F_{\rho\sigma})
\nonumber\\&&
+\underbrace{\Tr(D_\mu F_{\mu\sigma} D_\mu F_{\mu\sigma})}_{%
\textrm{Lorentz violating}}
+ \underbrace{(m^2,\mu^2) \Tr(F_{\mu\nu} F_{\mu\nu})}_{%
\textrm{$O(a^2 m^2, a^2\mu^2)$ so NNNLO}} \,,
\end{eqnarray}
where I use a schematic notation without coefficients,
and fermionic terms (obtained by generalizing
the analysis for Wilson fermions\cite{SheikWo,BRS})
\begin{eqnarray}
\mathcal{L}_q^{(6)} &\sim& 
\underbrace{\bar{\psi} D_\mu^3 \gamma_\mu \psi}_{%
\textrm{Euclidean non-invariant}}
+
\bar{\psi} D_\mu \Dslash D_\mu \gamma_\mu \psi + \dots
\nonumber\\
&&+ \underbrace{%
m \bar\psi \Dslash^2 \psi + \mu \bar\psi \Dslash^2 i\gamma_5\tau_3\psi}_{%
\textrm{$O(a^2 m,a^2\mu)$ so NNLO}}
+ \underbrace{%
m \bar\psi i\sigma_{\mu\nu} F_{\mu\nu}\psi 
+ \mu \bar\psi \sigma_{\mu\nu} F_{\mu\nu} \gamma_5\tau_3\psi}_{%
\textrm{$O(a^2 m,a^2\mu)$ so NNLO}}
\nonumber\\&&
+ \underbrace{%
(m^2,\mu^2)\bar\psi \Dslash \psi 
+ m\mu \bar\psi \Dslash i\gamma_5\tau_3\psi}_{%
\textrm{$O(a^2 m^2)$, etc. so NNNLO}}
+ \underbrace{%
(m^3, m \mu^2) \bar\psi \psi + (\mu^3,\mu m^2) i\gamma_5\tau_3\psi}_{%
\textrm{$O(a^2 m^3)$, etc. so NNNNLO}}
\nonumber\\&&
+ 
 (\bar\psi \psi )^2 + (\bar\psi\gamma_\mu\psi)^2 + \dots \,,
\end{eqnarray}
where the first ellipsis indicates other Euclidean invariant terms
with three derivatives and the second other four-fermion operators.
As can be seen, in the GSM regime, most of the fermionic operators
in $\CL^{(6)}$ are of at least NNLO. In particular, no flavor-parity
breaking terms appear, since they require a factor a $\mu$.
The net result is that the part of $\CL^{(6)}$ of NLO in the GSM
regime is the same as that for untwisted Wilson fermions
(with the ellipses having the same meaning as above):
\begin{eqnarray}
\mathcal{L}_{\rm NLO}^{(6)} &\sim& 
\Tr(D_\mu F_{\rho\sigma} D_\mu F_{\rho\sigma})
+ \Tr(D_\mu F_{\mu\sigma} D_\rho F_{\rho\sigma})
\nonumber\\&&
+ \bar{\psi} D_\mu \Dslash D_\mu \gamma_\mu \psi + \dots
+ (\bar\psi \psi )^2 + (\bar\psi\gamma_\mu\psi)^2 + \dots
\nonumber\\&&
+\underbrace{\Tr(D_\mu F_{\mu\sigma} D_\mu F_{\mu\sigma})
+\bar{\psi} D_\mu^3 \gamma_\mu \psi}_{%
\textrm{Euclidean non-invariant}} \,.
\label{eq:L6NLO}
\end{eqnarray}
Note that the only symmetry broken by $\mathcal{L}_{\rm NLO}^{(6)}$ 
that is not already broken by $\mathcal{L}_{\rm NLO}^{(5)}$
is Euclidean invariance. In fact, we will see that in the
PGB sector the Euclidean non-invariant terms pick up an additional
factor of $p^2$ and (given the overall $a^2$) are of NNLO.

This takes care of the action, but what about currents and densities?
The matching of these between the lattice theory and the EFT
can be worked out using symmetries\cite{Symanzik}, 
and I quote only the relevant results\cite{ALPHA,nondegenimp}:
\begin{eqnarray}
V_\mu^b &=& \bar\psi \gamma_\mu T^b \psi + a \tilde c_V \partial_\nu 
\bar\psi i \sigma_{\mu\nu} T^b \psi \,,
\label{eq:VSym}\\
A_\mu^b &=& \bar\psi \gamma_\mu\gamma_5 T^b \psi + a \tilde c_A \partial_\mu 
\bar\psi \gamma_5 T^b \psi \,,
\label{eq:ASym}
\\
S^0 &=& \bar\psi \psi + a \tilde g_S \Tr(F_{\mu\nu} F_{\mu\nu})\,,
\qquad
P^b = \bar\psi \gamma_5 T^b \psi
\label{eq:SPSym}
\end{eqnarray}
Here I work at NLO in the GSM regime, so $am\sim a\mu$ terms are dropped.
I also drop the mixing of $S^0$ with the identity operator, as it does
not contribute to connected matrix elements.
The coefficients $\tilde c_{V,A}$ and $\tilde g_S$ depend on $g(a)$
and on $\ln(a)$, just like the $b_i$ above.
The content of these equations is that the on-shell matrix elements of
the operators shown, evaluated in the EFT, will
reproduce those of the lattice currents and densities,
including the leading discretization error. 
These forms apply for any choice of lattice currents and densities
(e.g. ultra-local or smeared)  as long
as they have been multiplied by appropriate $Z$-factors
so as to be correctly normalized. The numerical values of the coefficients
$\tilde c_V$ etc. will, of course, depend on the form of the lattice operators,
and on the lattice action. In particular, if the action and operators
have been $O(a)$ improved, then these coefficients will vanish.
Note that the density $P^b$ is automatically improved.

When we map the operators in (\ref{eq:VSym}-\ref{eq:SPSym})
into \chpt\ we are free to do this for the $O(1)$ and $O(a)$ parts
separately and then combine at the end. It is straightforward
to see that the $\tilde c_V$ and $\tilde g_S$ terms map into operators
which are of NNLO and can be dropped. 
For the former, the argument is given by Wu and I\cite{ShWuII},
and follows because of the need to have three derivatives in
order to match the quark-level operator.
For the $\tilde g_S$ term the argument is even more simple: the
matching of chiral singlet $\Tr(F_{\mu\nu}F_{\mu\nu})$ gives the
LO chiral kinetic term $\Tr(\partial_\mu \Sigma \partial_\mu \Sigma^\dagger)$.
Thus the $\tilde g_S$ term is of size $a p^2$, and so of NNLO
 compared to $\bar\psi\psi$, which maps to
$\Tr(\Sigma+\Sigma^\dagger) \sim O(1)$.

I conclude that, at NLO in the GSM regime, 
the currents and densities in the Symanzik EFT have the
same form as in the continuum, aside from the $\tilde c_A$ term. 
Thus, as long as one treats the $\tilde c_A$ term separately,
one can include the currents and densities using sources
just as in continuum QCD, and the resulting theory will
have a {\em local} chiral symmetry.\footnote{%
There is a one subtlety here. After adding in the $O(1)$ parts
of the currents and densities, the local invariance is only
true for the continuum part of the Symanzik action, $\CL_{\rm tmQCD}$.
It can be extended to $\CL^{(5)}_{\rm NLO}$, however, by allowing
the corresponding spurion, called $\widetilde A$ below, to
transform like $\chi$ under the local chiral symmetry.
It can also be extended to $\CL^{(6)}_{\rm NLO}$, but this is
not actually necessary because the resulting contributions to
the currents are of NNLO. I should note that
there is some disagreement on the validity of this approach
for mapping currents and densities\cite{AokiBar04,AokiBar06}, 
which is why I have given here
a more detailed discussion than is present in the literature\cite{ShWuII}.}

\subsection{Mapping the Symanzik action into \chpt}\label{sec:mapSymintoChpt}

I now turn to the second step of the procedure---taking the 
Symanzik EFT and determining the chiral EFT which describes it
at long distances. This was done for continuum Lagrangian, $\CL_{\rm tmQCD}$,
in sec.~\ref{sec:chptcont}---a twisted mass was already included
by the generality of the formalism. The result is 
$\CL^{(2)}_\chi$ of eq.~(\ref{eq:L2}) at LO and $\CL^{(4)}_\chi$ of
eq.~(\ref{eq:L4}) at NLO.
The task here is to include the effects of 
$\CL^{(5,6)}$, as well as the $\tilde c_A$ term
in eq.~(\ref{eq:ASym}).
To do so systematically requires a power counting scheme,
to which I now turn.

\subsubsection{Power counting and terminology}\label{sec:regimes}

As anticipated above, discretization errors
introduce a new parameter into the power counting in \chpt.
In addition to the usual chiral expansion in powers of
{$p^2/\Lchi^2 \sim m_{NG}^2/\Lchi^2\sim m_q/\LQCD$},
we must include $a\LQCD$. 
Note that $\LQCD$ is the only scale available to balance dimensions
when we map quark-level operators in $\mathcal{L}_{\rm Sym}$ into \chpt.
I stress that, once one has the Symanzik EFT in hand, 
$m$ and $a$ are on a similar footing---both are simply small parameters in
a continuum Lagrangian. 

\begin{figure}[ht]
\centerline{\epsfxsize=4.1in\epsfbox{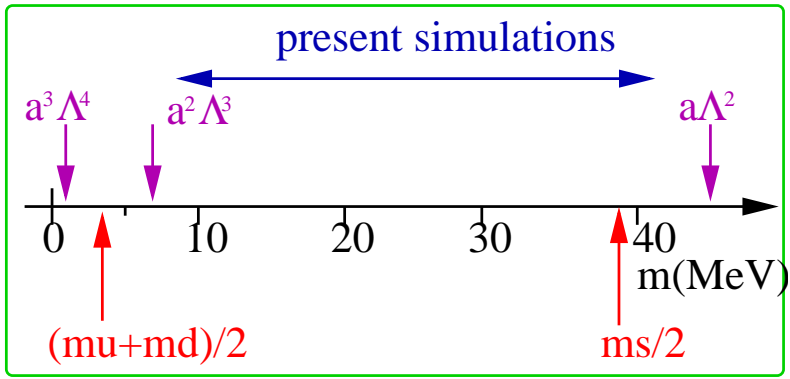}}
\caption{Comparing discretization errors to mass corrections. I use 
$a^{-1}=2\;$GeV and $\Lambda=300\;$MeV.}
\label{fig:power_cnt}
\end{figure}

How we should weight discretization errors relative to mass corrections?
The numerical comparison is shown in Fig.~\ref{fig:power_cnt}.
I have been conservative by having ``present simulations''
range down to $m_s/10$, 
which is yet to be achieved with Wilson-like fermions.\footnote{%
Note that the comparison is not precise: the relative coefficients
of mass and discretization effects in the EFT are expected to be
of $O(1)$, but could easily be $2-3$.}
I conclude from the figure
that the appropriate power counting for the coming decade is
{$a^2\LQCD^3 \ltapprox m_q \ltapprox a\LQCD^2$},
and that to disentangle quark mass dependence from discretization effects
we need to remove errors of $O(a)$ and understand 
those of $O(a^2)$.

I will consider in the following two regimes (sketched below):
\begin{center}
\begin{minipage}{2in}
\noindent
{\bf I.} The GSM regime, already introduced above, which I define more precisely
as  {$a\LQCD^2\ltapprox m_q \ll \LQCD$}, so that it
includes {$a\LQCD^2\ll m_q$} (where we would like to be)
as well as {$ a\LQCD^2\approx m_q$} (where we actually are).
This is the regime in which we want to learn how to remove
$O(a)$ errors. 
\end{minipage}
\hfill
\begin{minipage}{2.4in}
\centerline{\epsfxsize=2.4in\epsfbox{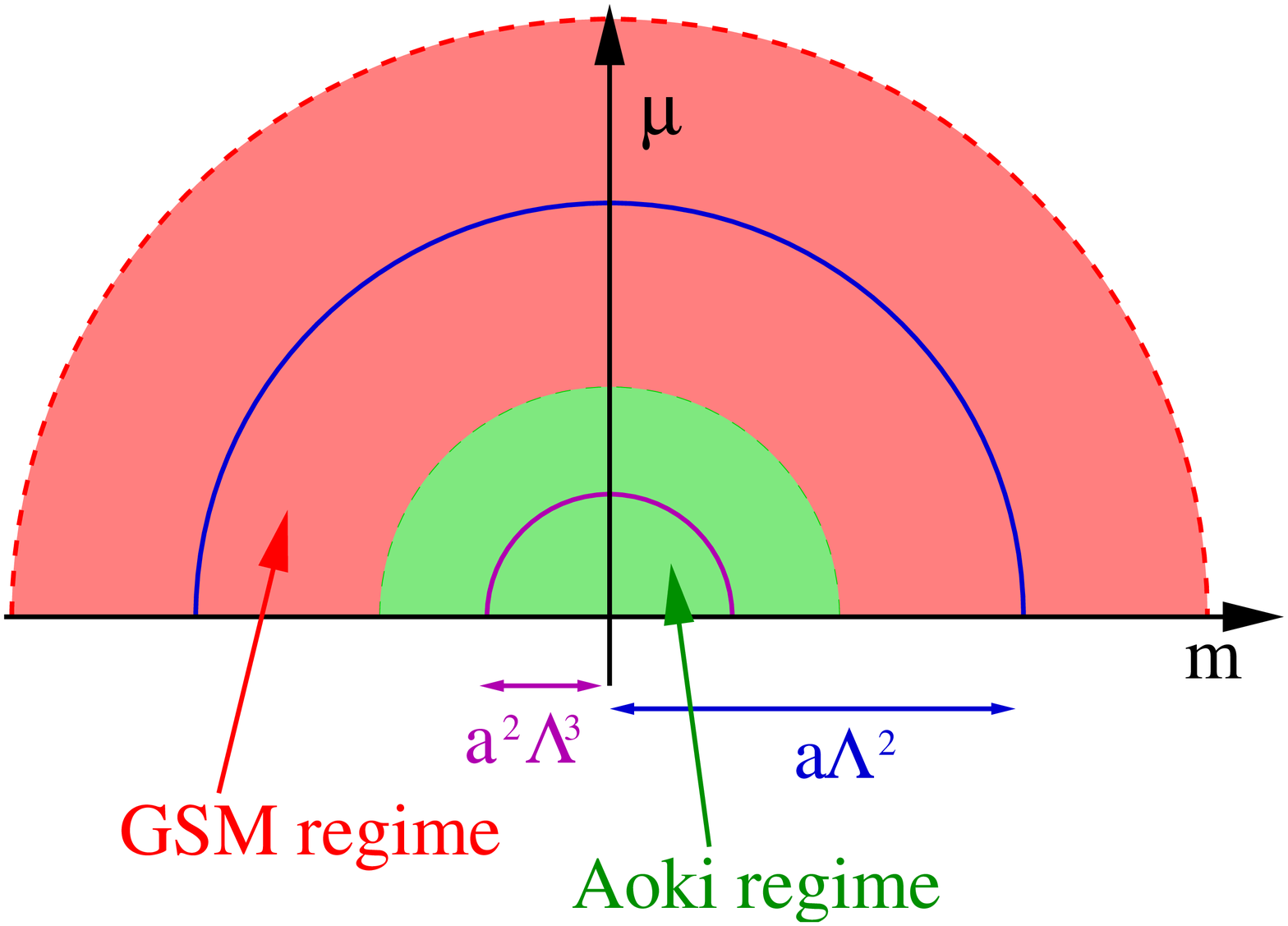}}
\end{minipage}
\end{center}
\noindent
{\bf II.} The ``Aoki regime'', in which 
{$m_q \ltapprox a^2\LQCD^3 \ll \LQCD$}
(including {$m_q \ll a^2\LQCD^3$}). This is where we cannot
avoid discretization errors, and where they lead to
non-trivial phase structure. 
Entering this regime changes the relative weight given
to operators in \chpt, but not the operators themselves.

I begin by working in the GSM regime at NLO. This requires keeping
terms of $O(a^2)$, and thus up to $\CL^{(6)}$ in the Symanzik expansion.
The enumeration will turn out to suffice also for a LO study of the
Aoki regime.

\subsubsection{Mapping $\mathcal{L}_{\rm NLO}^{(5)}$
and $\mathcal{L}_{\rm NLO}^{(6)}$ into \chpt}

{$\mathcal{L}_{\rm NLO}^{(5)}$} transforms just like a mass term
under $SU(2)_L\times SU(2)_R$ (recall that we set $N=2$, 
although the considerations are easily
generalized):
\begin{equation}
a \mathcal{L}_{\rm NLO}^{(5)} \sim 
 a \bar{\psi} i \sigma_{\mu\nu} F_{\mu\nu} \psi 
=  \bar{\psi}_L \widetilde{A} i \sigma_{\mu\nu} F_{\mu\nu} \psi_R 
+  \bar{\psi}_R \widetilde{A}^\dagger i \sigma_{\mu\nu} F_{\mu\nu} \psi_A 
\,.
\end{equation}
Here {$\widetilde A$} is a spurion transforming like {$M$},
i.e. {$\widetilde A \to U_L \widetilde A U_R^\dagger$},
so that $\mathcal{L}_{\rm NLO}^{(5)}$ is invariant.
At the end we set $\widetilde A =a$.
The enumeration of operators in the chiral Lagrangian is
like that for $M$, yielding the new terms\cite{RS,BRS,Munster,ShWuII}
\begin{equation}
\mathcal{L}^{(2)}_{\chi,A} = 
- \frac{{ f}^2}{4} \tr(\hat{A}^{\dagger} \Sigma + 
               \Sigma^\dagger\hat{A})
\label{eq:L2A} 
\end{equation}
\begin{eqnarray} 
\lefteqn{\mathcal{L}^{(4)}_{\chi,A} =
  { W_{45}} \tr(D_\mu \Sigma^\dagger D_\mu \Sigma)
         \tr(\hat{A}^{\dagger} \Sigma +
         \Sigma^{\dagger}\hat{A})
 - { W_{68}'} \big[\tr(\hat{A}^{\dagger} \Sigma +
       \Sigma^{\dagger}\hat{A})\big]^2}
\nonumber\\ &&
- { W_{68}} \tr(\chi^{\dagger} \Sigma + \Sigma^\dagger\chi) 
         \tr(\hat{A}^{\dagger} \Sigma + \Sigma^{\dagger}\hat{A})
+ { W_{10}} \tr(D_\mu \hat{A}^\dagger D_\mu \Sigma + 
          D_\mu \Sigma^\dagger D_\mu \hat{A})  
\nonumber\\ &&
- { H'_2} \tr(\hat{A}^\dagger \chi+ \chi^\dagger \hat{A})
- { H_3} \tr(\hat{A}^\dagger \hat{A}) 
\,, \label{eq:L4A}
\end{eqnarray}
where $ \hat{A}= 2{W_0}\widetilde A$ is the analog of 
$\chi=2 { B_0}(s+ip)$, with $W_0$ a new LEC at LO.
We will not need to use $\hat{A}$ as a source (since we already
have $\chi$), so we will always be set $\hat{A}\to\hat{a}=2 W_0 a$.
$SU(2)$ simplifications can then reduce the number of terms.
In particular
$(\hat{A}^\dagger \Sigma+\Sigma^\dagger \hat{A})$ is proportional
to the identity, allowing single trace terms involving this
combination to be rewritten with two traces. These simplifications
have been used in (\ref{eq:L4A}).

Note that, following the discussion
at the end of sec.~\ref{sec:SymtmLQCD},
I have included sources for currents and densities 
and enforced {\em local} chiral invariance by
using covariant derivatives $D_\mu$ 
[defined as in eq.~(\ref{eq:ChPTcovder})].
This incorporates all discretization effects in
lattice matrix elements except those due to the
$\tilde c_A$ term in eq.~(\ref{eq:ASym}), which I
will treat separately below.

${\CL}_{\rm NLO}^{(6)}$ contains three types of terms.
First, those that are invariant under Euclidean and chiral symmetries
[the gluonic terms on the first line of eq.~(\ref{eq:L6NLO}) and some
of the four-fermion operators]. These match into \chpt\ as follows:\cite{LeeSh}
\begin{equation}
a^2 \Tr(D_\mu F_{\rho\sigma} D_\mu F_{\rho\sigma})
+ \dots
+ a^2 \bar{\psi} D_\mu \Dslash D_\mu \gamma_\mu \psi + \dots
\longrightarrow a^2 \tr(D_\mu\Sigma D_\mu\Sigma^\dagger) \,,
\label{eq:L6matchI}
\end{equation}
i.e. one obtains the leading order continuum result multiplied
by $a^2$. This leads to an $O(a^2)$ correction to the LEC $f$, and is
present for any fermion discretization (including chirally invariant ones).

Second, there are
four-fermion operators which violate chiral symmetry
(e.g. those having LR-LR structure). Their matching
 can be analyzed using two $\hat{A}$ spurions, leading to\cite{BRS}
\begin{equation}
(\bar\psi \psi )^2 + (\bar\psi\gamma_\mu\psi)^2 + \dots
\longrightarrow
\tr\left[(\hat{A}^{\dagger} \Sigma +\Sigma^\dagger \hat{A})\right]^2 \,.
\end{equation}
This operator is already present in $\CL^{(4)}_{\chi,A}$, having
been produced by two insertions of $\CL^{(5)}_{NLO}$. This illustrates
that what is relevant for matching are
the symmetries broken by the operators (here, chiral symmetry), 
and not their detailed form.
The four-fermion operators simply change an unknown
coefficient, $W'_{68}$, by an unknown amount.
The only exception is if
one uses a non-perturbatively $O(a)$ improved quark action,
as discussed below,
when $W'_{68}$ would vanish were it not for the four-fermion operators.

Finally, there are the terms violating Euclidean symmetry.
These can be decomposed into Euclidean singlet and non-singlet
parts. The former match as in eq.~(\ref{eq:L6matchI}), while
the latter give rise to Euclidean non-invariant 
chiral operators\cite{LeeSh}
\begin{equation}
a^2\Tr(D_\mu F_{\mu\sigma} D_\mu F_{\mu\sigma})
+ a^2 \bar{\psi} D_\mu^3 \gamma_\mu \psi
\longrightarrow a^2 \tr(D_\mu^2\Sigma D_\mu^2 \Sigma^\dagger)
\,.
\end{equation}
Since one needs four factors of $D_\mu$ to make a non-invariant operator, 
the result, when combined with the two powers of $a$, 
is an operator of NNNLO in \chpt, two orders higher than we are working.

We thus find that ${\CL}_{\rm NLO}^{(6)}$ adds no new operators,
so the results (\ref{eq:L2A}) and (\ref{eq:L4A}) are complete.
They are to be added to $\mathcal{L}^{(2)}_\chi$ [eq.~(\ref{eq:L2})]
and $\mathcal{L}^{(4)}_\chi$ [eq.~(\ref{eq:L4})], respectively,
to obtain the
full LO and NLO contributions to the chiral Lagrangian in the
GSM regime.
Note that using a twisted mass had no impact
on the analysis of this subsection,
since $\mathcal{L}_{\rm NLO}^{(5,6)}$ have the same form
as for untwisted Wilson fermions.

I now return to the $\tilde c_A$ term in the
axial current, eq.~(\ref{eq:ASym}).
To obtain the full axial current in the EFT one must separately
match this term into \chpt\ and add it to the result obtained
by taking derivatives of the Lagrangian obtained
above with respect to sources. 
It is a simple exercise to show, however, that the result
is simply to change the coefficient $W_{10}$, since
the operator it multiplies is exactly of the form $a\partial_\mu P^b$,
at linear order in the sources.
Thus the final form of the previous paragraph remains complete,
albeit with somewhat changed (although still unknown) coefficients.

There are thus five new LECs introduced by discretization errors:\footnote{%
This becomes ten new LECs in $SU(3)$ or PQ theories\cite{BRS}.}
$W_0$ at LO, $W_{45}$, $W_{68}$, $W'_{68}$ and $W_{10}$ at NLO.
What do we know about their values? 
Of course this depends on the choice of fermion and gauge actions,
so we can only make order of magnitude estimates.
Perturbing in $a$ and $m$ after rotating to Minkowski
space, one finds
\begin{equation}
\frac{W_0}{B_0}\sim 
\frac{\langle\pi|\bar\psi i\sigma_{\mu\nu}F^{\mu\nu}\psi|\pi\rangle}
           {\langle\pi|\bar\psi\psi|\pi\rangle}
\sim \LQCD^2
\,.
\label{eq:W0byB0}
\end{equation}
The first relation is not an equality because there are unknown
coefficients multiplying numerator and denominator, while
the second is a dimensional estimate. One might be tempted to
include a loop factor because the matrix element in the numerator
requires at least one gluon loop, but this is a non-perturbative
matrix element so such counting is inappropriate.

The four NLO $W_i$ are dimensionless, because the dimensions needed
to balance powers of $a$ are provided by the $W_0$ residing in $\hat{A}$.
Like the Gasser-Leutwyler coefficients $L_i$, 
they depend on the renormalization scale, $\mu$. 
We will find that the combinations 
\begin{equation}
\widetilde W = W_{45}-L_{45}\,,\quad
W= W_{68}-2 L_{68}\,,\ \textrm{and}\
W' = W'_{68}-W_{68} +L_{68} 
\label{eq:shiftedLECs}
\end{equation}
are $\mu$ independent, so the scale dependence of the
$W_i$ themselves is comparable to that of the $L_i$.
Using the argument of sec.~\ref{sec:powercont},
we then estimate $|W_i|\sim |L_i| \sim 1/(4 \pi)^2$.
Another line of argument gives a similar estimate.
In continuum \chpt, $\bar\psi \psi$ maps into 
$\CO=f^2 2B_0 \tr(\Sigma+\Sigma^\dagger)/4$.
Thus we expect the mapping of $a^2(\bar\psi \psi)^2$ to contain
$O(1)\times a^2 \CO^2$ plus other operators. 
In fact, this four-fermion operator
(along with others) maps into
$a^2 \CO^2 16 W'_{68} W_0^2/(f^4 B_0^2)$.
Comparing, and using eq.~(\ref{eq:W0byB0}),
I find $W'_{68} \sim (1/16) (f/\LQCD)^4$.
It is reassuring that this estimate, albeit crude, agrees
with the order of magnitude of that above.

The analysis to this point has assumed that the fermion
action is not non-perturbatively improved. Since many
simulations now use such improvement, it is interesting
to ask how it changes the analysis. Improvement sets
$\CL^{(5)}_{NLO}$ to zero. Since this was the source of
almost all terms linear in $a$ in $\CL^{(2,4)}_{\chi,A}$ above,
the impact is to set $\CL^{(2)}_{\chi,A}=W_{45}=W_{68}=0$.\footnote{%
Note that the vagaries of the notation do
not allow one to set $W_0=0$, for then all discretization effects
would vanish.}
The term quadratic in $a$, with coefficient $W'_{68}$ survives,
since this also comes from matching with $\CL^{(6)}_{NLO}$.
Two coefficients of terms linear in $a$ do not vanish: $W_{10}$
and $H'_3$. Why? As already noted, 
the former multiplies a term which only
contributes if the source for the axial current is non-vanishing
(the part containing the vector current cancels). 
It thus represents discretization errors in the matrix
elements of this current. But these matrix elements require
improvement {\em additional to that of the action}, namely
the addition of the ``$c_A$ term'' to the axial current\cite{ALPHA}. 
If this improvement has not been implemented, then $W_{10}\ne 0$. 
Note that, following the discussion at the end of sec.~\ref{sec:SymtmLQCD},
the matrix elements of the vector current and the densities
are automatically improved (at NLO) if the action is improved.

The non-vanishing of $H'_3$ even with non-perturbative improvement
is a consequence of the improvement being on-shell.
Contact terms in correlation functions are not improved,
and this translates into the HECs of $O(a)$ being non-zero.
This is, however, an academic point, since the only
quantity of interest affected by $H'_3$ is the
scalar condensate\cite{ShWuII}, and this is very difficult
to calculate in practice on the lattice.

\subsection{Results for $m_q\sim a\LQCD^2$ (GSM regime)}\label{sec:GSMresults}

We are now ready to reap the benefits of 
the work we have done setting up \chpt\ for tmLQCD,
which I will call tm\chpt.
We can already see one benefit of the \chpt\ technology:
although there are several unknown LECs describing 
discretization errors, no more are needed for the entire
twisted mass plane than for the untwisted Wilson mass axis.

In this section I discuss the results in the GSM
regime, first at LO and then at NLO.

\subsubsection{Tm\chpt\ at LO}\label{sec:tmchptLO}

The complete LO Lagrangian include discretization errors is
\begin{eqnarray}
\mathcal{L}^{(2)}_{\chi,\GSM} &=& 
 \frac{{ f}^2}{4} \tr(D_\mu \Sigma D_\mu \Sigma^\dagger)
-\frac{{ f}^2}{4} \tr(\chi^{\dagger} \Sigma + \Sigma^\dagger\chi) 
- {
\frac{{ f}^2}{4} \tr(\hat{A}^{\dagger} \Sigma + 
               \Sigma^\dagger\hat{A})} 
\,.
\end{eqnarray}
It will be useful to change the notation for sources:
\begin{equation}
\chi = 2 B_0( s+ip) = 2 B_0(m+i\mu\tau_3) + \delta\chi\,,
\qquad
\delta\chi = 2 B_0(\delta s+ i \delta p)
\,.\label{eq:deltachi}
\end{equation}
We now make the simple but important observation that
the factors of $\hat{A}$ can be absorbed by using the
shifted variable\cite{ShSi} { $\chi'=\chi+\hat A$}:
\begin{eqnarray}
\mathcal{L}^{(2)}_{\chi,\GSM} &=& 
 \frac{{ f}^2}{4} \tr(D_\mu \Sigma D_\mu \Sigma^\dagger)
-\frac{{ f}^2}{4} \tr(\chi'^{\dagger} \Sigma + \Sigma^\dagger\chi') 
\,.
\end{eqnarray}
The result has exactly the same form as the LO continuum Lagrangian.
The shift $\chi\to\chi'$ corresponds to an
$O(a)$ shift in the untwisted mass, $m\to m'=m + a {W_0}/{B_0}$,
but leaves $\delta s$ (and, indeed, $\delta\chi$) unchanged..
Recalling the definition of $m$, eq.~(\ref{eq:mdefs}), this
is equivalent to a shift in the critical mass,
$\Delta m_c=- a^2 {Z_S W_0}/{B_0}$.
This shift is not measurable, however, since
$m_c$ is not known {\em a priori}. It must be determined
non-perturbatively from the simulation itself.
The traditional definition is that $m_c$ is the bare mass
at which $m_\pi^2\to 0$ on the Wilson axis.
Since at LO $\CL^{(2)}_{\chi,\GSM}$ predicts $m_\pi^2\propto m'$,
we discover that this ``$m_\pi^2$'' definition of $m_c$
{\em automatically} includes the $O(a^2)$ shift in $m_c$,
and chooses the untwisted quark mass to be $m'$.

The upshot is that, with the standard numerical definition of $m_c$,
the pion and vacuum sectors are automatically
$O(a)$ improved at LO in tm\chpt\ for any twist angle. 
The analysis of this theory
is as in the continuum and I sketch it quickly.
The twist angle $\omega_0$ is defined by\footnote{%
This is another unfortunate notation\cite{ShWuII}.
This is a renormalized twist angle with a good continuum limit, because
of the $Z$-factors in the definitions of $m'$ and $\mu$. The subscript
on $\omega_0$ does not indicate a bare quantity.}
\begin{equation}
(m'+i\tau_3 \mu) \equiv
m_q e^{i\omega_0 \tau_3}\,,\quad (m_q\ \textrm{real and positive})
\end{equation}
giving the geometry shown below (note the renormalized axes).
\vspace{-.5cm}
\begin{figure}[ht]
\centerline{\epsfxsize=3.1in\epsfbox{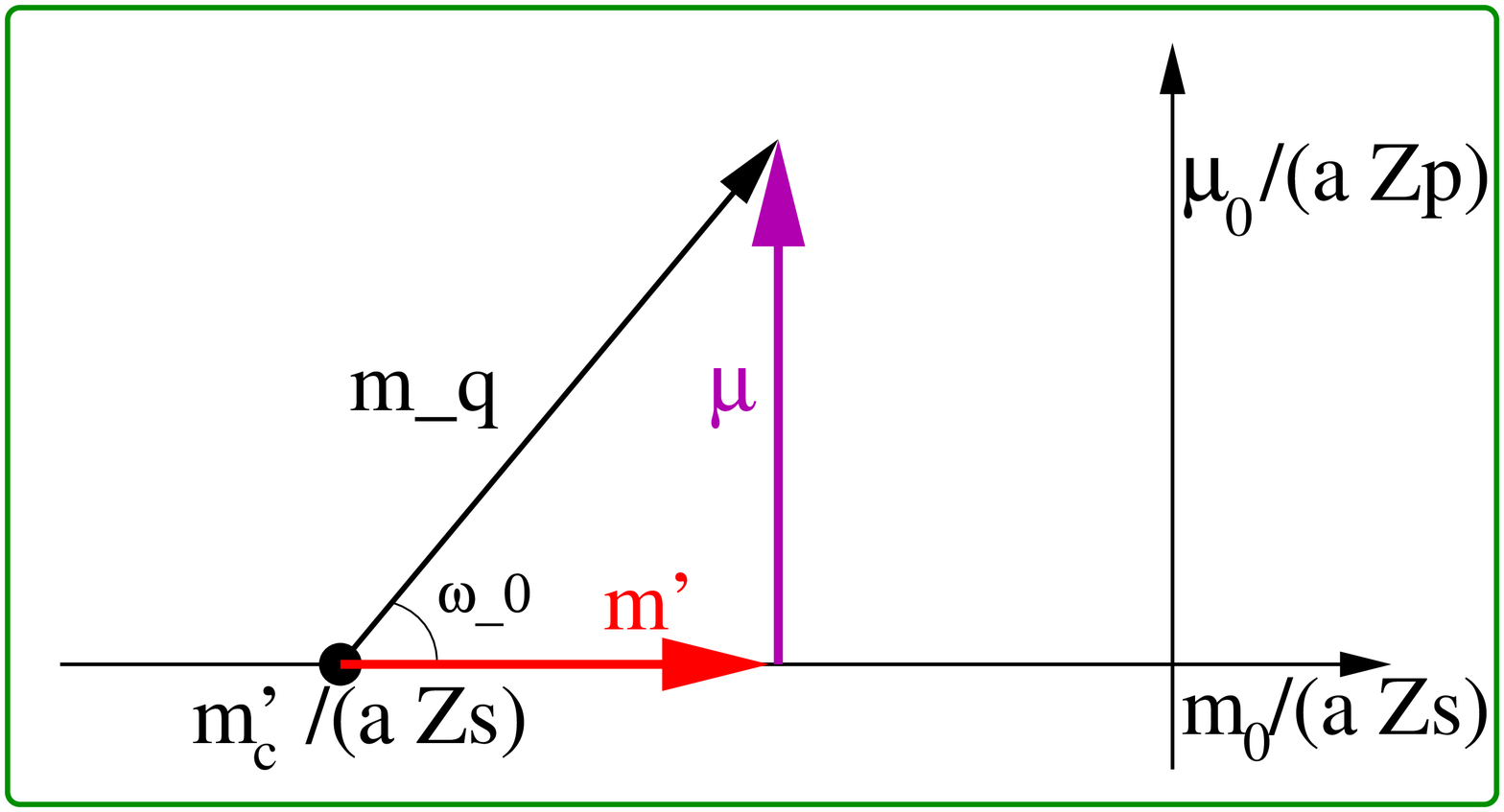}}
\end{figure}
\vspace{-.5cm}

\noindent
The potential is minimized when $\Sigma$ is aligned with $\chi'$:
{$\langle\Sigma\rangle = \exp(i\omega_0\tau_3)$}.
It is conventional to
expand $\Sigma$ about the condensate in a symmetric way,
\begin{equation}
\Sigma = \xi_0 \;
\Sigma_{ph}\;\xi_0\,,\quad
\xi_0\equiv exp(i\omega_0 \tau_3/2)\,,\quad   
\Sigma_{ph}\equiv \exp(i\vec{\pi}\cdot\vec{\tau}/f)
\,,
\end{equation}
since this corresponds diagonalizing the mass matrix with
an axial transformation.\footnote{%
One could equally well use, say, 
a LH transformation to diagonalize $M$,
and a correspondingly asymmetric form for $\Sigma$.
The difference is a $U(1)$ vector transformation that has
no effect on the physics, although there will be extra phases
associated with operators} 
The resulting theory is
\begin{eqnarray}
\mathcal{L}_{\chi,\rm LO} &=& 
 \frac{{ f}^2}{4} \tr(D_\mu \Sigma_{ph} D_\mu \Sigma_{ph}^\dagger)
-\frac{{ f}^2}{4} 
\tr\left\{\left[\widehat{m_q} + (\delta\chi)_{ph}^\dagger\right] \Sigma_{ph} + 
\textrm{h.c.}\right\}
\,,
\end{eqnarray}
where {$\widehat{m_q} \equiv 2 B_0 m_q$} and ``h.c.'' is hermitian conjugate.
The masses and interactions of the pions are 
manifestly independent of $\omega_0$, 
since the mass term has been ``untwisted''.
At the same time, however, sources have been twisted:
$(\delta\chi)_{ph}^\dagger=\xi_0 (\delta\chi)^\dagger \xi_0$ and
\begin{equation}
D_\mu \Sigma_{ph} = \partial_\mu \Sigma_{ph}
- i l_{\mu}^{ph}\Sigma_{ph}
+ i \Sigma_{ph}r_\mu^{ph} \,,\ \textrm{with}\ 
l_\mu^{ph} = \xi_0^\dagger l_\mu \xi_0\,,\
r_\mu^{ph} = \xi_0 r_\mu \xi_0^\dagger \,.
\end{equation}
The operators used to determine physical matrix
elements (obtained by functional derivatives
with respect to the physical sources)
are thus related to those
in the lattice theory (obtained using the original sources)
by an $\omega_0$-dependent twist.
This reproduces the twisting one finds at the quark level:
fields in the original (or  ``twisted'')  basis
[that with $M=m_q\exp(i\omega_0\tau_3)$] are
related to those in the ``physical basis''
[$M=m_q$]  by an axial transformation
[$\psi_{ph} = \exp(i\omega_0\tau_3\gamma_5/2) \psi$], 
so that operators are also transformed\cite{tm}, e.g.
\begin{equation}
\bar u_{ph} \gamma_\mu\gamma_5 d_{ph}
= \cos\omega_0 (\bar u \gamma_\mu\gamma_5 d) 
- i \sin\omega_0 (\bar u \gamma_\mu d)
\,.
\end{equation}
Thus, at maximal twist ($\omega_0=\pi/2$) the charged pion
should be created with the lattice {\em vector current}.

\subsubsection{tm\chpt\ at NLO}

The NLO Lagrangian, rewritten in terms of {$\chi'$}, and dropping HECs, is
\begin{eqnarray}
\lefteqn{\mathcal{L}^{(4)}_{\chi,\GSM} = \CL^{(4)}_\chi + \CL^{(4)}_{\chi,A} }
\nonumber\\
&&=
- { L_{13}} \tr(D_\mu \Sigma D_\mu \Sigma^\dagger)^2
-  L_{2} \tr(D_\mu \Sigma D_\nu \Sigma^\dagger)
      \tr(D_\mu \Sigma D_\nu \Sigma^\dagger) 
\nonumber \\
&&
+ { L_{45}} \tr(D_\mu \Sigma^\dagger D_\mu \Sigma)
         \tr(\chi'^{\dagger} \Sigma +  \Sigma^\dagger\chi')
- { L_{68} }
\big[\tr(\chi'^{\dagger} \Sigma + \Sigma^\dagger\chi')\big]^2
\nonumber\\ &&
{ + {\widetilde W} \tr(D_\mu \Sigma^\dagger D_\mu \Sigma)
         \tr(\hat{A}^{\dagger} \Sigma +
         \Sigma^{\dagger}\hat{A})
- { W} \tr(\chi'^{\dagger} \Sigma + \Sigma^\dagger\chi') 
         \tr(\hat{A}^{\dagger} \Sigma + \Sigma^{\dagger}\hat{A})}
\nonumber\\ &&
{ - { W'} \big[\tr(\hat{A}^{\dagger} \Sigma +
       \Sigma^{\dagger}\hat{A})\big]^2
+ { W_{10}} \tr(D_\mu \hat{A}^\dagger D_\mu \Sigma + 
          D_\mu \Sigma^\dagger D_\mu \hat{A})  } \,.
\end{eqnarray}
Here I used $SU(2)$ relations to combine terms,
eq.~(\ref{eq:L4}), with $L_{13}=L_1+ L_3/2$, $L_{45}=L_4+L_5/2$
and $L_{68}=L_6+L_8/2$.
I have also used the shifted $W$'s defined
in eq.~(\ref{eq:shiftedLECs}). The latter, as noted above, 
turn out to be independent of the renormalization scale.
If we are using a non-perturbatively $O(a)$ improved action 
then $W=\widetilde W=0$, and if we also improve the axial current
then $W_{10}=0$.

Subsequent results are simplified by
the observation that $W_{10}$ is redundant\cite{ShWuII}. 
A change of variables,
{$\delta \Sigma = ({2}{W_{10}}/{f^2}) 
\left(\Sigma \hat{A}^\dagger \Sigma -\hat{A}\right)$}, which keeps
$\Sigma\in SU(3)$ up to NNLO corrections, cancels the $W_{10}$ term
while shifting the other LECs:
$W\to W+W_{10}/4$ and $\widetilde W\to\widetilde W + W_{10}/2$.
Because of this, I set $W_{10}=0$ henceforth.
Note that if one uses a non-perturbatively
$O(a)$ improved action but an unimproved axial current, then after
this change of variables $W$ and $\widetilde{W}$ no longer vanish,
but are related by  $2W-\widetilde W=0$.


I now present a sampling of NLO 
results\cite{Munster,Scorzato,ShWuII,AokiBar04}.
These require as a first step the determination of the condensate
about which to expand, and this is realigned by NLO terms.
In the continuum there is no realignment, because the $L_{68}$ term
is already extremized by the LO condensate,
$\langle\Sigma\rangle =\exp(i\omega_0\tau_3)\propto \chi'$. 
The ``discretization terms'' (those proportional to $W$ and $W'$)
do, however, lead to a realignment, because they ``pull'' the condensate
either towards or away from the identity direction. The result is:
\begin{equation}
\langle\Sigma\rangle = e^{i (\omega_0+\epsilon) \tau_3}\,,
\quad
\epsilon = -\frac{16 \hat{a} \sin\omega_0}{f^2}
\left( {W} + 2 { W'} 
         \cos\omega_0 \frac{\hat{a}}{\widehat{m_q}} \right) 
\end{equation}
Note that $\epsilon$ vanishes
on the Wilson axis ($\omega_0=0,\pi$), and that the $W'$ contribution
 is enhanced
if $\widehat{m_q} \ll \hat{a}$ (in which case one enters the Aoki regime,
where an $O(1)$ vacuum realignment is possible, as discussed below).

While $\epsilon$ is not measurable in simulations (as it is
defined within \chpt\ and not in terms of observables), 
it illustrates the typical magnitude of NLO effects.
Given that we expect $|W|,|\widetilde{W}|,|W'|\sim 1/(4\pi)^2$ and $W_0\sim \LQCD^3$,
it follows that $\epsilon \sim a\LQCD$.
To implement these expectations, it is useful to use rescaled 
variables\cite{observations}
\begin{equation}
{\delta_W} =  \frac{16 \hat{a} W}{f^2}\sim a\LQCD,\ 
 {\delta_{\widetilde W}} =  \frac{16 \hat{a} \widetilde W}{f^2}\sim a\LQCD,\
w' = \frac{16 \hat{a}^2 W'}{f^2} \sim a^2\LQCD^4.
\label{eq:deltaWdef}
\end{equation}

The general form of NLO results for observables is illustrated by
\begin{eqnarray}
m_{\pi^{\pm}}^2 &=& \widehat{m_q}\left[1 + \frac12 L_\pi + 
\frac{16}{f^2} \widehat{m_q} (2L_{68}-L_{45}) \right]
 \nonumber \\
&&+ \widehat{m_q} \cos\omega_0 { (2 \delta_W-\delta_{\widetilde W})}
+ 2 (\cos\omega_0)^2 w'
\,.\label{eq:mpisqGSM}
\end{eqnarray}
The first line is the continuum NLO result 
[the $SU(2)$ version of eq.~(\ref{eq:mpipm}), with chiral
logs defined in (\ref{eq:chirallog})],
while the second shows the impact of discretization.
Recall that $\widehat{m_q}$ is defined to be positive.
The scale dependence of the chiral log
is absorbed by $2 L_{68}-L_{45}$, leaving $2\delta_W-\delta_{\widetilde W}$
scale invariant. 
Chiral logs do not contain discretization corrections at this order 
because the LO discretization errors can be absorbed into $\chi'$.

The result (\ref{eq:mpisqGSM}) shows the different possibilities 
for removing $O(a)$ errors.
\begin{itemize}
\item
Non-perturbatively $O(a)$ improve the quark action,
in which case $2\delta_W-\delta_{\widetilde W}=0$ and the $O(a)$ term
vanishes.
\item
Use ``mass averaging''\cite{FR}
in which one averages over $\omega_0$ and $\omega_0+\pi$
at fixed $\widehat{m_q}$. This flips the sign of both
$m'$ and $\mu$, and thus of $\cos\omega_0$,
and cancels the $O(a)$ term. It has been shown to lead to $O(a)$
improvement for all physical quantities, including matrix
elements\cite{FR}.
\item
Work at maximal twist, $\omega=\pm\pi/2$ (both choices of sign
are equivalent, and I use the positive sign henceforth).
This removes the $O(a)$ term
(as it must since it is a special case of mass averaging
where the average is automatic\cite{FR}),
and, in this case though not in general, also the $O(a^2)$ term.
\end{itemize}

\begin{figure}[ht]
\begin{center}
\includegraphics[angle=-90,width=14cm]{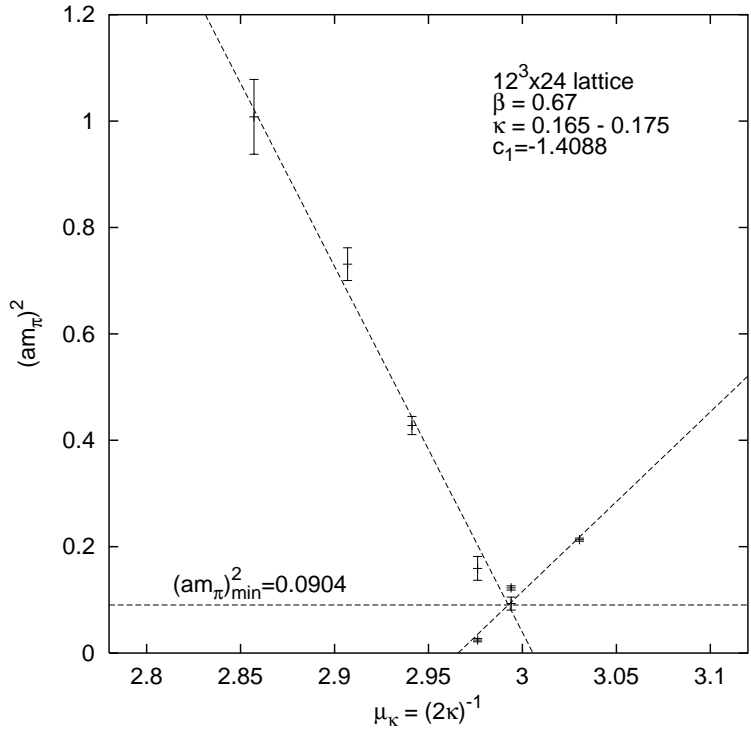}
\end{center}
\caption{Unquenched results for $(am_\pi)^2$ as a 
function of $(2\kappa)^{-1}= m_0+4$ 
for $\mu=0$ and with $a^{-1}\approx 0.2\;$fm\protect\cite{Farchioni0410031}.
Straight lines are to guide the eye.}
\label{fig:mpi_mu}
\end{figure}

I note two further features of the result (\ref{eq:mpisqGSM}). 
The $O(a)$ errors
are determined throughout the twisted-mass plane by the combination
$2\delta_W-\delta_{\widetilde W}$. In particular, on the Wilson axis, this term
predicts an asymmetry in the slopes on the two sides of $m_c$.
This has recently been observed numerically, as part of the initial
studies of the properties of tmLQCD. I show an example of the
results in Fig.~\ref{fig:mpi_mu}. 
Defining the asymmetry in a quantity $Q$ as\cite{observations}
\begin{equation}
{\rm AS}(Q) \equiv \frac{Q(m',\mu) - (-)^p Q(-m',\mu)}
                  {Q(m',\mu)  + (-)^p Q(-m',\mu)}
\,,
\end{equation}
(with $p$ a parity which is +1 for most quantities),
one finds
\begin{equation}
{\rm AS}(m_{\pi^\pm}^2) 
=  ({m'}/{m_q}){(2\delta_W \!-\! \delta_{\widetilde W})}
\stackrel{\mu=0}{\longrightarrow} 
{(2\delta_W - \delta_{\widetilde W})}
\,.
\label{eq:ASmpisq}
\end{equation}
The observed asymmetry in fig.~\ref{fig:mpi_mu} is $\sim 0.3$,
consistent with the expected size 
($a^{-1}\approx 1\;$GeV and $\LQCD\approx 0.3\;$GeV).

The second feature of (\ref{eq:mpisqGSM}) that I want to emphasize is
that the $w'$ term (proportional to $a^2$) gives an {\em additive}
correction to $m_\pi^2$. This is indicative of the breaking
of chiral symmetry. It can be of either sign. If $w'>0$ then one
expects a minimum pion mass, as observed in fig.~\ref{fig:mpi_mu},
while if $w'<0$ the pion mass-squared can become negative, indicating
an instability. This leads to the well-known Aoki 
phase\cite{Aokiphase}.\footnote{%
Since these phenomena occur when $m_q \sim a^2\LQCD^3$,
i.e. in the Aoki regime, there are corrections to 
(\ref{eq:mpisqGSM}), which will be discussed below.}

A graphic demonstration of the impact of discretization errors
is provided by the contours of $m_\pi^2$ in the twisted mass
plane. Examples are shown later in fig.~\ref{fig:contour}.
At LO the contours are circles (as in the continuum), 
indicating that $\omega_0$ is redundant, 
but they are significantly deformed by the discretization errors
that enter at NLO.

Discretization errors linear in $a$ can be separated from
the continuum parts of the predictions by using the
asymmetries defined above.
These are pure discretization effects, while the corresponding
symmetries are $O(a)$ improved because symmetrization is the
essentially the same as mass averaging.
The asymmetries for quantities 
accessible with simulations are\cite{ShWuII,observations}
\begin{eqnarray}
\!\!\!\!\!\!\!\! 2\, {\rm AS}(\langle 0|P^\mp|\pi^\pm\rangle)
&=& AS(\langle\pi|S^0|\pi\rangle) = {\rm AS}(m_{\pi^\pm}^2)
= ({m'}/{m_q})
{(2\delta_W \!-\! \delta_{\widetilde W})}
\label{eq:ASI}\\
{\rm AS}(f_\pi) &=& (m'/m_q)
{\delta_{\widetilde W}}/2\,,
\qquad
{\rm AS}(m_{\rm PCAC}) = ({m_q}/{m'})
{\delta_{W}} \,. \label{eq:ASII}
\end{eqnarray}
What is predicted here are the form of the mass dependence,
and the relations between asymmetries in different quantities.
The PCAC mass is defined below in eq.~(\ref{eq:mPCACdef}).
Note that the asymmetries on the first line vanish for a non-perturbatively
improved action, while those on the second do not unless the axial current
is also non-perturbatively improved.

TmLQCD explicitly breaks flavor and parity symmetries,
and tm\chpt\ can be used to see how such breaking
effects physical quantities. Of particular interest
is flavor breaking, e.g. the splitting of the pion multiplet
or the $\Delta$-baryon multiplet. The good news is that such
breaking is automatically of $O(a^2)$ for any twist angle. 
This is because $\mu\gamma_5\tau_3$ structure of the $O(a)$ flavor breaking 
implies that it contributes at linear order
only to parity-violating matrix elements. 
To obtain a contribution to masses one needs two 
parity-breaking insertions
(to bring one back to the original parity).
An example which can be studied using tm\chpt\ at NLO is
\begin{equation}
m_{\pi^0}^2 -  m_{\pi^\pm}^2 = 
-2 w' (\sin\omega_0)^2  
= -2 w' \frac{\mu^2}{m'^2 + \mu^2} \,.
\label{eq:deltampisq}
\end{equation}
(Recall that $w'\sim a^2$.)
The splitting must vanish on the Wilson axis (as there is then no
flavor breaking), and is necessarily even in $\mu^2$
from the argument above.
Not surprisingly, the splitting is maximized for $\omega_0=\pi/2$.
Similar results hold for $\Delta$ baryons\cite{WalkerLoudWu}.
Which pion is heavier depends on the sign of $w'$, which, as already
noted, also determines the nature of the phase structure in the
Aoki regime\cite{Scorzato}. 

Flavor-parity breaking of $O(a)$ does occur for unphysical
parity-violating matrix elements.
This is true for any non-zero twist, including maximal twist---the
argument for automatic $O(a)$ improvement\cite{FR} does not
hold for such quantities.
Results for the following parity-flavor violating
form factors are available in tm\chpt\cite{ShWuII}:
$\langle \pi_b |{A}_{ph,\mu}^{b},P_{ph}^{b}|\pi_3\rangle $,
$\langle \pi_b |{A}_{ph,\mu}^{3},P_{ph}^3|\pi_b\rangle $ and
$\langle \pi_3 |{A}_{ph,\mu}^{3},P_{ph}^3|\pi_3\rangle$,
where $b=1,2$. Here is one example:
\begin{eqnarray}
\langle \pi_b(p_2) |{P}_{ph}^{3}|\pi_b(p_1)\rangle &=&
i B_0 \sin\omega_0 
\left[
  \delta_W -  \delta_{\widetilde W}
- \frac{{(2 \delta_W-\delta_{\widetilde W})} q^2}{2(q^2 + m_{\pi_3}^2)} 
\right] \nonumber \\
&&+ \frac{i B_0 \sin(2\omega_0) w'}{q^2 + m_{\pi_3}^2}\,,
\label{eq:parityviolating}
\end{eqnarray}
where $q=p_1-p_2$ is the momentum transfer.
These quantities provide an interesting window into the workings
of tmLQCD, and are predicted once one has determined the LECs,
but are difficult to study numerically because they involve
quark-disconnected contractions. 

I am aware of one detailed comparison of the results from simulations
of dynamical tmLQCD with tm\chpt\ formulae\cite{Farchioni05fit}.
(There is also a detailed fit to quenched data\cite{AokiBarlat05}
which I discuss below.)
The simulations are for relatively coarse lattices ($a\approx 0.13$ and
$0.18\;$fm), yet find reasonable agreement with the NLO forms
sketched above, with the continuum LECs consistent with continuum results,
and the magnitudes of the ``lattice'' LECs consistent
with expectations (although poorly determined). 
The asymmetries discussed above are clearly present as a function of $m'$.
The authors do their fits versus the PCAC mass, however, 
and for these the asymmetries are reduced, leading to the poor
determination of the lattice LECs.

These fits give one confidence that tm\chpt\ is a useful tool, 
and is likely to become more so as lattice spacings and quark masses
are reduced.
Thus it can aid extrapolations, and, perhaps more importantly,
guide the investigation of the properties of tmLQCD, in particular
its phase diagram and the issue of defining maximal twist.
I return to these issues shortly.

Other simulations have studied the pion mass 
splitting\cite{pionsplittingI,pionsplittingII,pionsplittingIII}.
This involves both quark-connected and disconnected contributions,
with the latter hard to calculate accurately.
Nevertheless, this is a key quantity to determine as it both
sets the scale for isospin breaking and the size and nature of
the phase structure. 
TmLQCD has some similarity to staggered fermions (although not the
need to use rooting) because a desired continuum symmetry  is broken
(flavor for twisted mass fermions, taste for staggered), in both cases
at $O(a^2)$. Fits to staggered fermions require this taste-breaking
to be treated at {\em leading order} in ``staggered'' \chpt\
(i.e. the equivalent of the Aoki regime here), because 
taste-splittings are large. Numerically,
the splitting are $\sim a^2\Lambda^4$ with $\Lambda\approx 1\;$GeV.
The hope for tmLQCD is that the flavor-breaking is smaller,
and can be treated as a NLO effect (the GSM regime).
One can also tune its size by varying
the quark and gluon actions.
Quenched results for the mass splitting 
(with Wilson fermion and gluon actions and $a\approx 0.1\;$fm)
find $\Lambda\approx 0.7\;$GeV
\cite{pionsplittingI} 
(i.e. four times smaller splittings than with staggered fermions), 
and there are
indications of a significant reduction if one uses dynamical quarks
and improved gluon actions\cite{pionsplittingII} 
or the non-perturbatively improved quark action\cite{pionsplittingIII}.
Thus the situation is promising.

\subsection{Defining $m_c$ and the twist angle}\label{sec:maxtwist}

The critical mass plays a central role in tmLQCD,
providing the origin about which one defines the twist angle
(see the figure in sec.~\ref{sec:tmchptLO}).
As noted above, it must be determined non-perturbatively
as part of the simulation (and recalculated for each lattice
spacing and choice of actions). 
The questions I address in this section are these:
How accurately does one
need to determine $m_c$ in order that automatic
$O(a)$ improvement holds at maximal twist?
What methods allow this accuracy to be achieved?
I will discuss these questions
using the framework of tm\chpt, rather than than
use the Symanzik EFT as in the original
treatment\cite{FR} and subsequent 
extensions\cite{AokiBar04,FRMP,AokiBar06}.
Tm\chpt\ is less general (referring only to the vacuum
and pion sectors), but more powerful in its domain
of applicability (as it starts from the Symanzik EFT
and includes further non-perturbative information).

The form of the $O(a)$ correction in the result for
$m_\pi^2$, eq.~(\ref{eq:mpisqGSM}), is generic, namely
that it is proportional to $a \cos\omega_0$.
To obtain automatic $O(a)$ improvement one 
needs $\cos\omega_0=O(a)$ and thus $\omega_0=\pi/2 + O(a)$.
In other words, maximal twist can mean ``maximal up to
$O(a)$''.
What does this imply for the required accuracy in the
determination of $m_c$? This depends on the relative
size of quark masses and discretization effects.
In the GSM regime, with $\mu\sim a\LQCD^2$,
an $O(a)$ accuracy in $\omega_0$ requires $m'= O(a^2)$
(and thus a determination of the dimensionless
critical mass $m_c$ with an accuracy of $O(a^3)$).
This is relatively straightforward to achieve, as we will see.
Once one enters the Aoki regime, $\mu\sim a^2\LQCD^3$,
the required accuracy increases to $m'=O(a^3)$.
Since simulations are likely to need to enter this
regime, it is important to know how to achieve this 
greater precision.

The traditional definition of $m_c$
used with Wilson fermions is to extrapolate $m_0$ to the point
where $m_\pi^2=0$. This method might be adequate in the
GSM regime, but fails in the Aoki regime\cite{AokiBar04}.
This failure has its origin in the 
$a^2$ ($w'$) term in eq.~(\ref{eq:mpisqGSM}) and will be discussed
in detail in the next section.
Either the pion mass does not
vanish but reaches a non-zero minimum at $m_c$
(at a first-order phase boundary), in which case extrapolating
to $m_\pi^2=0$ overshoots,
or it vanishes over a {\em range}
of $m'$ of width $a^2$ (the Aoki phase),
with the correct choice of $m_c$ being in the middle of the
range\cite{AokiBar04} but the extrapolation giving one of the end-points.
In either case, $m'\sim O(a^2)$ at the putative critical point,
which is not accurate enough for the Aoki regime.
There are also practical issues with this method, reflecting
the difficulty in doing accurate extrapolations, but
I will not belabor them as this method is no longer being
used in practice.

To do better one can adapt the method
used to determine the normalization of currents 
and improvement coefficients, i.e. 
enforce the symmetries that are broken by discretization.
Here, parity and flavor are broken explicitly, but are
restored in the continuum limit. Enforcing this restoration
in particular correlators
for $a\ne 0$ gives a non-perturbative determination of the twist angle,
which can, if desired, be tuned to maximal twist.
Since flavor and parity {\em are} broken, different choices of correlator
lead to $O(a)$ differences in the twist angle, but all choices
lead to automatic $O(a)$ improvement.

I will need to use
the relation between twisted and physical bases discussed
in sec.~\ref{sec:tmchptLO}. Using this, a simple calculation 
finds the following relations
between currents and densities in the two bases:
\begin{eqnarray}
{A}_{ph,\mu}^b &=& 
\cos\omega \, A_\mu^b + \epsilon^{3bc}\sin\omega\, V_\mu^c\,,\quad
{A}_{ph,\mu}^3 = A_\mu^3\,, 
\label{eq:Aph}\\
V_{ph,\mu}^b &=& 
\cos\omega\, V_\mu^b + \epsilon^{3bc}\sin\omega\, A_\mu^c\,,\quad
{V}_{ph,\mu}^3 = V_\mu^3\,, 
\label{eq:Vph}\\
{P}_{ph}^3 &=& \cos\omega\, P^3 + i \sin\omega\, S^0/2\,,\quad
{P}_{ph}^b = P^b\,,
\label{eq:Pph}\\
{S}_{ph}^0 &=& \cos\omega\, S^0 + 2 i \sin\omega\, P^3\,,
\label{eq:Sph}
\end{eqnarray}
where the flavor label $b=1,2$. The flavor non-singlet scalar
density vanishes at LO in $SU(2)$ \chpt\ and I do not discuss it.
What these relations mean is that, if you are working in
the twisted basis (as one usually does on the lattice), then
to construct the physical axial current with, say, $b=1$,
you must take a linear combination of the lattice
currents $A^1_\mu$ and $V^2_\mu$. These relations assume
that the currents have been correctly normalized by multiplying by
their corresponding Z-factors. 
Note that $A_\mu^3$ and $P^b$ do not rotate, and so
are good choices to create physical pions.

The idea is now to take either (\ref{eq:Aph},\ref{eq:Vph}) or
(\ref{eq:Pph},\ref{eq:Sph}) as a {\em definition} of $\omega$,
and enforce parity-flavor restoration in a particular correlator.
Two examples are [method (ii) will be explained later]:
\begin{itemize}
\item Method (i)\cite{omegaAref} (``$\omega_A$ method'') 
$\langle{V}^2_{ph,\mu}(x) {P}_{ph}^1(y) \rangle
\propto \langle 0| {V}^2_{ph,\mu} | \pi^1\rangle = 0$ .
\item Method (iii)\cite{ShWuII} (``$\omega_P$ method'') 
$\langle {S}_{ph}^0(x) {A}^3_{ph,\mu}(y) \rangle
\propto \langle 0| {S}_{ph}^0 | \pi^3\rangle = 0$ .
\end{itemize}
The correlators are to be evaluated for $x\ne y$,
and the long-distance contribution is as indicated.
Using (\ref{eq:Aph}-\ref{eq:Sph}) one can
manipulate these criteria into results for the twist angle
in terms of correlators in the twisted basis:
\begin{equation}
\tan\omega_A \equiv \frac{\langle V_\mu^2(x) P^1(y) \rangle}
                       {\langle A_\mu^1(x) P^1(y) \rangle} 
\,,\quad
\tan\omega_P \equiv \frac{i \langle S^0(x) A_\mu^3(y)\rangle}
                         {2 \langle P^3(x) A_\mu^3(y)\rangle}\,.
\label{eq:omegaAP}
\end{equation}
Maximal twist occurs when the denominators vanish, i.e
\begin{equation}
\omega_A=\pi/2 \Rightarrow \langle A_\mu^1(x) P^1(y) \rangle = 0\,,
\quad
\omega_P=\pi/2 \Rightarrow \langle A_\mu^3(x) P^3(y) \rangle = 0\,.
\label{eq:maxtwistAP}
\end{equation}
While superficially similar, the two criteria differ because
of flavor breaking. The correlator in method (iii)
includes quark-disconnected contractions and is much more difficult to
calculate in practice. I include it for illustrative reasons.
Method (i) is used in practice.\footnote{%
It has also been called the parity-violating method, or the PCAC method.}
One fixes $\mu_0$
and varies $m_0$ until $\omega_A=\pi/2$.
The resulting $m_0(\mu_0)$ depends on the choice of
discretization of the axial current (e.g. $O(a)$ improved or not),
and, in general, upon the separation $x-y$. 
At large distances, which are used in practice,
the pion contribution dominates and the resulting $m_0$ becomes
independent of separation.
At such distances, method (i) is
equivalent to the vanishing of the PCAC mass:
\begin{equation}
m_{PCAC} =
{\langle \partial_\mu A_\mu^b(x) P^b(y) \rangle}/
     {2 \langle P^b(x) P^b(y) \rangle}=0\,.
\label{eq:mPCACdef}
\end{equation}
One nice feature of either criterion is that
knowledge of $Z$-factors is not required, unlike the determination
of $\omega$ at non-maximal twist.\footnote{%
A generalization of method (i) allows the determination of
$\omega_A$ for any twist angle
without {\em a priori} knowledge of $Z_{A,V}$\cite{omegaAref}.}

\begin{figure}[ht]
\begin{center}
\includegraphics[width=7cm]{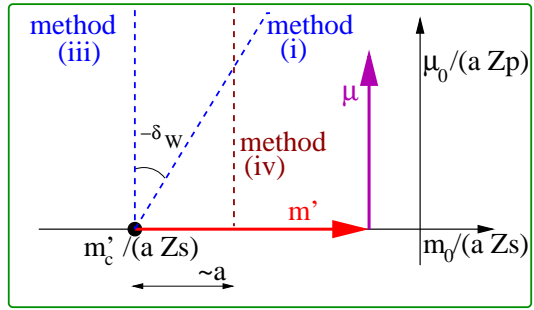}
\caption{Results from different methods of defining maximal twist}.
\label{fig:maxtwist}
\end{center}
\end{figure}

Both methods (i) and (iii) can be implemented 
in tm\chpt. With the technology developed above,
we can work at NLO in the GSM regime, and
calculate the correlators at long distances,
for then the pion contribution dominates.
I quote only the results at maximal twist:\cite{ShWuII}
\begin{equation}
\textrm{(i):}\ 
 \omega_A=\pi/2 \  \Rightarrow\ 
\omega_0 = \pi/2 + {\delta_W}\,,
\quad
\textrm{(iii):}\ 
\omega_P=\pi/2 \ \Rightarrow\ 
\omega_0 =  \pi/2\,,
\end{equation}
[with $\delta_W$ defined in eq.~(\ref{eq:deltaWdef})].
These results are shown in fig.~\ref{fig:maxtwist}.
In method (i) one finds a line at an angle $\delta_W\sim a$
to the vertical. Thus one is not at maximal twist in terms of
$\omega_0$, but $\cos\omega_0\sim a$ so automatic improvement still holds.
It turns out that method (iii) leads to a vertical approach to
the Wilson axis. The methods come together at the critical mass.
Thus one could implement method (iii) by using method (i) for $\mu\ne0$,
extrapolating to $\mu=0$ to determine $m_c$, and then working at fixed $m_0=m_c$.

Quenched simulations at $1/a\approx 2\;$GeV with unimproved Wilson fermions
find an offset angle{ $\delta_W\approx -0.35$}\cite{Lewis,Jansen05}. 
This is a direct measure of discretization errors.
Written as $a\Lambda^2$ this gives { $\Lambda\approx 0.7\;$GeV}, 
a large but not unreasonable value.

As noted above, the pion mass method can also be used, in principle, within
the GSM regime. In practice the extrapolation can miss the critical mass,
so it is useful to define a ``straw-man'' method (iv) in which $m'$ is held fixed at
a value of $O(a)$. As is clear from the figure, $\omega_0$ varies from
$\sim 1$ to $0$ as $\mu$ varies from $\sim a$ to $0$. 
Thus one is never near maximal twist in the GSM regime,
and automatic improvement is lost.

\subsection{Results for $m_q\sim a^2\LQCD^3$ (Aoki regime)}\label{sec:Aokiregime}

As discussed above, to describe simulations we need to extend
the analysis into the Aoki regime. This requires a change in
the power counting:\cite{AokiBarlat05,observations,AokiBar06}
 at LO we keep terms of size $m'\sim\mu\sim a^2$, 
\begin{equation}
\!{\CL}_{\chi,Aoki}^{\rm LO} = 
 \frac{{ f}^2}{4} \tr(D_\mu \Sigma D_\mu \Sigma^\dagger)
\!-\!\frac{{ f}^2}{4} \tr(\chi'^{\dagger} \Sigma \!+\! p.c.) 
\!-\! { W'} \big[\tr(\hat{A}^{\dagger} \Sigma \!+\! p.c.)\big]^2,
\label{eq:LAokiLO}
\end{equation}
while at NLO we keep those of size $m a\sim \mu a\sim a^3$,
\begin{eqnarray}
\lefteqn{\! {\CL}_{\chi,Aoki}^{\rm NLO} =
- \frac{W_{3,1}}{f^2}
\tr(\hat{A}^\dagger\hat A)\Tr(\hat{A}^\dagger\Sigma+p.c.)
- \frac{W_{3,3}}{f^2}
\left[ \tr(\hat{A}^\dagger\Sigma)^3 + p.c. \right]
}
\nonumber\\ && \!\!\!\!\!
+ {\widetilde W} \tr(D_\mu \Sigma^\dagger D_\mu \Sigma)
         \tr(\hat{A}^{\dagger} \Sigma\! + p.c.)
- { W} \tr(\chi'^{\dagger} \Sigma + p.c.) 
         \tr(\hat{A}^{\dagger} \Sigma\! + p.c.).
\label{eq:LAokiNLO}
\end{eqnarray}
The two $a^3$ terms (those on the first line) are new, while
previous NLO terms $\propto m^2$ become of NNLO.
In fact, the $W_{3,1}$ term can be absorbed into $\chi'$
by a further shift of $O(a^4)$ in $m_c$. 
This leaves one new LEC at NLO, $W_{3,3}$. 
Note that only the source parts of the
$W$ and $\widetilde W$ terms enter at NLO.

In the following I will describe the analysis of the
phase structure at LO, and then mention some modifications
caused by NLO terms.


The orientation of $\langle\Sigma\rangle$ is now determined by
an {\em equal} competition between the mass and $a^2$ terms in
$\CL_{\chi, Aoki}^{\rm LO}$. The former is minimized when 
$\langle\Sigma\rangle\propto \chi'\propto \exp(i\omega_0\tau_3)$,
while the latter either favors
$\langle\Sigma\rangle=\pm 1$ ($W'>0$) or 
$\langle\Sigma\rangle=\exp[i(\pi/2)\hat{n}\cdot\vec\tau]$ 
with $\hat{n}^2=1$ ($W'<0$).
The analysis along the Wilson axis is simple\cite{ShSi},
and one finds two cases. Either the condensate jumps discontinuously from 
$\langle\Sigma\rangle=+1$ to $\langle\Sigma\rangle=-1$ at a
first-order transition ($W'>0$),
or it swings between these values continuously, as is possible within $SU(2)$
($W'<0$). In the latter case, the condensate breaks flavor,
so there are exact lattice Goldstone bosons.
This is the Aoki phase\cite{Aokiphase}.
The presence of two possible phase structures was also predicted by 
Creutz\cite{Creutzaokiphase}.

Moving into the twisted-mass plane the analysis becomes more complicated
as one must minimize a quartic\cite{Munster,Scorzato,ShWuII}.
I show below how the two scenarios on the Wilson axis extend 
into the mass plane.
\begin{center}
\includegraphics[width=4cm]{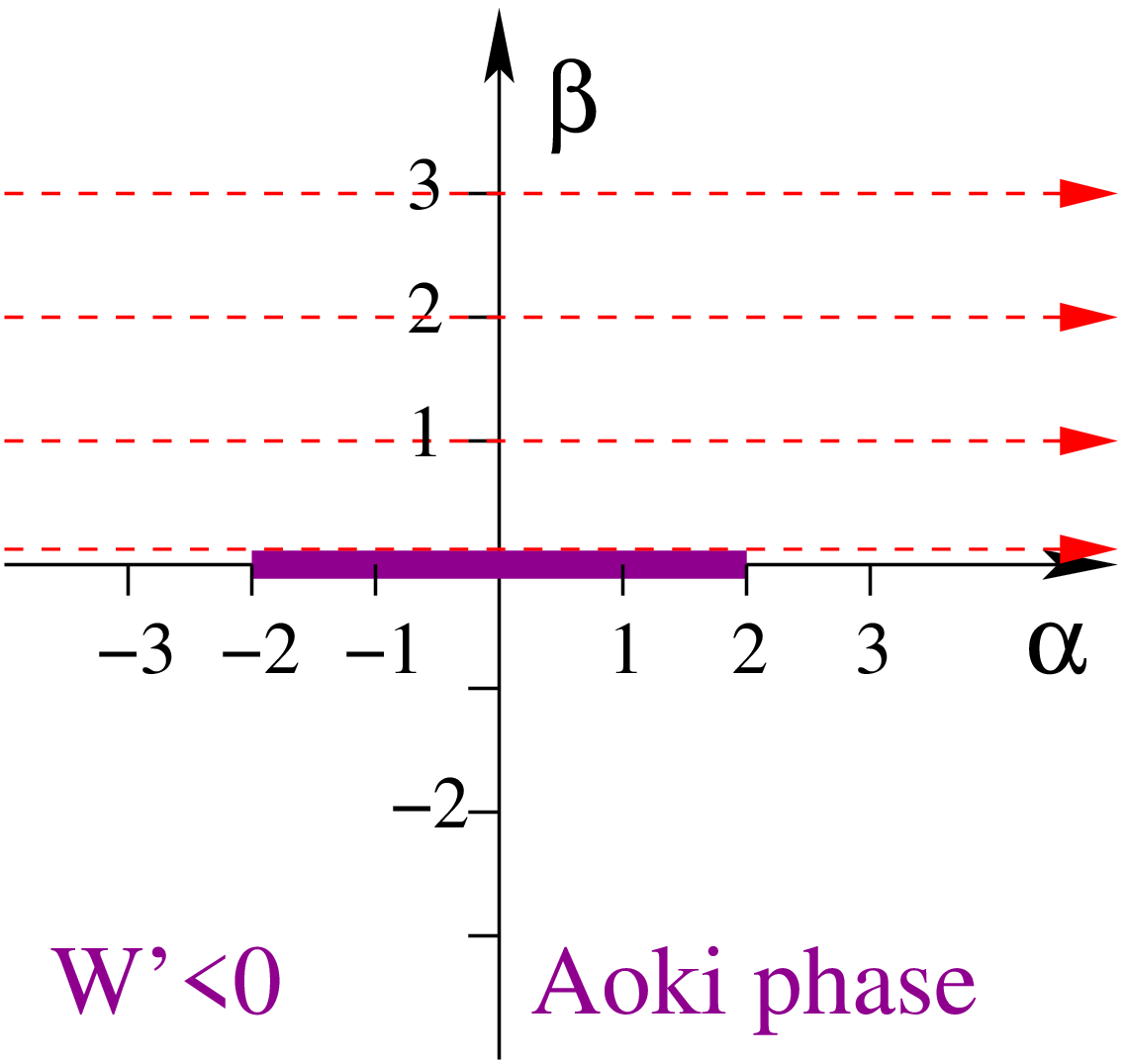}
\hspace{1truecm}
\includegraphics[width=4cm]{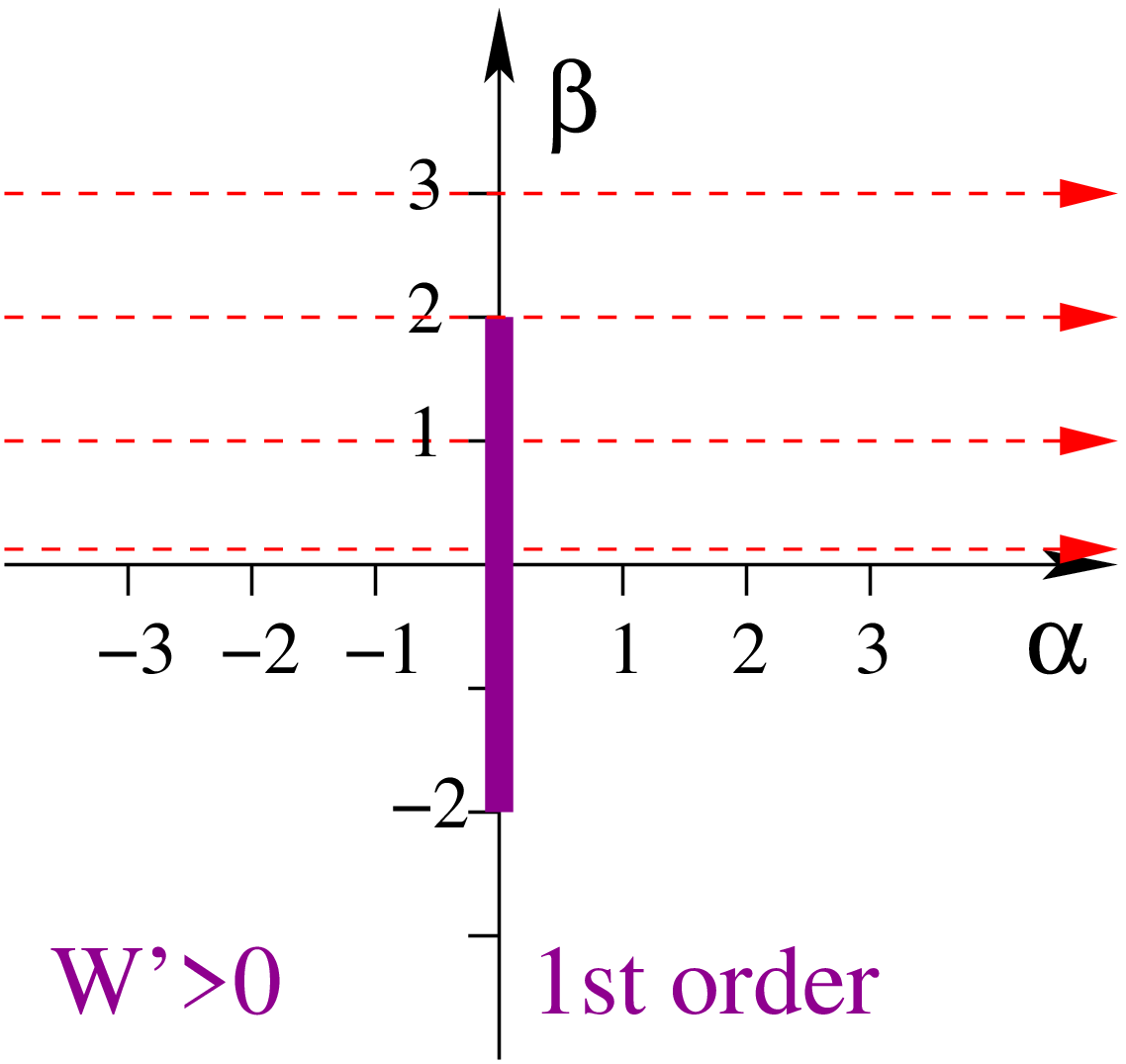}
\end{center}
Here I use mass variables rescaled by $\sim a^2$:
\begin{equation}
\alpha= \frac{2 B_0 m'}{(16 { |W'|} \hat{a}^2/f^2)}=\frac{\hat{m'}}{|w'|}
\,,\quad
\beta= \frac{2 B_0 \mu}{(16 { |W'|} \hat{a}^2/f^2)}=\frac{\hat{\mu}}{|w'|}\,.
\label{eq:alphabeta}
\end{equation}
Both scenarios have a first-order transition boundary, indicated by
the solid line, with second-order end-points.
Note again that the parameter $w'$, whose sign determines which scenario
applies, and whose magnitude gives the size of the phase boundaries,
is the same parameter as appears in the pion mass splitting\cite{Scorzato},
eq.~(\ref{eq:deltampisq}). Thus a calculation of this splitting
{\em in the GSM regime} (which, as noted above, has been attempted) predicts
the phase structure {\em in the Aoki regime}.

\begin{figure}[htb]
\begin{center}
\includegraphics[width=5.5cm]{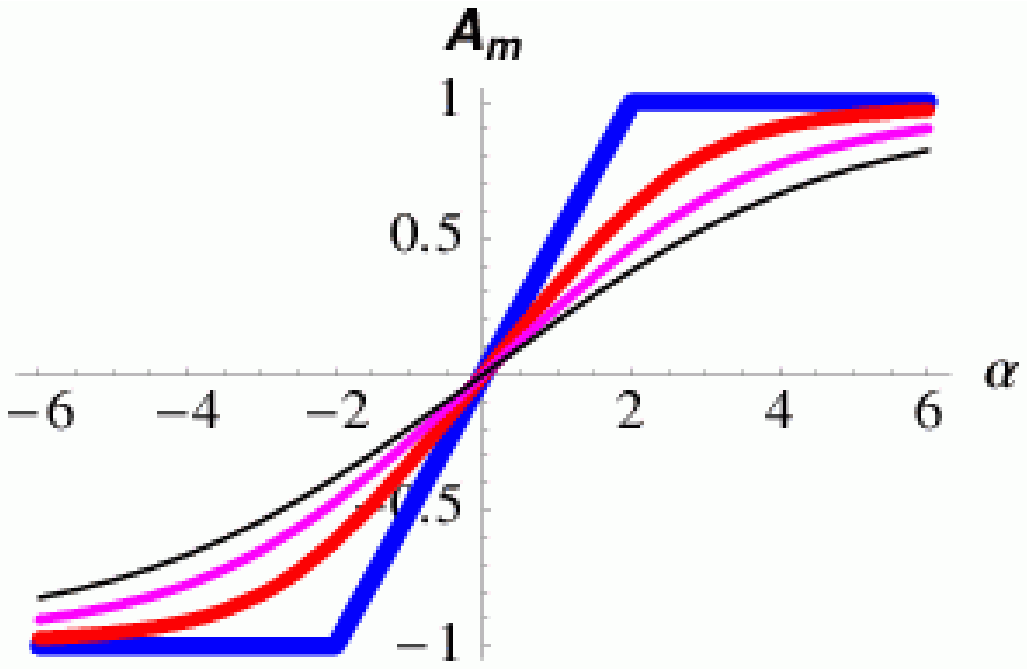}   
\includegraphics[width=5.5cm]{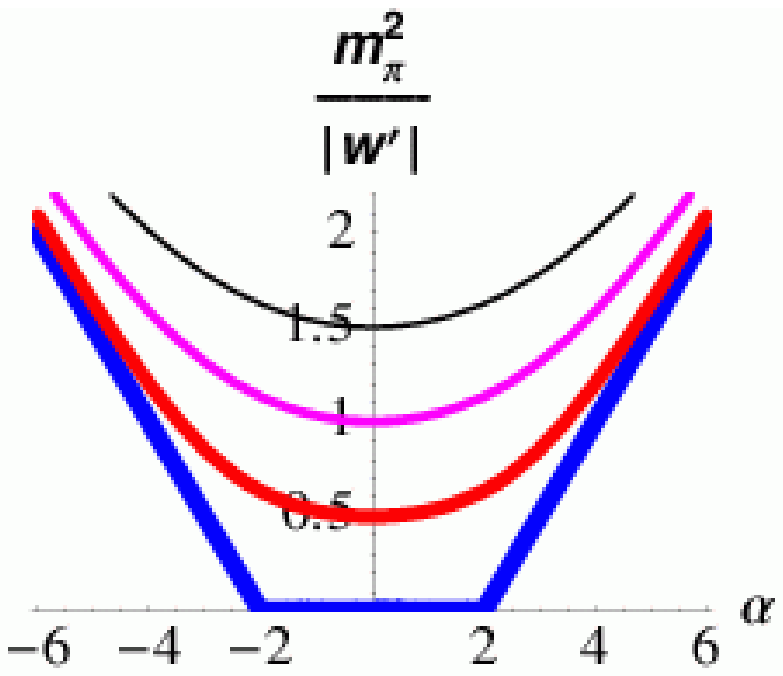} 
\end{center}
\caption{Condensate and charged pion masses for $W'<0$, for four
choices of twisted mass ($\beta=0,1,2,3$).
The thickness of the lines decreases with increasing $\beta$.}
\label{fig:Aokicond}
\end{figure}

Wu and I have given detailed plots of the condensate and pion masses along
the dashed horizontal lines in the phase diagrams above\cite{ShWuII}.
I show here only a sample. I write the condensate as
$\langle \Sigma\rangle = A_m + i B_m \tau_3$, and plot the
scalar component $A_m$, as well as the charged and neutral pion mass-squareds.
Figure~\ref{fig:Aokicond} shows results for $W'<0$.
The ``swinging'' of the
condensate between $\pm 1$, described in words above, is here the
$\beta=0$ curve. The charged pion masses vanish when $|A_m|< 1$, for then
$|B_m|> 0$ and flavor is spontaneously
broken. Moving away from the Wilson axis ($\beta=1,2,3$),
the Aoki-phase is washed out by the explicit breaking of flavor.
The neutral pion (not shown) is always heavier than the charged pions.

If the mass is purely twisted ($\alpha=m'=0$) then one finds 
for all $\beta$ that $A_m=0$,
so the condensate points in the $\tau_3$ direction. Thus, despite
the large discretization errors, the condensate ends up pointing in
the same direction as it would without them. This is one way
of understanding why automatic
$O(a)$ improvement continues to work, as discussed further below.

\begin{figure}[htb]
\begin{center}
\includegraphics[width=5.5cm]{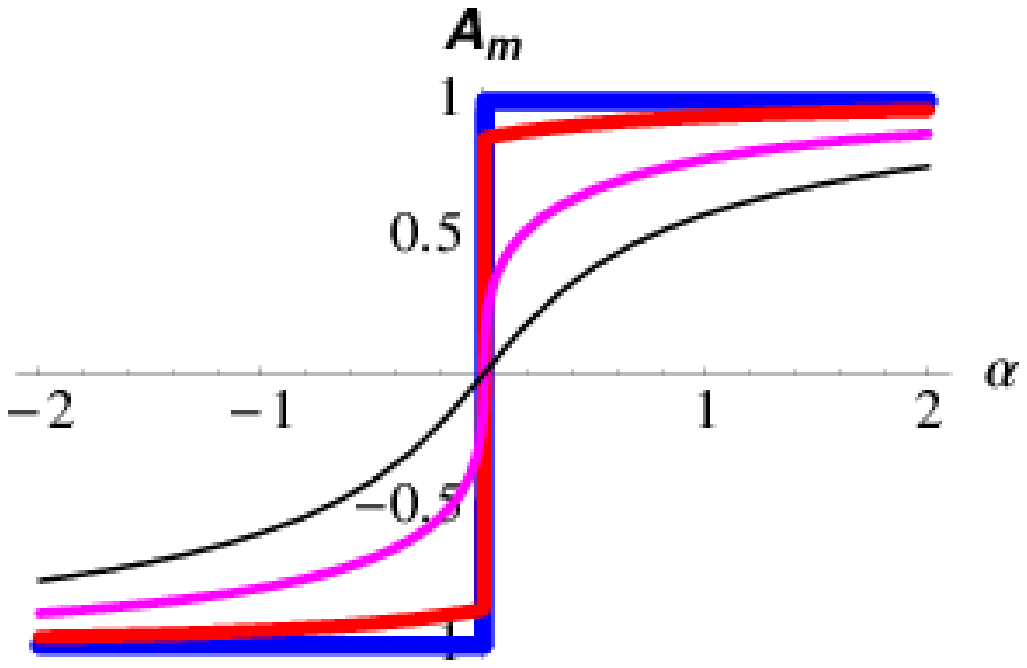} 
\includegraphics[width=5.5cm]{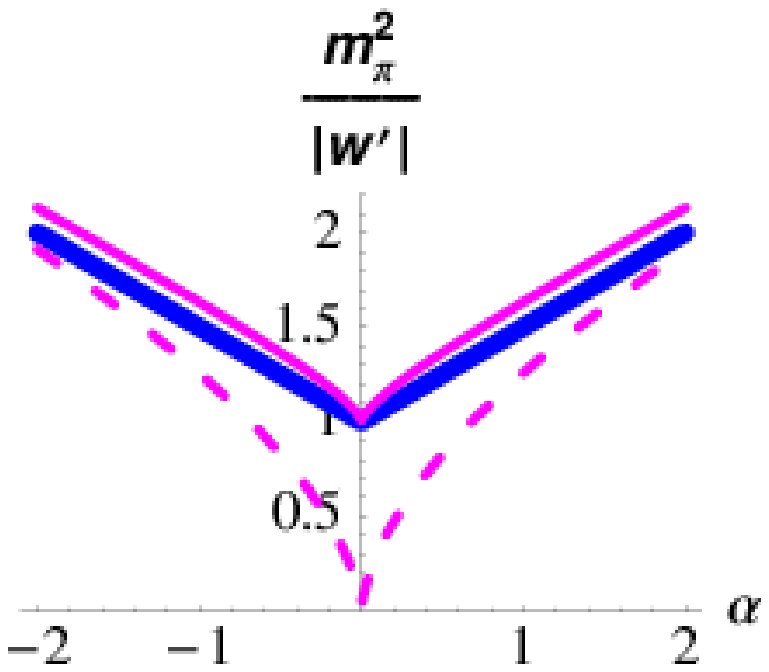} 
\end{center}
\caption{Condensate and pion masses
for $W'>0$. Notation as in fig.~\protect\ref{fig:Aokicond},
except that pion masses are shown only for $\beta=0,2$.
For $\beta=0$ charged and neutral pions are degenerate. 
For $\beta=2$, the neutral mass is shown dashed.}
\label{fig:firstorder}
\end{figure}

I illustrate the results for the first-order scenario ($W'>0$)
in fig.~\ref{fig:firstorder}. The left panel shows the discontinuity
in the condensate for $|\beta|<2$, transforming into
a smooth crossover for $|\beta|> 2$.
Above the transition, note again that $A_m=0$ when $\alpha=0$.
The right panel shows that the pion mass on the Wilson axis
($\beta=0$) has a non-zero minimum, 
while for $\beta=2$ the neutral mass goes down to zero at the
second-order end-point. 
In this scenario, the charged pions are
always heavier than the neutral once one moves off the Wilson axis.
Unlike the Aoki-phase scenario, however, no pions are massless along
the phase boundary away from the end-points. This is because no lattice
symmetry is broken along this transition.

\subsubsection{Applications to lattice simulations}

\noindent {\bf (I)}
The most important lesson we learn from tm\chpt\ is
to expect non-trivial phase structure when $m_q\sim a^2$,
with two possible scenarios.
The prediction of an Aoki-phase was actually made long 
before the tm\chpt\ analysis\cite{Aokiphase},
in order to understand how $m_\pi^2$ could vanish without an
underlying chiral symmetry,
and quenched studies in the 1980's and 90's found
evidence for such a phase.
What has happened in the last few years is the beginning
of detailed dynamical studies of the twisted-mass plane.
Those with unimproved Wilson gauge and fermion actions
found evidence for the first-order scenario.
I think it is fair to say this was a surprise, and that
it was predicted by \chpt\ I view as a significant success.
It is amusing that for decades we have been 
assuming that $m_\pi^2$ extrapolates all the way to zero, whereas
(at least with the simplest actions) it actually never makes it.

\begin{figure}[hbt]
\begin{center}
\includegraphics[width=5.5cm]{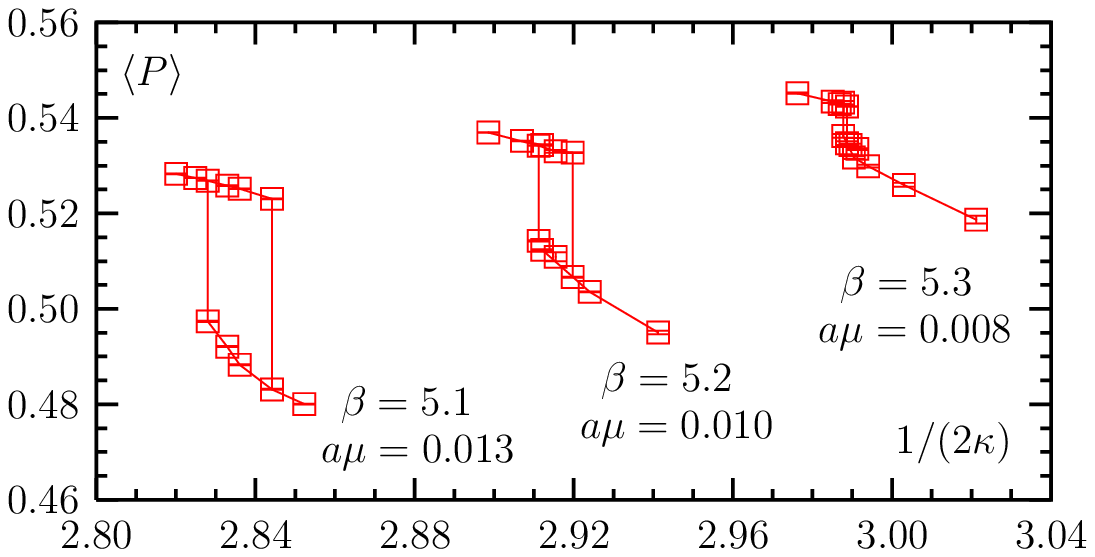}
\includegraphics[width=5.5cm]{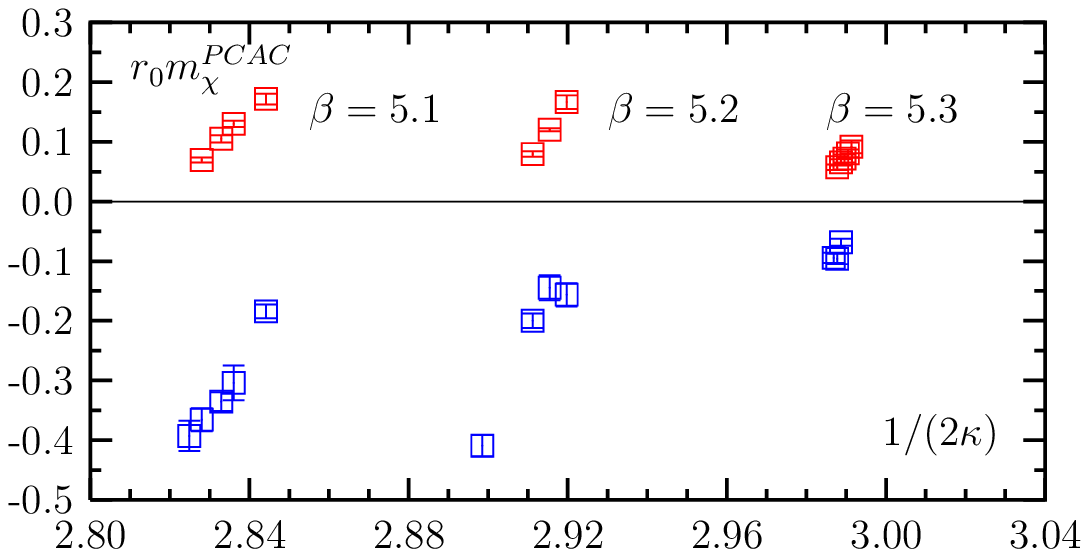}
\end{center}
\caption{Average plaquette and $m_{\rm PCAC}$ plotted vs.
$(2\kappa)^{-1}=m_0+4$ for fixed $\mu_0$\protect\cite{firstorder}.}
\label{fig:firstorderresults}
\end{figure}

I show an example of the evidence for
the first-order scenario in Fig.~\ref{fig:firstorderresults}.
This shows scans at fixed twisted mass 
(with $\mu$ roughly fixed in physical units)
for three lattice spacings ($a$ decreasing as $\beta$ increases).
The plaquette and $m_{\rm PCAC}$ both have a discontinuity,
and show hysteresis. Both effects 
decrease as one approaches the continuum limit,
qualitatively consistent with expectations. 
The fact that $m_{\rm PCAC}$ has a minimum away from zero
is a manifestation of the non-zero minimum in the 
pion masses.

\noindent {\bf (II)} 
The predictions for physical quantities that I sketched above
in the GSM regime have been extended to NLO in the Aoki 
regime\cite{ShWuII,AokiBarlat05,observations,AokiBar06}.
This is not trivial to implement, since minimization
of the potential (with the $W_{3,3}$ term included)
now involves a sextic equation.
Detailed fits to quenched data have, however, been done,
and find reasonable values for the resulting LECs\cite{AokiBarlat05}.

\begin{figure}[hbt]
\begin{center}
\includegraphics[width=5.5cm]{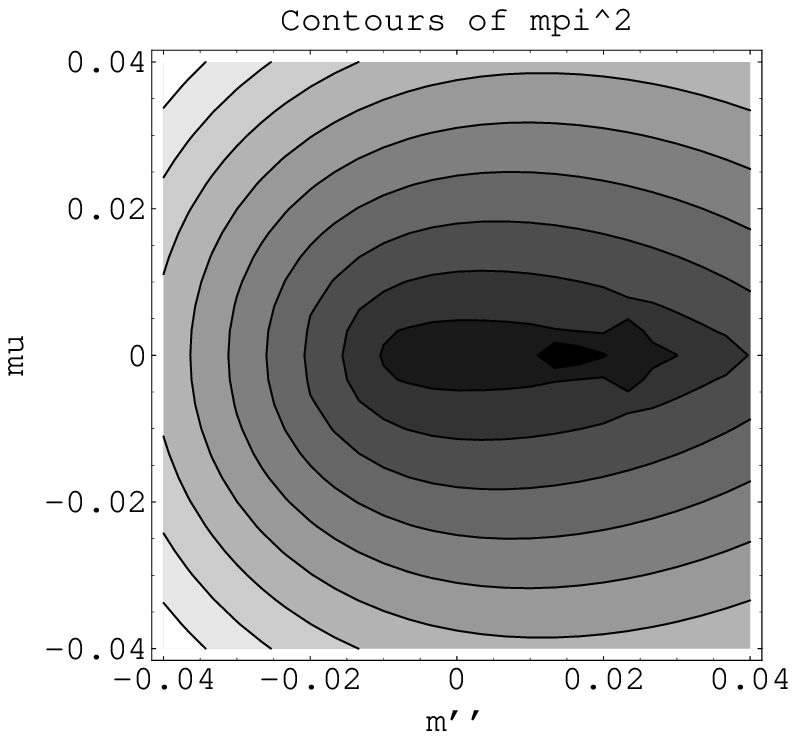}
\includegraphics[width=5.5cm]{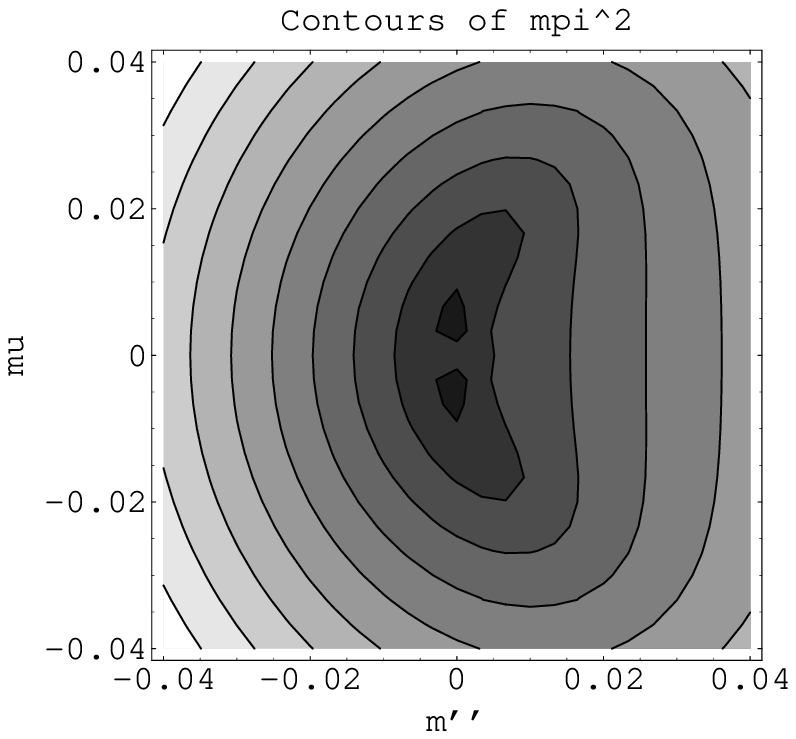}
\end{center}
\caption{Contour plots of $m_{\pi^\pm}^2$ in twisted mass plane,
for Aoki-phase (left) and first-order (right) scenarios, 
using a representative
parameter set for $1/a\approx 2\;$GeV. Quark masses are in GeV.
Raggedness in contours is due to numerical errors.}
\label{fig:contour}
\end{figure}

I cannot resist showing one NLO prediction.
Discretization effects distort the contours
of constant $m_{\pi^\pm}^2$ from circles into the forms
shown in fig.~\ref{fig:contour}.
Here $m''$ is just the shifted mass $m'$ with the 
$O(a^3)$ term proportional to $W_{3,1}$ absorbed as well.
These plots are illustrative
of the size of the expected effects, not the result of
details fits. I have taken standard LO continuum LECs,
and set the NLO continuum LECs and chiral logs
to zero for simplicity.
For the lattice LECs I use
$\delta_W=\delta_{\widetilde W}=-0.3$,
$|w'|= (250\,{\textrm MeV})^2$ and $W_{3,3}=0$,
values which are not unreasonable for $a^{-1}\approx 2\;$GeV.
Quark masses range up to $\sim m_s/2$.
Clearly the impact of discretization errors
and phase structure could be very significant for
actual simulation parameters.

\begin{figure}[htb]
\begin{center}
\includegraphics[width=5.4cm]{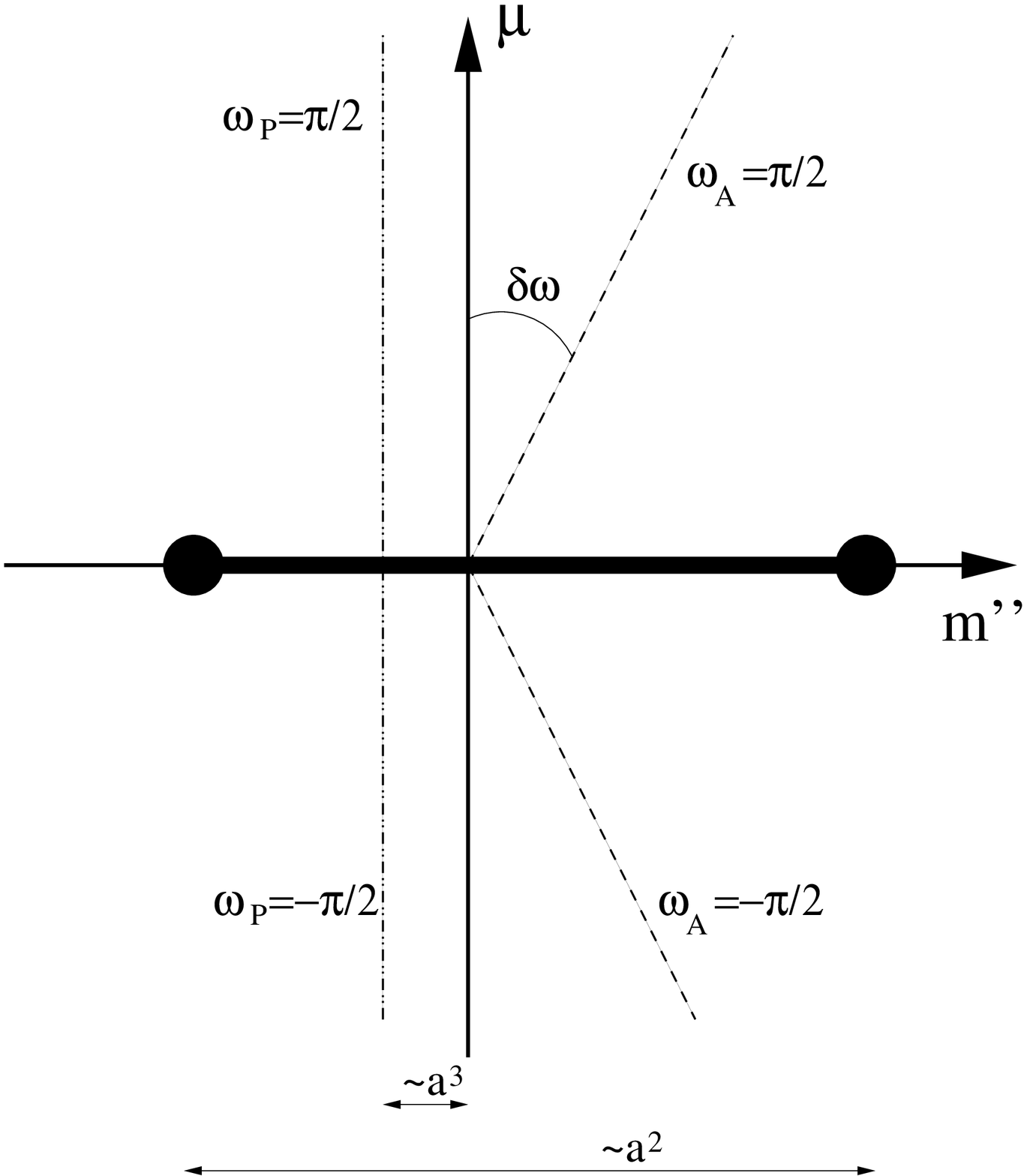}
\hspace{.2truecm}
\includegraphics[width=5.4cm]{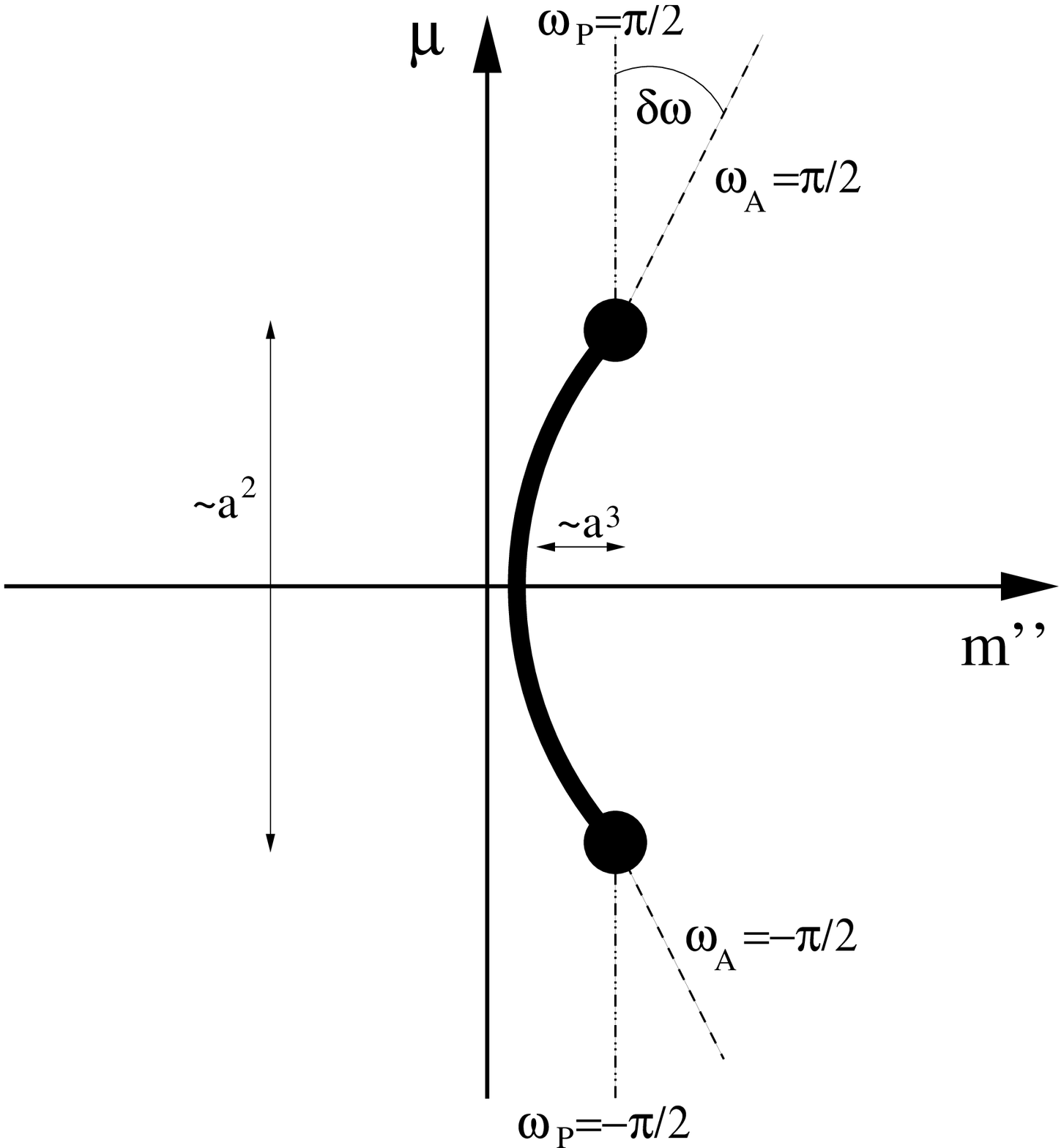}
\end{center}
\caption{Phase diagram at NLO for Aoki-phase
(left) and first-order (right) scenarios. Method (i) corresponds to $\omega_A=\pi/2$,
method (iii) to $\omega_P=\pi/2$.}
\label{fig:phasediagramNLO}
\end{figure}

\noindent {\bf (III)}
Another important question that can be studied using tm\chpt\
in the Aoki regime is whether automatic $O(a)$ improvement at
maximal twist still holds using the criteria introduced
in sec.~\ref{sec:maxtwist}. 
The answer is
positive\cite{AokiBar04,ShWuII,observations,AokiBar06},
aside from a caveat I will explain.
In fig.~\ref{fig:phasediagramNLO} I show
the impact of NLO corrections on the
phase boundaries and the lines of maximal twist\cite{observations}.
These plots allow one to understand the qualitative features
of the contours in fig.~\ref{fig:contour}.

The lines of maximal twist shown in the figure are 
those for methods (i) and (iii).
Along either line the physical quark mass is proportional to $\mu$.
If one simulates on one of these lines, reducing
$a$ while holding $\mu$ fixed (i.e. working at fixed physical quark mass),
then the dependence of physical quantities
on $a$ will be quadratic or of higher order.
This is automatic $O(a)$ improvement.
In the Aoki-phase scenario it holds even for $\mu\ll a^2$, i.e. all
the way to the Wilson axis. For the first-order scenario the result
breaks down, however, when one runs into the end-point with $\mu\approx |w'|\sim a^2$.
This is the caveat mentioned above.

It is perhaps surprising that one can work in a regime where the quark mass
is much smaller than the leading $O(a)$ effects of discretization,
and yet be able to tune parameters so these 
effects do not enter into physical quantities. 
It is worth understanding this qualitatively.
The point is that the tuning has to be {\em very} fine. 
If $\mu\sim a^2$, then, as fig.~\ref{fig:phasediagramNLO}
shows, one must tune $m''$ to an accuracy of $a^3$. This is what
the criteria in eq.~(\ref{eq:maxtwistAP}) accomplish. 
In terms of the Symanzik Lagrangian, this amounts to canceling
the untwisted mass (so that it is $O(a^3)$ or smaller). The
result is
\begin{eqnarray}
{\CL}^{(4+5)} &=&
\bar \psi \Dslash \psi + \mu \bar\psi i \gamma_5\tau_3\psi
+a c \bar\psi i\sigma_{\mu\nu}F_{\mu\nu}\psi + O(a^2)
\\
&=&
\bar \psi_{\rm phys} \Dslash \psi_{\rm phys} 
+ \mu \bar\psi_{\rm phys}\psi_{\rm phys}
+ a c \bar\psi_{\rm phys} \gamma_5\tau_3\sigma_{\mu\nu}F_{\mu\nu}\psi_{\rm phys}
\,,
\end{eqnarray}
where in the second line I have rotated to the physical basis.
The absence of a $\bar\psi\psi$ term on the first line means that
the $O(a)$ term is purely flavor-parity breaking in the
physical basis, and does not contribute
to physical matrix elements however large it is.\footnote{%
A similar argument can be made directly
at the chiral Lagrangian level, and relies on the
result, noted above, that the condensate points
near the $\tau_3$ direction at maximal twist~\protect\cite{Munster,ShWuII}.}
This is nothing more than a summary of the original argument for automatic
improvement\cite{FR} (although simplified, as I am not including operators).
The only new point here is that the argument goes through for
arbitrarily small $\mu$ as long as $m''$ is tuned accurately enough.

It is now time to unveil what I call ``method (ii)'', which works as follows.
One first defines a critical mass by extrapolating results from
method (i) to the Wilson axis, 
and then keeps $m_0$ fixed at this critical mass for all $\mu$.
It turns out to correspond to working at $m''=0$.
In the GSM regime (recall fig.~\ref{fig:maxtwist}) method (ii) 
 is equivalent to method (iii), but in the Aoki regime the line
it defines is offset from that of method (iii) by $O(a^3)$.
Such a difference might seem too small to be significant,
but it need not be\cite{AokiBar06}. To see this, consider the
Aoki-phase scenario. Then from fig.~\ref{fig:phasediagramNLO}
one sees that method (ii) interpolates between method (i)
($\mu < a^2$) and method (iii) ($\mu \gg a^2$). Now, it turns
out that the condensate has a fixed orientation as one
moves along the lines of either method,
but the orientations for the two methods differ by an angle
of $O(a)$. Thus for method (ii) the condensate varies
direction rapidly for $\mu\ltapprox a^2$, with a gradient of $\sim a/a^2$.
This can disrupt extrapolations as the resulting form is not a
simple polynomial\cite{AokiBar06}.
I raise this worry because method (ii) is being used in practice
by some groups,
and it would be preferable to use method (i).

\begin{figure}[htb]
\begin{center}
\includegraphics[width=7cm]{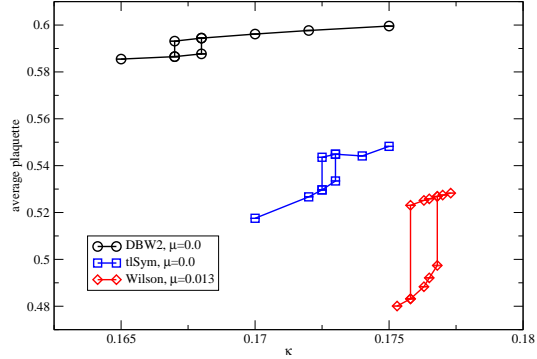}
\end{center}
\caption{Hysteresis of plaquette at $a\approx 0.17\;$fm
for DBW2 (top), tree-level improved Symanzik (middle)
and Wilson (bottom) gauge actions.\protect\cite{Farchionilat05}.}
\label{fig:DBW2}
\end{figure}

\noindent {\bf (IV)}
A further lesson from tm\chpt\ is that
both the size of the phase boundaries and the isospin splitting
for pions are determined by the same parameter, $w'$.
It thus  makes sense
to try and tune the gauge and fermion actions to reduce $|w'|$.
Note that this tuning is not the same as a systematic improvement
program, but it is nevertheless very important.
Initial results in this direction are very encouraging.
For example, fig.~\ref{fig:DBW2} shows that the discontinuities
in the plaquette are significantly reduced by alternative
gauge actions.\footnote{%
In fact, the reduction is greater than appears, as the
results with the Wilson action are for $\mu\ne0$, 
where the transition is weaker than for $\mu=0$,
the value used for the other actions.}
It is also found that using an improved {\em fermion}
action reduces the pion isospin splitting and thus 
$w'$\cite{pionsplittingIII}.
This is an active area of present research.

\begin{figure}
\begin{center}
\includegraphics[width=5.5cm]{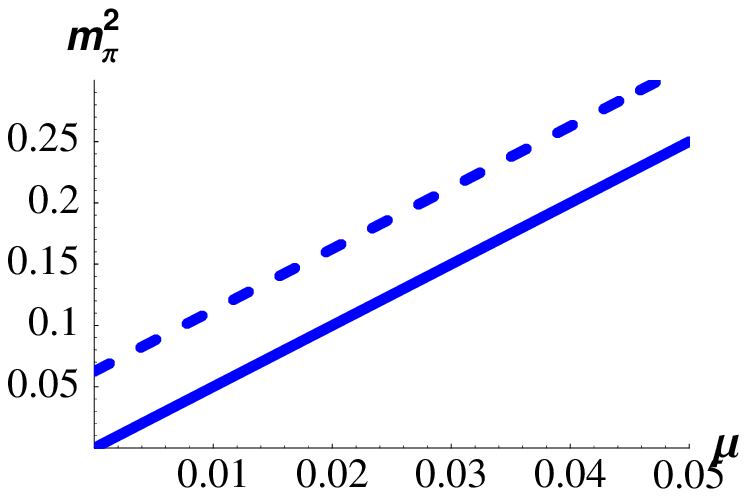}
\includegraphics[width=5.5cm]{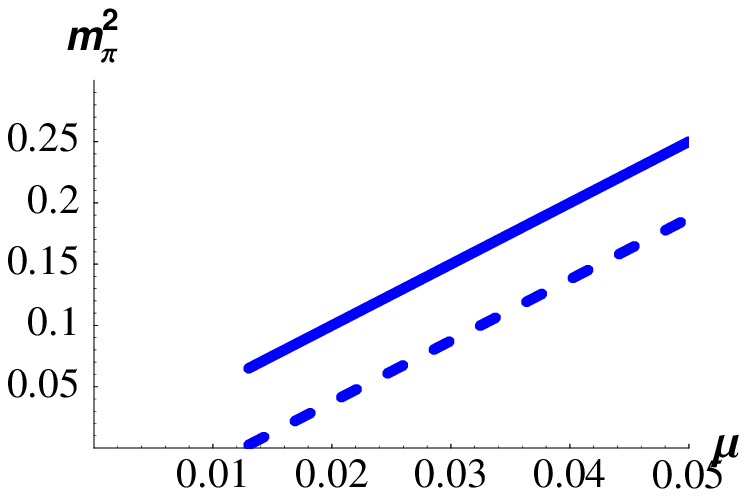}
\end{center}
\caption{$m_{\pi^\pm}^2$ (solid) and $m_{\pi^0}^2$ (dashed) in
GeV$^2$ versus $\mu$ in GeV for method (i).
Left panel: Aoki-phase scenario; right panel: first-order scenario.}
\label{fig:whichscenario}
\end{figure}

\noindent {\bf (V)}
It might appear from fig.~\ref{fig:phasediagramNLO} that,
if we had to choose, we would prefer the
Aoki-phase scenario because we can work at maximal twist down to $\mu=0$.
I think, however, that the choice is not so clear. 
To illustrate why, I compare the results
for pion masses in the two scenarios in fig.~\ref{fig:whichscenario},
using the same ``reasonable'' choices of LECs as above.
The role of the charged and neutral pions is 
interchanged in the two scenarios, and there is no clear
advantage to using one over the other. The figure also emphasizes
the significance of the $O(a^2)$ effects.
I conclude that it is much more important to reduce the
magnitude of $w'$ than it is to end up in one or other scenario.

\subsubsection{Bending near maximal twist}\label{sec:bending}

The discussion to this point makes clear the importance of accurate
tuning to maximal twist. Initial studies of tmLQCD did not always attain
this accuracy and observed a phenomenon called ``bending''.
Although this is largely a closed chapter in the history of
tmLQCD, it is worth learning the appropriate lessons, particularly as tm\chpt\
played an important role in elucidating the phenomenon.

\begin{center}
\begin{minipage}{5.5cm}
The sketch to the right shows the Aoki-phase scenario with three choices
of the approach to the chiral limit shown by arrows.
They are all nominally at maximal twist.
These are method (i) [thick lines];
fixing $m_c$ at an endpoint where {$m_\pi=0$} [medium thickness],
and missing the endpoint  by $\sim a^2$ [thin lines].
\end{minipage}
\hfill
\begin{minipage}{5cm}
\begin{center}
\includegraphics[width=5cm]{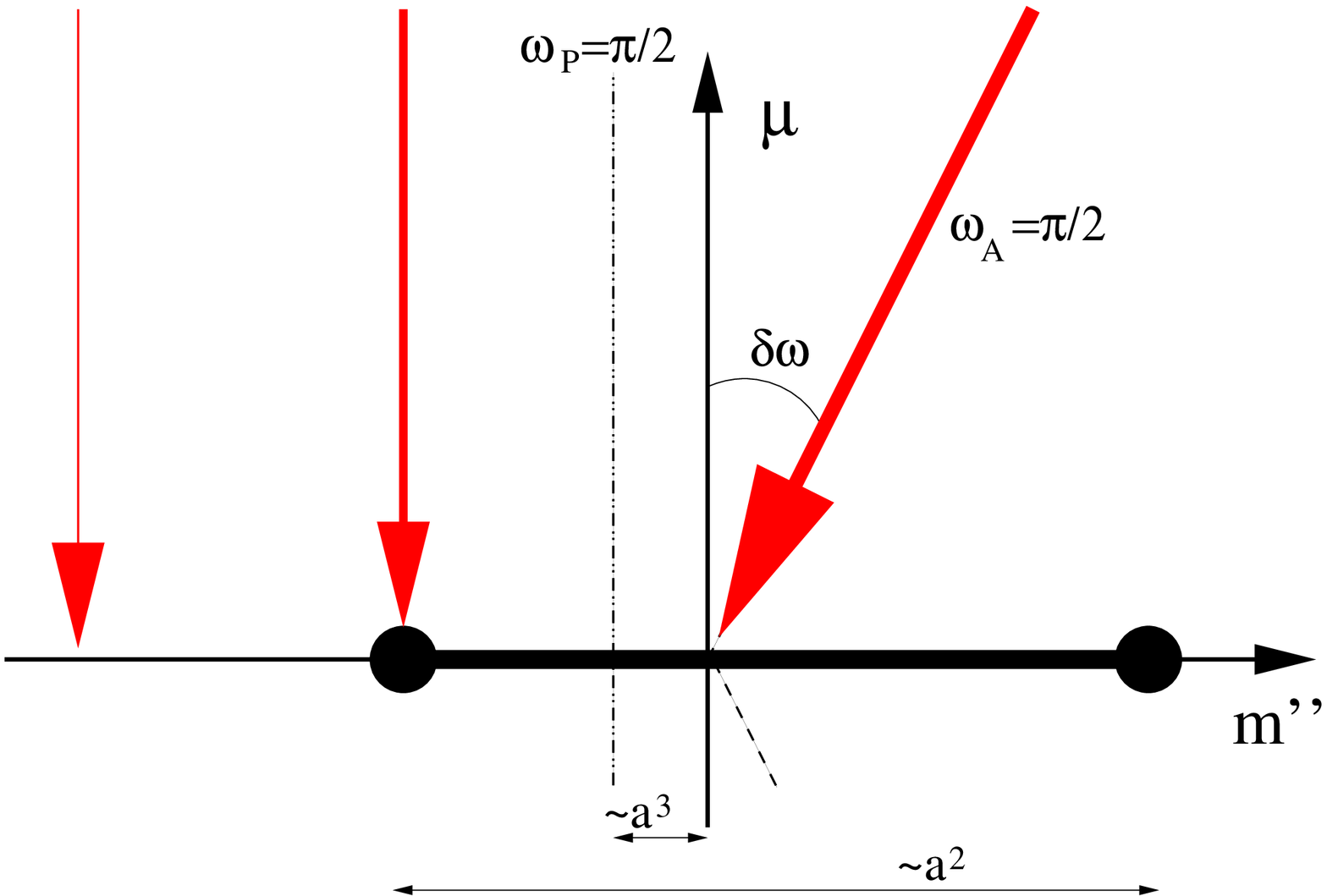}
\end{center}
\end{minipage}
\end{center}

\begin{figure}
\begin{center}
\includegraphics[width=5cm]{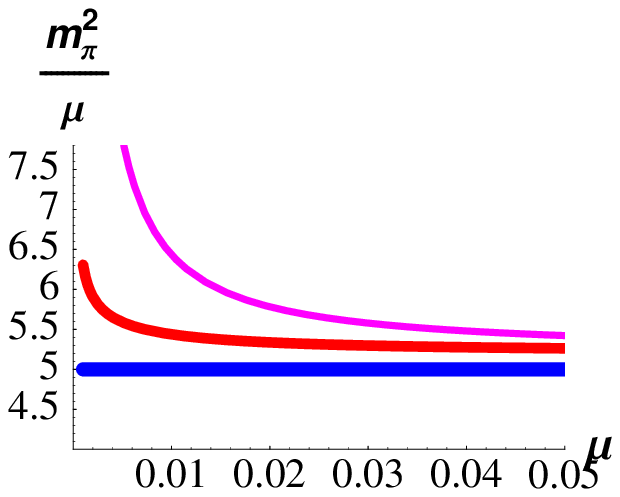}
\includegraphics[width=6cm]{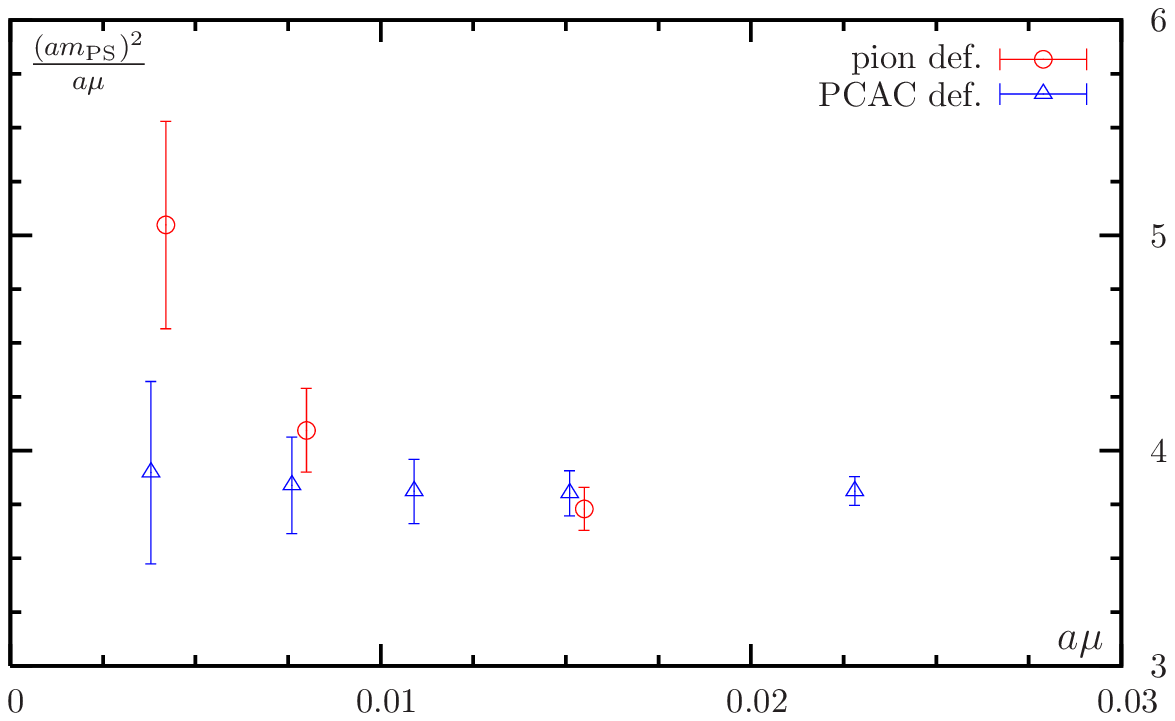}
\end{center}
\caption{``Bending'' in $m_\pi^2/\mu$ vs. $\mu$ from 
tm\chpt\protect\cite{observations} (left---both axes in GeV) 
and quenched simulations\protect\cite{Shindler} (right---both axes in lattice
units at $1/a\approx 2\;$GeV using bare $\mu$).}
\label{fig:bend}
\end{figure}

I show in fig.~\ref{fig:bend} the NLO tm\chpt\ results 
(using the same parameter set and assumptions as above) for 
$m_{\pi^\pm}^2/\mu$ for
the three methods (with the thickness of the lines matching those above),
along with results from simulations.
The horizontal axis is $\mu$, which is the quark mass when working
at maximal twist.
For my parameters (i.e. physical LECs and chiral logs dropped)
the correct result is a constant,
and is illustrated by the $O(a)$ improved result from method (i)
[thick line].
The results from the other two choices show a noticeable
bending away from the continuum result at small masses,
i.e. a breakdown of $O(a)$ improvement. 
The numerical results are for a similar range of quark masses,
but the axes are in different units and so the two plots can
only be compared qualitatively.
The numerical results are
for method (i) (``PCAC definition''---triangles) 
and the ``$m_\pi=0$ definition'' (circles),
and are qualitatively consistent with tm\chpt.
A detailed fit of these and other quenched data to
quenched tm\chpt\ has been done, and finds
quantitative agreement\cite{AokiBarlat05}.

The most important conclusion from this figure is that one should
use a good definition of maximal twist, such as method (i). 
In particular, the failure of the $m_\pi=0$ method is apparent. 
This can be understood qualitatively as follows. The condensate at maximal twist
should lie in the $\tau_3$ direction, $\langle\Sigma\rangle = i\tau_3 + O(a)$.
This is the case with method (i). In the other cases, however, the condensate
starts in this direction for $\mu\gg a^2$ but rotates away when $\mu\sim a^2$
and ends up at 
$\langle\Sigma\rangle=-1$ when $\mu=0$. Thus automatic $O(a)$ improvement
is lost.\footnote{%
If one treats deviations from {$\langle\Sigma\rangle=i\tau_3$} as small
perturbations, then discretization errors
have $\pi^0$ poles and are infrared enhanced\cite{FRMP}. 
This appears superficially like a breakdown of the Symanzik expansion,
but, in fact, the divergences are summed automatically in tm\chpt\
if one expands about the correct vacuum\cite{observations}.}

\subsubsection{Does tmLQCD work in practice?}

Tm\chpt\ has been a very useful guide to the exploratory
numerical studies of tmLQCD, but the most important question
is whether working at maximal twist is practical. In particular,
is the requisite fine tuning possible, and cheap enough, in practice?
Are results for physical quantities really improved?
These questions are being actively studied and 
the answers appear to be 
positive\cite{Shindler,Farchionilat05,quenchedtmscaling}.

A different question is how best to include the strange quark,
as one needs an even number of quarks to implement maximal twisting.
One could just add a single non-perturbatively $O(a)$ improved
untwisted strange quark to a maximally twisted up-down pair,
but this would require non-perturbative improvement of each
new operator involving the strange quark, which is exactly
what we are trying to avoid with automatic improvement at maximal twist.
An alternative is to introduce the charm quark
as well, and treat the strange-charm pair as a twisted doublet\cite{Shindler}.
Non-degenerate masses can be incorporated while maintaining maximal 
twist\cite{nondegenerate}. The disadvantages are that
flavor breaking is more complicated (occurring in
both $\tau_3$ and $\tau_1$ directions in the strange-charm sector),
and that $(a m_c)^2$ effects may not be small.
This option is being actively studied\cite{2+1+1}.

Finally, I think it likely that isospin breaking in pion masses
will need to be included in loop calculations, requiring one
to work at NNLO in the GSM power counting. First steps in
this direction have been taken\cite{WChPT}.

For further possible uses of tmLQCD, see the lectures by Sint.

\section{Partial quenching and PQ\chpt}\label{sec:PQPQChPT}
My final lecture is devoted to introducing partially quenched
QCD and the corresponding EFT, PQ\chpt. 
I will consider PQQCD only in the continuum, and focus on
conceptual issues. 
I will not discuss the extension to
include discretization errors---this is straightforward 
following the methods of the previous lecture---nor
the related topic of mixed action \chpt. 
Nor will I discuss quenched QCD
and quenched \chpt\ (Q\chpt), except in passing, as it is not useful for
obtaining quantitative physical information about QCD.

\subsection{What is PQQCD and why is it useful?}

I have given an overview of the meaning of partial
quenching in the introduction. Here I give a concrete example
for the pion correlator in QCD. If we write out how this
is calculated in detail,
\begin{eqnarray}
\lefteqn{C_\pi(\tau)=- \left\langle
\sum_{\vec x} \bar u \gamma_5  d(\vec x,\tau)\ 
              \bar d \gamma_5 u(0)\right\rangle }
\label{eq:CpiQCD}\\
&\equiv& - \frac1Z \int DU \prod_q Dq D\bar q
e^{-S_{\rm gauge} - \int_x \sum_q \bar q (\Dslash + m_q) q} 
\sum_{\vec x} \bar u  \gamma_5 d(\vec x,\tau)\ 
              \bar d \gamma_5 u(0) \nonumber\\
&=&\frac1Z\int DU \prod_q \det(\Dslash\! +\! m_q) e^{-S_{\rm gauge}} 
\sum_{\vec x} \tr\left[
\gamma_5 { \left(\frac1{\Dslash\!+\!m_d}\right)_{x0}}
\gamma_5 { \left(\frac1{\Dslash\!+\!m_u}\right)_{0x} } \right]
\nonumber
\end{eqnarray}
\vspace{-.5truecm}
\begin{minipage}{5cm}
\begin{center}
\includegraphics[height=1.5cm]{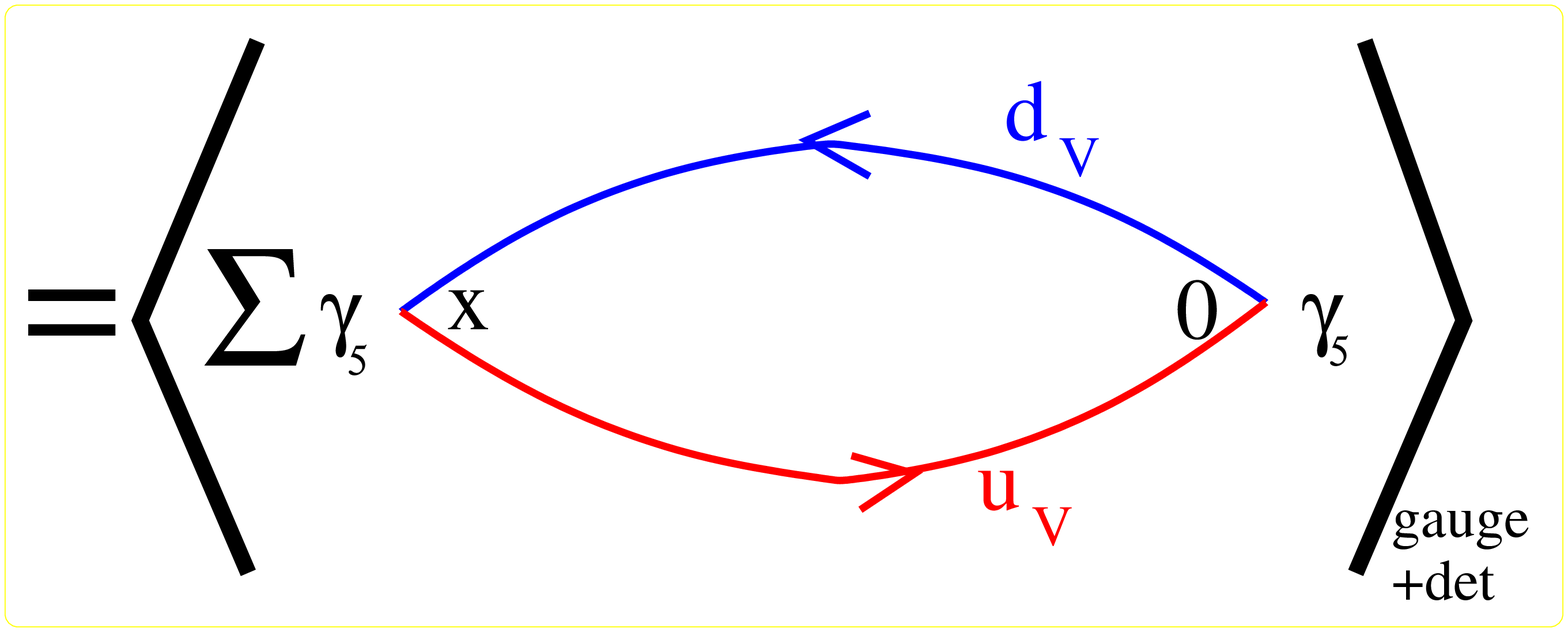}
\end{center}
\end{minipage}
\begin{minipage}{6cm}
$\propto  f_\pi^2 e^{-m_\pi \tau} + \textrm{exponentially suppressed}$.
\end{minipage}
\vskip 1.0cm
\noindent
we see that, in practice,
we are free to use different masses in
the determinant (``sea'' or ``dynamical'' quark masses)
and in the propagators (``valence'' quark masses).
Doing so is called partial quenching.
It is relatively cheap to do so, since the calculation of propagators
is usually a small overhead on the generation of configurations,
thanks to the determinant, and because the rate of increase of CPU time 
as $m_q$ decreases is less severe for propagators.
The big question is whether we can make use of the extra
information provided by working with $m_{\rm val} \ne m_{\rm sea}$.

At this point it is appropriate to interject a comment on 
nomenclature. In the bad old days we used the quenched approximation,
in which $m_{\rm sea} \to \infty$, so the quark determinant
could be ignored. This allowed simulations to be done, but
at the heavy cost of ignoring quark loops and thus considering
a theory which is not unitary and had various sicknesses.
QQCD is at best a model of QCD---there is no quantitative relation
between the two theories. Partial quenching is in one sense
a less extreme version of quenching,  and thus the name.
If $m_{\rm sea} \gg \LQCD$ then PQQCD, like QQCD,
is only qualitatively related to QCD, and the name is appropriate.
What I consider here, however, is PQQCD with sea quarks light enough,
$m_{\rm sea}\ll\LQCD$,
that (as I will argue) we can use \chpt\ to relate the
results {\em quantitatively} to physical QCD.
For such theories a more appropriate name,
with less negative connotations, would be ``partially unquenched''.
But I will not try and create a naming revolution and will
stick with the canonical nomenclature.

One thing that having light sea quarks does not change is that
PQQCD is unphysical.
To see this consider a contribution to $\pi^0\pi^0$ scattering in QCD:
the two pions first annihilate into glue, 
from which is created, say, a $\pi^+\pi^-$ intermediate state, 
which subsequently annihilates back into glue, 
and then finally a new $\pi^0$ pair is created.
In the corresponding process in PQQCD, the intermediate $\pi^+\pi^-$
pair must be composed of sea quarks, and thus have different masses
from the valence $\pi^0$ pairs. Having different intermediate 
and external states means that the theory is not unitary.
Another unphysical feature is the appearance of double poles
in propagators. This is not possible if one can insert a complete
set of physical states. The fact that there are double poles
can be seen in PQ\chpt\cite{BGPQ}, but can also be seen more
generally from the properties of the quark-level theory\cite{ShShphi0}.

Because PQQCD is unphysical, it is essential 
to have a quantitative method relating its properties
to those of QCD.
This is provided by PQ\chpt. 
Thus we must simulate in the regime of quark
masses where \chpt\ is valid if PQQCD is to be a useful tool.
This is illustrated above in fig.~\ref{fig:PQQCD}.
Note that the presence of 
the physical subspace with $m_{\rm val}=m_{\rm sea}$ 
within PQQCD is crucial in order to pin down the
relation to QCD.

\subsection{A field theoretic formulation of PQQCD}\label{sec:Morel}

As we have seen, the construction of an EFT
makes essential use of symmetries.
Thus to develop PQ\chpt\ we need a formulation of PQQCD
which makes its symmetries manifest. One way to do this is to
use Morel's trick\cite{Morel} of introducing commuting spin-$1/2$
fields (ghost quarks labeled $\widetilde q$), whose determinant
can cancel that from the valence quarks:
\begin{equation}
\!\!\int\!\! D\bar q Dq  e^{-\bar q(\Dslash\!+\!m_q) q} 
\!=\!
\mathrm{det}(\Dslash\!+\!m_q)\,,\
\int\!\! D\widetilde q^\dagger D\widetilde q 
e^{-\widetilde q^\dagger(\Dslash\!+\!m_q) \widetilde q} 
\!=\!
\frac{1}{\mathrm{det}(\Dslash\!+\!m_q)}\,.
\end{equation}
The formulation then includes three types of quark:
valence quarks 
$q_{V1}$, $q_{V2}$, \dots $q_{VN_V}$, ($N_V=2,3,\dots$),
sea quarks
$q_{S1}$, $q_{S2}$, \dots $q_{SN}$  ($N=2,3$), and
ghost quarks 
$\widetilde q_{V1}$, 
$\widetilde q_{V2}$, \dots $\widetilde q_{VN_V}$ ($N_V=2,3,\dots$).
The ghosts are degenerate with corresponding valence quarks.
It is notationally convenient to package these fields
into {$(N+2 N_V)$-dim} vectors:
\begin{equation}
Q^{tr} = \bigg( 
\underbrace{{ q_{V1}}, 
\dots, { q_{VN_V}}}_{\rm valence},\,
\underbrace{
{ q_{S1}},
\dots, { q_{SN}}}_{\rm sea},\,
\underbrace{{\widetilde q_{V1}}, 
\dots,  { \widetilde q_{VN_V}}}_{\rm ghost}\bigg) \,,
\end{equation}
\begin{equation}
\overline{Q} = 
\bigg( 
\underbrace{{ \bar q_{V1}}, 
\dots, { \bar q_{VN_V}}}_{\rm valence},\,
\underbrace{
{\bar q_{S1}}, 
\dots, { \bar q_{SN}}}_{\rm sea},\,
\underbrace{{\widetilde q_{V1}^\dagger}, 
\dots,  { \widetilde q_{VN_V}^\dagger}}_{\rm ghost}\bigg)\,,
\end{equation}
and similarly for the masses
\begin{equation}
{\mathcal M} =
\bigg( 
\underbrace{{m_{V1}}, 
\dots, {m_{VN_V}}}_{\rm valence},\,
\underbrace{
{m_{S1}}, 
\dots, {m_{SN}}}_{\rm sea},\,
\underbrace{{m_{V1}}, 
\dots, {m_{VN_V}}}_{{\rm ghost} ={\rm valence}}\bigg)\,.
\end{equation}
The action of PQQCD is a simple extension of that of QCD
\begin{eqnarray}
S_{\rm PQ} &=& S_{\rm gauge} + \int \overline Q (\Dslash + {\mathcal M}) Q 
\\
\overline Q (\Dslash + {\mathcal M}) Q &=&
\sum_{i=1}^{N_V}{ \bar q_{Vi} (\Dslash+m_{Vi}) q_{Vi} } 
+\sum_{j=1}^{N}{\bar q_{Sj} (\Dslash+m_{Sj}) q_{Sj}}
\nonumber\\
&&\mbox{}\qquad
+\sum_{k=1}^{N_V}{
\widetilde q_{Vk}^\dagger (\Dslash+m_{Vk}) \widetilde q_{Vk}} \,.
\end{eqnarray}
If we similarly define an extended measure as
\begin{equation}
D\overline Q DQ
\equiv
\prod_{i=1}^{N_V} \left(
{ D\bar q_{Vi} D q_{Vi}}
{ D\widetilde q_{Vi}^\dagger D\widetilde q_{Vi}}
\right)
\prod_{j=1}^{N}\left({ D\bar q_{Sj} D q_{Sj}}\right)\,,
\end{equation}
then we can write down a 
partition function containing valence quarks
that nevertheless reproduces that of QCD:
\begin{eqnarray}
Z_{\rm PQ} &=& \int DU D\bar Q DQ\ e^{-S_{\rm PQ}}
\\
&=& \int DU e^{-S_{\rm gauge}}
\prod_{i=1}^{N_V} \left(
\frac{{ \mathrm{det} (\Dslash+m_{Vi})}}
     {{ \mathrm{det} (\Dslash+m_{Vi}) }}\right)
\prod_{j=1}^{N}{ \mathrm{det}(\Dslash+m_{Sj})} \\
&=& \int DU e^{-S_{\rm gauge}}
\prod_{j=1}^{N}{ \mathrm{det}(\Dslash+m_{Sj})}
= Z_{\rm QCD}
\end{eqnarray}

So far this is rather trivial. The power of the method is
that it provides field-theoretic expressions for PQ correlation functions,
e.g.
\begin{eqnarray}
\!\!\!\!\!\!\!C_\pi^{\rm PQ}(\tau)
&\equiv&
-Z_{\rm PQ}^{-1} \int DU D\overline Q D Q\
 e^{-S_{\rm PQ}} 
\sum_{\vec x} { \bar u_V }\gamma_5 { d_V}(\vec x,\tau)\ 
              {\bar d_V} \gamma_5 { u_V}(0) \\
&=&
Z_{\rm PQ}^{-1} \int DU \prod_{j=1}^{N} {\det(\Dslash + m_{Sj})}
 e^{-S_{\rm gauge}} \nonumber\\
&&\mbox{}\qquad
\times \sum_{\vec x} \tr\left[
\gamma_5 { \left(\frac1{\Dslash+m_{V d}}\right)_{x0}}
\gamma_5 { \left(\frac1{\Dslash+m_{V u}}\right)_{0x} } 
\right]  \,.
\end{eqnarray}
This is exactly the pion correlator 
with which we started, eq.~(\ref{eq:CpiQCD}),
but with differing valence and sea quark masses.\footnote{%
This formulation works as well for
quenched QCD---one just omits the sea quarks\cite{Morel}.}

As noted above, PQQCD is ``anchored'' to QCD
(or, more precisely, to physical, QCD-like, theories).
Let us see how this works in our new formulation.
Consider the PQ pion correlator, but now set the valence
quark masses equal to two of the sea quark masses, i.e.
${m_{Vu}}={ m_{Sj}}$ and ${m_{Vd}}={ m_{Sk}}$.
Then the PQ correlator is equal to a physical QCD correlator:
\begin{eqnarray}
\!\!\!\! C_\pi^{\rm PQ}(\tau)
&=&
Z_{\rm PQ}^{-1} \int DU D\overline Q D Q\
 e^{-S_{\rm PQ}} 
\sum_{\vec x} {\bar u_V }\gamma_5 { d_V}(\vec x,\tau)\ 
              {\bar d_V} \gamma_5 { u_V}(0) \\
&=&
Z_{\rm PQ}^{-1} \int DU D\overline Q D Q
 e^{-S_{\rm PQ}} 
\sum_{\vec x} {\bar q_{Sj} }\gamma_5 { q_{Sk}}(\vec x,\tau)\ 
              {\bar q_{Sk}} \gamma_5 { q_{Sj}}(0)  \\
&=&
Z_{\rm QCD}^{-1} \int DU \prod_{i=1}^{N}{ D\overline q_{Si} Dq_{Si}}\
 e^{-S_{\rm QCD}} 
\nonumber\\
&&\mbox{}\qquad\qquad\qquad\qquad \times
\sum_{\vec x} {\bar q_{Sj} }\gamma_5 {q_{Sk}}(\vec x,\tau)\ 
              {\bar q_{Sk}} \gamma_5 { q_{Sj}}(0) \\
&=&
C_\pi^{\rm QCD}(\tau)\,.
\end{eqnarray}
To obtain the second line I used the result that the
propagators obtained by doing the Wick contractions
are the identical to those from the first line. 
This is an example of the enlarged symmetry of the PQ theory.
Having removed the valence quarks from the operators,
the valence and ghost integrals cancel, leaving a
QCD correlation function.
This analysis generalizes to any correlation function containing
$N$ or less different valence quark-antiquark pairs
(recall that we can add any number of valence quarks).
If there are more than $N$ such pairs, e.g. a two point
function containing
$\overline q_{V1} q_{V2} \overline q_{V3} q_{V4}$ and its conjugate
in the presence of 3 light sea quarks,
then there is no corresponding QCD correlator even if valence
masses are all equal to sea masses.
This is an example, developed further below, of how PQQCD gives
one access to combinations of Wick contractions that do not
occur in QCD itself.

The field theoretic formulation shows that PQQCD is
a well-defined statistical system.
In particular, the ghosts do not present a theoretical problem
in the Euclidean functional integral.
As long as quark masses are positive, the functional
integrals over the ghost quarks converge
(since $\Dslash$ has imaginary eigenvalues).
Of course, the theory remains unphysical, and indeed we 
can now see this more directly.
If we rotate to Minkowski space we will
violate the spin-statistics theorem, 
and thus have an unphysical theory.
Put another way, PQQCD does not satisfy reflection positivity
(as can be seen, for example, from the fact
that the ghost pion correlator
has the opposite sign to that for the normal pion),
and so one cannot construct a physical Hilbert space
with a positive Hamiltonian.
But the unphysical nature need not be a problem if
we use PQQCD in Euclidean space as a tool to gain information about
QCD.

\subsection{Developing PQ\chpt}\label{sec:developPQChPT}

Before diving into the theoretical details let me
give some qualitative motivation for what follows.
We will need to assume that PQQCD is described by
an effective theory close to that for QCD.
I think it is fair to say that our confidence in this
assumption is based in part on results from simulations.
In particular, the charged pion correlator, $C_\pi^{\rm PQ}(\tau)$,
has essentially the same form at long distances in PQQCD (and also QQCD)
as in QCD:
it falls as $\exp(-m_\pi \tau)$ at long times,
and $m_\pi^2 \propto (m_{Vu}+m_{Vd})$ to good approximation.
There is no sign of unphysical effects, e.g. double poles
[which fall as $t\exp(-m_\pi t)$], 
or negative residues.
It is only when one looks in detail that one finds deviations
from QCD expectations, such as the enhanced chiral
logarithms in QQCD. There are also other correlators where partial
quenching has a dramatic effect: double poles do appear
in the $\eta'$ correlator,
and there are negative contributions to the scalar-isovector correlator.
Nevertheless, the apparent closeness of the infrared physics of PQQCD
to that of QCD provides 
important motivation for the development of (P)Q\chpt.

This development has been done using two methods.
The first is the ``graded-symmetry'' method\cite{BGPQ} 
based on Morel's trick
and which I use here.
This is an extension of earlier work on the quenched theory\cite{BGQ}.
The main issue, as we will see,
is the whether the usual ``derivation'' of
an EFT goes through when the theory is unphysical.
The second method uses the ``replica-trick''\cite{replica}, 
in which one removes the valence determinant by sending $N_V\to 0$
rather than using ghosts.\footnote{%
This method gives a formalization of the ``quark-line''
method which I used to develop quenched \chpt\cite{SSchirallogs}.}
This has the advantage that the theories with integer $N_V$,
from which one is extrapolating, are physical, and so one
expects their long-distance physics to be described by \chpt.
Its disadvantage is that extrapolating in $N_V$ is, in general,
theoretically uncontrolled.

For the purposes of the present lecture I could use either method,
since they are known to give the same results at one-loop in PQ\chpt,
and it is plausible that this holds to all orders\cite{replica}.
I choose the graded-symmetry method as I am more 
familiar with it.\footnote{%
The equivalence is less well established in contexts
where EFT calculations are non-perturbative,
such as in the $\epsilon-$regime.
Here calculations are harder in both approaches, but 
agree where they overlap\cite{damgaardlat01}.}

\subsubsection{Symmetries of PQQCD}

In the notation developed above, 
$S_{\rm PQQCD} = S_{\rm gauge} + \overline Q (\Dslash + {\mathcal M}) Q$
looks just like $S_{\rm QCD}$,
and appears to have a graded extension of chiral symmetry
when  $\CM\to0$, involving rotation of quarks into ghosts and vica-versa:
\begin{equation}
Q_{L,R} \longrightarrow {U_{L,R}} Q_{L,R}\,,
\ 
\overline Q_{L,R} \longrightarrow \overline Q_{L,R}{U_{L,R}^\dagger}\,,
\ 
{U_{L,R} \in SU(N_V+N|N_V)}.
\end{equation}
The apparent symmetry is
{$SU(N_V+N|N_V)_L\times SU(N_V+N|N_V)_R\times U(1)_V$}.
In fact, there are subtleties in the ghost sector:
the transformations are inconsistent with the requirement
that $\overline Q$ contain $\widetilde q_V^\dagger$ in the
ghost sector (and thus is related to $Q$, unlike in the quark sectors).
This is necessary for convergence of the ghost
functional integral\cite{damgaard}.
I do not have space to discuss this technical detail,
and I refer the interested reader to the 
literature\cite{damgaard,zirnbauer,ShShphi0,GSS}. The bottom line
is that, for perturbative calculations in the EFT, one gets
the same answer using the apparent symmetry group.

\subsubsection{Brief primer on graded Lie groups}\label{sec:primer}

Since these are lectures, I recall a few basic properties
of graded Lie groups. Graded means that
the group matrices, $U$,  contain both commuting and anticommuting elements
\begin{equation}
U= \left(\begin{array}{cc} A & B \\ 
\underbrace{C}_{N_V+N} &\underbrace{D}_{N_V} \end{array}\right)\quad
\textrm{{$A,D$} commuting; {$B,C$} anticommuting} \,.
\end{equation}
$U$ is unitary [i.e. {$U\in U(N_V+N|N_V)$}] if
$U U^\dagger = U^\dagger U= 1$, as for normal matrices,
as long as one complex conjugates anticommuting variables
as $(\eta_1 \eta_2)^* \equiv \eta_2^* \eta_1^*$.
Trace is generalized to ``supertrace'', defined
to maintain cyclicity:
\begin{equation}
\str\;U  \equiv \tr A-\tr D \quad \Rightarrow \quad
\str(U_1 U_2) = \str(U_2 U_1) \,.
\end{equation}
Determinants generalize to ``superdeterminants'',
\begin{equation}
\sdet\;U  \equiv \exp[\str(\ln U)] 
={\det(A - B D^{-1} C)}/{\det(D)}\,,
\label{eq:sdet}
\end{equation}
which satisfy $\sdet(U_1 U_2) = \sdet(U_1)\sdet(U_2)$.
Using this one can define
{$U \in SU(N_V+N|N_V)$} as unitary graded matrices
with unit superdeterminant.

To get a feel for the subtleties of the graded groups it is
useful to consider examples of $SU(N_V+N|N)$ matrices:
\begin{eqnarray}
\!\! U_I&=& \left(\begin{array}{cc} SU(N_V+N)& 0 \\ 
                            0 & SU(N_V) \end{array}\right)
\quad\Rightarrow\quad \sdet\; U_I = 1\,,
\\
\!\! U_{II}&=& \left(\begin{array}{cc} e^{i\theta N_V}& 0 \\ 
                            0 & e^{i\theta(N+N_V)}\end{array}\right)
\ \ \Rightarrow\ \
\sdet\; U_{II} = \frac{(e^{i\theta N_V})^{N+N_V}}{(e^{i\theta(N+N_V)})^{N_V}}
        = 1 \,.
\end{eqnarray}
$U_I$ looks just like an $SU(2N_V+N)$ matrix, while $U_{II}$
does not (having a {\em determinant} differing from unity).
Its superdeterminant is unity thanks to the $\det(D)$ in
the denominator of (\ref{eq:sdet}).

One feature of $U(N_V+N|N_V)$ which is unchanged from ungraded groups
is that one can pull out a commuting $U(1)$ factor,
{$U(N_V+N|N_V) = [SU(N_V+N|N_V)\otimes U(1)]/Z_N$},
with the $U(1)$ being a phase rotation:\footnote{%
Note that $\sdet(U_{III})=1$ in the quenched theory ($N=0$),
so that $U_{III}$ lies in $SU(N_V|N_V)$,
indicating that the quenched group structure is more complicated.}
\begin{equation}
U_{III}= \left(\begin{array}{cc} e^{i\theta }& 0 \\ 
                            0 & e^{i\theta}\end{array}\right)
\quad\Rightarrow\quad
\sdet\; U_{III} = \frac{e^{i\theta (N+N_V)}}{e^{i\theta N_V}}
        = e^{i\theta N} \,.
\end{equation}

\subsubsection{Chiral symmetry breaking}

We now follow the same steps as we did for
QCD in sec.~\ref{sec:brokenchiral}, noting differences along the way. 
We expand PQQCD about {${\mathcal M}=0$}, where the
chiral symmetry group
{${ G}=SU(N_V+N|N_V)_L\times SU(N_V+N|N_V)_R$}
is exact.\footnote{%
{\em A posteriori} we will find that we must take the chiral limit
with {$m_V$} and {$m_S$} in fixed ratio, because there
are divergences if {$m_V\to0$} at fixed {$m_S$}\cite{SSenhanced}. 
This non-analyticity is not a barrier to the construction of the EFT.
Non-analyticities are present in physical quantities in continuum
\chpt\ as well (e.g. the chiral logs), but arise from
infrared physics, just as the divergences here. There is no
reason to think that the coefficients in the EFT,
which result from integrating out ultraviolet physics, are
non-analytic in $m_V,m_S$.}
We know that this symmetry is broken,
because it is broken in the massless QCD contained
in massless PQQCD.
To study the symmetry breaking we introduce a
graded generalization of the order parameter (\ref{eq:Omega}) 
used for QCD:
\begin{equation}
\Omega_{ij} = \langle Q_{L,i,\alpha,c} \overline Q_{R,j,\alpha,c} \rangle_{\rm PQ}
\ {\lower0.6ex\hbox{$\stackrel{G}{\longrightarrow}$}}\ 
U_L\, \Omega \,U_R^\dagger\,.
\label{eq:OmegaPQ}
\end{equation}
Next we assume that the graded vector symmetry, $SU(N_V+N|N_V)_V$,
is not spontaneously broken.
If {${\mathcal M}$} is diagonal, real and positive,
this  follows from an extension\cite{ShShphi0} of
the Vafa-Witten theorem for QCD\cite{VafaWitten}.
The assumption is thus that nothing singular happens
as $\CM\to 0$. Given this, we know that
$\Omega = \omega \times {\bf 1}$, and, furthermore,
we know from the QCD sub-theory that 
$\omega={\langle q_S \bar q_S\rangle}\ne 0$.
Thus the symmetry breaking is 
{${ G} \to { H}=SU(N_V+N|N_V)_V$},
a simple graded generalization of that in QCD.

We can now derive Goldstone's theorem using Ward identities 
for two-point Euclidean correlators, which remain exact
in the PQ theory\cite{ShShphi0}.
The result is that there are massless, spinless poles which couple
to $\overline Q\gamma_\mu\gamma_5 T^a Q$ for all
$(N+2 N_V)^2-1$ traceless generators, $T^a$, of $SU(N_V+N|N_V)$.
Of these, $2(N+N_V)N_V$ are fermionic (quark-ghost particles),
$(N+N_V)^2-1$ are bosonic with normal sign two-point functions
(quark-quark),
and $N_V^2$ are bosonic with unphysical sign two-point functions
(ghost-ghost). 

\subsubsection{Constructing the EFT}

It would appear that we have the ingredients to construct an EFT,
just as in QCD:
we know the symmetries, and we have a separation of scales.
In fact, we know much less than in QCD, because PQQCD is not physical.
In QCD, the poles in two-point functions correspond to particle
states in a physical Hilbert space, and thus we know that
correlation functions of arbitrary order will also have 
these poles, and from their residues we can extract the S-matrix,
which will be unitary. Furthermore, since QCD involves local
interactions, the S-matrix will satisfy cluster decomposition.
Then we can invoke Weinberg's ``theorem'' and write an effective,
local field theory for the light particles.
In PQQCD, by contrast, we {\em do not know} that the poles we have found
in two-point functions also appear
in higher order correlators, and we {\em do know} that there is neither a physical
Hilbert space nor S-matrix.  Indeed, it can be shown that neutral correlators
have double-poles if $m_V\ne m_S$\cite{ShShphi0}.
So we cannot rely on Weinberg's argument.
Instead we must simply {\em assume} that there is a local EFT 
containing the Goldstone modes and satisfying the symmetries.
In other words we assume a minimal change from the EFT for QCD.

This is not as {\em ad hoc} as it might sound. Let me give three arguments
in favor of this assumption. First, we know that there is a
local EFT for the QCD sub-theory, and that this describes the
long distance behavior of correlators. 
What PQQCD allows one to do is to separate individual
Wick contractions contributing to QCD correlators.
It seems implausible that, for example, the description of
these individual contractions would require a non-local interaction
(leading to a different pole structure)
whose effects cancel when one adds them to form QCD correlators.
Second, one can derive chiral Ward identities in PQQCD
for arbitrary order correlation functions, 
that are generalizations of those in QCD.
PQ\chpt\ satisfies these identities by saturating them with
Goldstone pole contributions. In this regard, there is numerical
evidence from simulations that in, say, four-point functions the
Goldstone poles dominate when one pulls one of the operators far from
the others, as predicted by PQ\chpt.
Finally, one can imagine
carrying out a Wilsonian renormalization group program
in a Euclidean theory, in which one successively integrates out
``shells'' of high-momentum modes. This automatically leads
to a local interaction, and symmetries are preserved.
One can think of Symanzik's EFT for lattice QCD as an example.
We cannot actually do the calculation here, given the non-perturbative
physics of QCD, but if we could, and if the two-point functions
correctly tell us the appropriate low-energy degrees of freedom,
it is plausible that we would end up with PQ\chpt.

Having assumed the nature of the EFT we continue following
the same steps as for QCD. We ``promote'' the condensate into a field,
\begin{equation}
\Omega/\omega \to \Sigma(x) \in SU(N_V+N|N)\,,
\qquad \Sigma 
\stackrel{G}{\longrightarrow}
U_L\, \Sigma \,U_R^\dagger \,,
\end{equation}
and, assuming standard masses so that $\langle\Sigma\rangle=1$, 
we define NG particles by
\begin{equation}
\Sigma=\exp\left[\frac{2i}{f} \Phi(x)\right] \,,\qquad
\Phi(x) = \left( \begin{array}{cc} \phi(x) & \eta_1(x) \\
         \eta_2(x) & \widetilde\phi(x)\end{array}\right)\,.
\label{eq:SigmaPQ}
\end{equation}
Here $\phi$ are the quark-quark ``normal'' NG bosons,
$\widetilde\phi$ are the ghost-ghost NG bosons, and
$\eta_{1,2}$ are quark-ghost NG fermions.
The constraint {$\sdet\;\Sigma=1$} implies 
{$\str\;\Phi = \tr\;\phi-\tr\;\widetilde\phi=0$}.
The QCD part of $\Sigma$ is
\begin{equation}
\Phi(x) = \left(\begin{array}{ccc} 0 & 0 & 0 \\
                                  0 & \pi(x) & 0\\
\underbrace{0}_{N_V} & \underbrace{0}_{N} & \underbrace{0}_{N_V}
\end{array}\right)
\quad\Rightarrow\quad
\Sigma = \left(\begin{array}{ccc} 1& 0 & 0 \\
                                  0 & \Sigma_{\rm QCD} & 0\\
0 & 0 & 1
\end{array}\right) \,.
\label{eq:QCDrestriction}
\end{equation}

Next we construct the most general local, Euclidean-invariant,
$G$-invariant Lagrangian 
built out of $\Sigma$, $D_\mu \Sigma$ and $\chi$.
Here the covariant derivative and $\chi$ are graded generalizations of
the corresponding terms in \chpt. In particular
$\chi = 2B_0 (s+ip)\to U_L \chi U_R^\dagger$, with the sources
set to $s=\CM$ and $p=0$ at the end.
We can use generalizations of the same building blocks as in \chpt, e.g.
\begin{equation}
L_\mu = \Sigma D_\mu \Sigma^\dagger \to U_L L_\mu U_L^\dagger\,,\quad
\str(L_\mu) = 0 \,,\quad
\end{equation}
The power counting (which is independent of the nature of the fields)
is the same as in \chpt.

In this way we arrive at the PQ chiral Lagrangian through NLO:
\begin{eqnarray}
{\mathcal L}_{\rm PQ}^{(2)} &=& 
\frac{f^2}{4} \str\left(D_\mu\Sigma D_\mu \Sigma^\dagger\right)
-
\frac{f^2}{4} \str(\chi \Sigma^\dagger + \Sigma \chi^\dagger )
\label{eq:L2PQ}
\\
{\mathcal L}_{\rm PQ}^{(4)} &=& 
- {L_1}\, \big[\str(D_\mu \Sigma D_\mu \Sigma^\dagger)\big]^2
- {L_2}\, \str(D_\mu \Sigma D_\nu \Sigma^\dagger)
      \str(D_\mu \Sigma D_\nu \Sigma^\dagger) 
\nonumber\\ &&
+ {L_3}\, \str(D_\mu \Sigma D_\mu \Sigma^\dagger
      D_\nu \Sigma D_\nu \Sigma^\dagger)
\nonumber\\ &&
+ {L_4}\, \str(D_\mu \Sigma^\dagger D_\mu \Sigma)
         \str(\chi^{\dagger} \Sigma\! +\!  \Sigma^\dagger\chi)
+ {L_5}\, \str(D_\mu \Sigma^\dagger D_\mu \Sigma
         [\chi^{\dagger} \Sigma \!+\!  \Sigma^\dagger\chi])
\nonumber\\ &&
-\! {L_6} 
\big[\str(\chi^{\dagger} \Sigma \!+\! \Sigma^\dagger\chi)\big]^2
\!-\! {L_7} \big[\str(\chi^{\dagger} \Sigma \!-\! \Sigma^\dagger\chi)\big]^2
\!-\! {L_8}\, \str(\chi^\dagger\Sigma\chi^\dagger\Sigma \!+ \!p.c.)
\nonumber\\ &&
+ i{L_{9}}\, \str(L_{\mu\nu} D_\mu \Sigma D_\nu \Sigma^\dagger \!+\! p.c.)
+ {L_{10}}\, \str(L_{\mu\nu} \Sigma R_{\mu\nu} \Sigma^\dagger) 
\nonumber\\ &&
+ {H_1}\, \str(L_{\mu\nu} L_{\mu\nu} + p.c.)
- {H_2}\, \str(\chi^\dagger\chi)
+ \CL_{\rm WZW,PQ}
\nonumber\\ && { + L_{\rm PQ} {\mathcal O}_{PQ} } \,.
\label{eq:L4PQ}
\end{eqnarray}
These are almost carbon copies of the corresponding results
in \chpt\ [eqs.~(\ref{eq:L2},\ref{eq:L4})], except that
{$\tr \to \str$}, and there is an additional term in $\CL_{\rm PQ}^{(4)}$
(the {${\mathcal O}_{\rm PQ}$ term})\cite{ShVdWunphysical}.
To my  knowledge, no-one has worked out the structure of the PQ
WZW term in detail, though its contributions 
to $\pi^0\to\gamma\gamma$ vertices have been 
analyzed\cite{WillAndre}.

Thus we find that the number of LECs in PQ\chpt\ is the same as
in \chpt\ at LO, and that there is only one more, $L_{\rm PQ}$,
at NLO. But how are these LECs related to those of \chpt?
The answer is simple\cite{ShSh}: {\em they are the same!} 
This can be seen by considering correlation functions created
by sources $s,p,l_\mu,r_\mu$ restricted to the QCD sub-space.
At the quark level, these are QCD correlators, since valence
and ghost contributions cancel identically. Thus they
are described by \chpt. At the EFT level, one can show
diagramatically in PQ\chpt\ that a similar cancellation occurs,
and that one can do the calculation using
$\Sigma$ restricted to the QCD subspace
as in (\ref{eq:QCDrestriction}). Inserting this form into
$\CL_{\rm PQ}^{(2,4)}$ one finds ($i=1,10$, HEC ignored):
\begin{equation}
{\mathcal L}_{{\rm PQ}}^{(2,4)}(\Sigma,L_i,L_{\rm PQ}) = 
{\mathcal L}^{(2,4)}(\Sigma_{\rm QCD},L_i)  \,.
\label{eq:PQtoUQ}
\end{equation}
In words, the calculation one does with QCD sources in PQ\chpt\
is exactly that one would do in \chpt. For the results to be
equal it must be that the $L_i$ are equal.
This is the key result in PQ\chpt, for it means that
the predictions of PQ\chpt\ involve only slightly more LECs than
those of \chpt.

\subsubsection{What about ${\mathcal O}_{\rm PQ}$?}

Back when we were constructing ${\CL}^{(4)}$ in \chpt, I noted that
one possible four-derivative term was not independent for $N\le 3$.
This is due to Cayley-Hamilton relations between traces of
finite matrices. Such relations do not hold for graded matrices, and
so the term is independent in PQ\chpt. It is convenient to write it as
\begin{eqnarray}
{\mathcal O}_{\rm PQ} &=& 
\str(D_\mu\Sigma D_\nu\Sigma^\dagger D_\mu\Sigma D_\nu\Sigma^\dagger)
+2\;\str(D_\mu\Sigma D_\nu\Sigma^\dagger D_\mu\Sigma D_\nu\Sigma^\dagger)
\nonumber\\&&\mbox{}
-\str(D_\mu\Sigma D_\mu \Sigma^\dagger)^2/2
-\str(D_\mu\Sigma D_\nu\Sigma^\dagger)\str(D_\mu\Sigma D_\nu\Sigma^\dagger)\,,
\label{eq:OPQ}
\end{eqnarray}
for then it vanishes if $\Sigma$ is restricted to its QCD 
subspace as in (\ref{eq:QCDrestriction}). 
This is why $L_{\rm PQ}$ does not appear on
the right-hand side of eq.~(\ref{eq:PQtoUQ}).
${\mathcal O}_{\rm PQ}$ does not vanish for general $\Sigma$, however,
and thus appears in {${\mathcal L}_{{\rm PQ}}^{(4)}$} with 
a new LEC.
This additional LEC also appears in 
standard \chpt\ if $N\ge 4$, when the Cayley-Hamilton relations become
less restrictive.
As one goes to higher order, the number of additional such operators
in PQ\chpt\ increases\cite{ShVdWunphysical}.

How does this new operator enter into results for measurable quantities
in PQQCD? It can only contribute to PQ quantities, for the considerations
above show that it vanishes when restricted to the QCD sub-space.
It turns out to contribute to PQ $\pi\pi$ scattering at NLO,
but to PQ $m_\pi$ and $f_\pi$ only at NNLO\cite{ShVdWunphysical}.
Thus its practical impact is small.

It is worthwhile, however, understanding its origin more deeply.
As I have repeatedly mentioned, PQQCD allows one to separate
individual Wick contractions, unlike QCD.
For example, $\pi^+ K^0$ scattering in QCD has two contractions
(thin [blue] is $u$, medium [red] is $d$, and thick [brown] is $s$):
\begin{center}
\includegraphics[height=2cm]{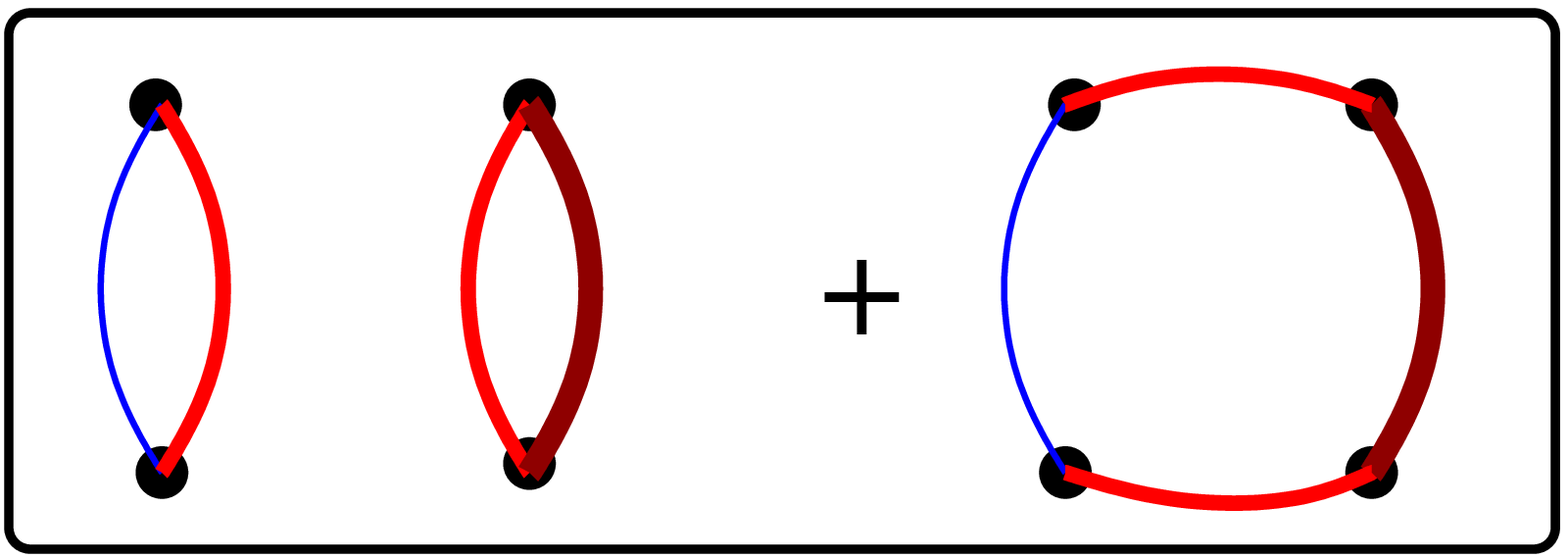}
\end{center}
and {${\mathcal O}_{\rm PQ}$} makes no contribution to this process.
We can separate these contractions in PQQCD using, for example,
scattering involving ghost quarks:
\begin{center}
\includegraphics[height=2cm]{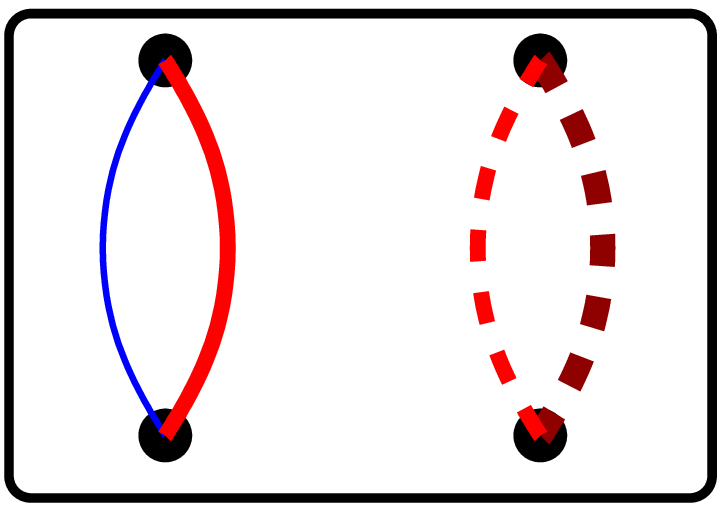}
\end{center}
This is the same as the first contraction contributing to QCD scattering, 
up to a sign.
{${\mathcal O}_{\rm PQ}$} does contribute to this correlator.
Thus $L_{\rm PQ}$ contains information about the relative size of
two contractions in the QCD process.

\subsection{PQ\chpt\ at LO}\label{sec:PQChPTLO}

With the Lagrangian in hand it is straightforward to develop perturbation
theory.
Inserting the expansion (\ref{eq:SigmaPQ}) into $\CL_{\rm PQ}^{(2)}$,
we find
\begin{eqnarray}
\!\!\!{\mathcal L}_{\rm PQ}^{(2)} &=& 
\str(\partial_\mu\Phi \partial_\mu\Phi) + \str(\chi\Phi^2) + \dots
\\
&=& \tr(\partial_\mu\phi\partial_\mu\phi + \partial_\mu\eta_1\partial_\mu\eta_2
-\partial_\mu\eta_2\partial_\mu\eta_1 
- \partial_\mu\widetilde\phi\partial_\mu\widetilde\phi)
\nonumber\\&&\mbox{}\!\!\!
+\tr\left[(\phi^2+\eta_1\eta_2)
\left(\begin{array}{cc} \chi_V & 0 \\ 0 & \chi_S \end{array}\right)\right] 
-\tr(\widetilde\phi^2 \chi_V) -\tr(\eta_2\eta_1 \chi_V) .
\label{eq:L2PQexpand}
\end{eqnarray}
Here $\chi_{V,S}$ are the mass matrices in the valence
and sea sectors, respectively, multiplied by $2 B_0$.
{$\phi$} is like the pion field in \chpt, except 
that it includes both valence and sea quarks.
The propagator for ``charged'' mesons with flavor 
$\bar q_1 q_2$ (which can be VV, VS or SS) is
\begin{equation}
(p^2 + m_{12}^2)^{-1}\,,  \qquad m_{12}^2= (\chi_1+\chi_2)/2\,.
\end{equation}
On the other hand, the terms involving the ``ghost-ghost'' boson
{$\widetilde\phi$} have unphysical signs. It appears that we are
expanding the ghost-ghost sector of $\Sigma$ about the wrong point,
since the potential is maximized. Furthermore, the kinetic term will
not give a convergent functional integral. Both these problems result
from our earlier decision to use the symmetry group ${G}$,
even though it was inconsistent with convergence.
A more careful treatment\cite{damgaard,ShShphi0,GSS} 
shows that one should have changed 
$\widetilde\phi \to i \widetilde \phi$, which solves the convergence
problem, at the cost of introducing $i$'s into vertices.\footnote{%
Obtaining a positive ghost propagator does not imply a restoration
of reflection positivity. 
This is now violated by the $i$'s in the vertices.}
In perturbation theory we can reshuffle the $i$'s by hand, and work
with the naive propagator one gets from (\ref{eq:L2PQexpand}).
For ``charged'' ghost mesons with flavor
$\bar{\widetilde{q}}_1 \widetilde q_2$  one has
\begin{equation}
-(p^2 + m_{12}^2)^{-1}\,, \qquad m_{12}^2= (\chi_1+\chi_2)/2\,.
\end{equation}
Finally, the NG fermion propagators 
can have either sign. There are no convergence issues for fermions,
but signs are important for cancellations.

What about the ``neutral'' fields ($\bar q_1 q_{1} $, etc.)?
Here we have to implement the constraint
{$\str(\Phi)=\tr(\phi)-\tr(\widetilde\phi)=0$}.
There are two ways to do this.
The first is simply use a basis of generators which is straceless:
{$\Phi=\sum_a \Phi_a T^a$}  with {$\str(T^a)=0$}.
This is analogous to excluding the $\eta'$ in \chpt,
but is more complicated in PQ\chpt.
In the second method, we remove the constraint
by including a singlet field,
$\Phi \to \Phi + \Phi_0/\sqrt{N}$,
adding a mass term to the action,
\begin{equation}
{\mathcal L}_{{\rm PQ}\chi} \to {\mathcal L}_{{\rm PQ}\chi} 
+ m_0^2\; \str(\Phi)^2/N \,,
\end{equation}
and then integrating out $\Phi_0$ by sending {$m_0^2\to\infty$}.
This is just a trick to project out the singlet.
To make it formally correct, 
we must regularize the theory with a cut-off so
that $m_0^2$ always exceeds any loop momenta.
This is the method mostly used in practice as it is simple to implement.

Using this method, the neutral propagator is obtained from:
\begin{eqnarray}
\!\!\!\!{\mathcal L}^{(2)} 
&=& \sum_{j=1}^{N+2N_V}
\epsilon_j(\partial_\mu\Phi_{jj}\partial_\mu\Phi_{jj} 
+ m_j \Phi_{jj}^2)
+ (m_0^2/N)(\sum_j \epsilon_j \Phi_{jj})^2 + \dots
\\
\epsilon_j &=& \left\{\begin{array}{l} +1\quad \textrm{valence or sea quarks} \\
                                       -1 \quad \textrm{ghosts} \end{array}\right.
\end{eqnarray}
The $m_0$ term couples all the $\Phi_{jj}$, so that, in particular,
neutral sea-quark states can contribute to neutral valence propagators.
The inversion of the kernel is not trivial for general quark masses, 
but can be accomplished
using linear algebra tricks\cite{BGPQ,ShSh}. 
I show an example of the result for $N=3$ non-degenerate sea quarks,
after having sent {$m_0^2\to\infty$} 
\begin{equation}
\langle\Phi_{ii} \Phi_{jj}\rangle =
 \frac{\epsilon_i \delta_{ij}}{p^2 + \chi_i} 
-\frac1N
 \frac{1}{(p^2+\chi_i)(p^2+\chi_j)}
{\frac{(p^2+\chi_{S1})(p^2+\chi_{S2})(p^2+\chi_{S3})}
      {(p^2+M_{\pi^0}^2)(p^2+M_\eta^2)} } \,.
\end{equation}
Here $M_{\pi^0}$ and $M_\eta$ are the masses of the sea-sector
neutrals, after inclusion of $\pi^0-\eta$ mixing 
evaluated at LO in \chpt\ (as discussed in the first lecture).
If we take $i,j$ to be valence labels, and set $\chi_i=\chi_j$
(or simply consider $i=j$), we see the infamous
double pole in the second term. It is reduced to a single
pole if the valence mass equals that of one of the sea quarks,
$\chi_i=\chi_{Sj}$. The form simplifies if the sea quarks
are degenerate,
\begin{equation}
\langle\Phi_{ii} \Phi_{jj}\rangle =
 \frac{\epsilon_i \delta_{ij}}{p^2 + \chi_i} 
-\frac1N
 \frac{{(p^2+\chi_S)}}{(p^2+\chi_i)(p^2+\chi_j)}\,.
\label{eq:neutraldegen}
\end{equation}
The residue of the double pole for {$\chi_i=\chi_j$}
is then $({\chi_i}-{ \chi_S})/N$, showing how it
vanishes in the physical subspace. Setting $\chi_i=\chi_j=\chi_S$
we obtain:
\begin{equation}
\langle\Phi_{SS} \Phi_{SS}\rangle =
 \frac{1}{p^2 + \chi_S} \left(1 - \frac1N\right)\, .
\end{equation}
This is the correct result in the sea sector, with
the $1/N$ term projecting out the $\eta'$.

Introducing $\Phi_0$ has allowed us 
to use the basis {$\Phi_{ij} \sim Q_i \overline Q_j$}
for neutral as well as charged states.
This means that one can follow the flavor indices in an
unambiguous way through any PQ\chpt\ diagram, resulting in
``quark-line diagrams''. This is a useful qualitative
tool in thinking about calculations.
Charged particles propagators are simple:\footnote{%
The sign of the propagator is to be determined from
eq.~(\ref{eq:L2PQexpand}).}
\begin{minipage}{7cm}
\hfill
$\langle\Phi_{ij} \Phi_{ji}\rangle = \pm \frac{1}{p^2 + (\chi_i+\chi_j)/2} =$
\end{minipage}
\begin{minipage}{3cm}
\begin{center}
\includegraphics[width=2cm]{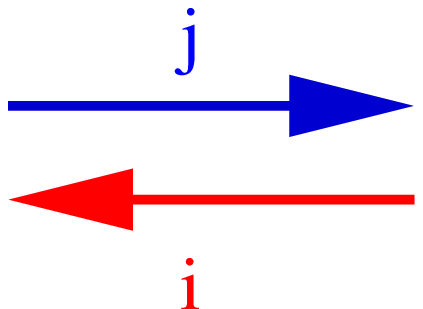}
\end{center}
\end{minipage}
\\
Typically one uses solid lines for quarks
(distinguishing valence and sea by a label), 
and dashed lines for ghosts.
In the diagram above I chose $i,j$ to be quark labels.
With this notation the neutral propagator is, schematically,
\begin{center}
\includegraphics[width=6cm]{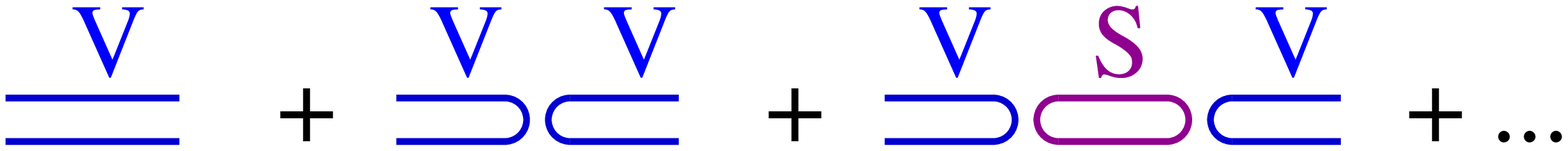}
\end{center}
where the ``hairpin'' is the $m_0$ vertex. 
The first term
corresponds to the single pole in (\ref{eq:neutraldegen}),
while the remaining diagrams all have a double pole (if $\chi_i=\chi_j$), 
and their
summation gives the second term in (\ref{eq:neutraldegen}).
Note that the valence and ghost contributions cancel exactly
between two hairpin vertices.

\subsection{NLO calculations in PQ\chpt\ and outlook}\label{sec:PQLs}

With the machinery developed above, perturbative calculations
in PQ\chpt\ are straightforward extensions of those in standard \chpt.
(I stress again that this is not true for non-perturbative calculations
like those in the $\epsilon-$regime.)
Let me sketch an example, that of the mass of a pion
composed of valence quarks {$V1,V2$}.
The types of quark-line diagrams corresponding to the one-loop diagram
are:
\begin{center}
\includegraphics[width=6cm]{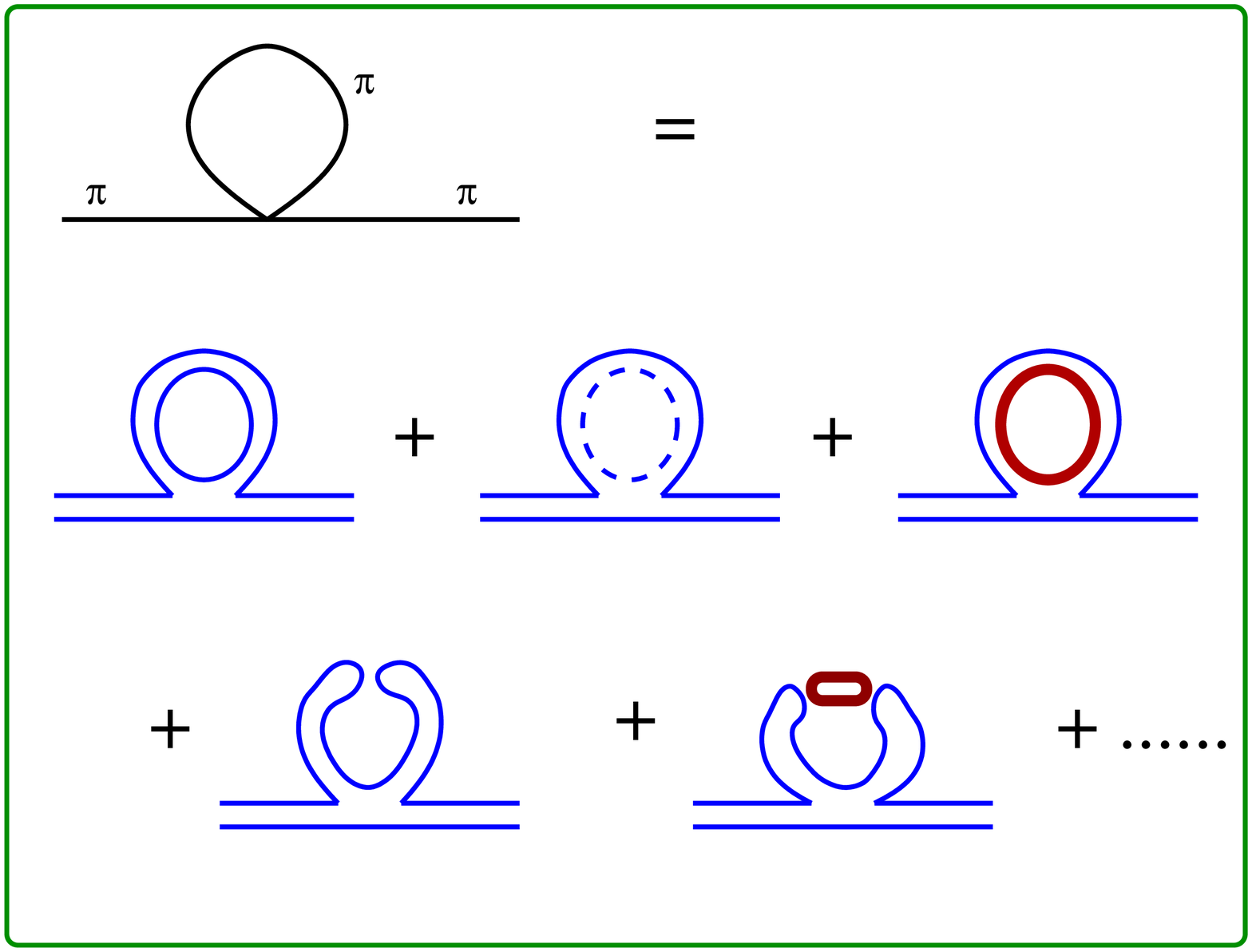}
\end{center}
Here thin (blue) lines are valence quarks, dashed lines
are ghosts, and thick (brown) lines are sea quarks.
I have used the result that the four-pion vertices from $\CL^{(2)}_{\rm PQ}$
involve a single supertrace and so give ``connected'' quark-line
vertices. The contributions of loops of valence-valence bosons
cancel those of valence-ghost fermions, as expected from the
underlying theory. This leaves only loops of
valence-sea bosons on the second line (which turn out to
cancel for $m_\pi$ but not for $f_\pi$), while the third line
shows the hairpin contributions.

The result (simplified by assuming degenerate 
sea quarks and $m_{V1}=m_{V2}\equiv m_V$)
is\cite{SSenhanced}
\begin{eqnarray}
m_{VV}^2 &=& \chi_V\bigg(1 + \frac1N \frac{2\chi_V-{\chi_S}}{\Lchi^2} \ln(\chi_V/\mu^2)
+ \frac{\chi_V-{\chi_S}}{N\Lchi^2}
\nonumber\\
&&\mbox{} + \frac8{f^2}\left[ (2 L_8-L_5)\chi_V + (2L_6 -L_4) {N\chi_S}\right]\bigg)\,.
\label{eq:mVVPQ}
\end{eqnarray}
The terms proportional to $1/N$ arise from the loops involving hairpins,
while the second line shows the analytic terms from the NLO Lagrangian.
The unphysical nature of the double poles in the hairpin contribution 
gives rise to
the ``enhanced logarithm'', proportional to
 $\chi_S\ln(\chi_V)$, which diverges when $m_V\to 0$ at fixed $m_S$,
and leads to a breakdown of PQ\chpt.
This is the divergence noted above which requires that one take the chiral limit
with $m_V/m_S$ fixed. The divergence occurs only at extremely small valence quark
masses, however, and should not be a problem in practice\cite{ShSh}.

The analytic terms in (\ref{eq:mVVPQ}) show the utility of partial quenching.
By varying $m_V$ and $m_S$, and including the chiral logs of the first line
in the fit, one can separately determine $2L_8-L_5$ and $2L_6-L_4$.
This should be compared to the setting $\chi_V=\chi_S$, so that one is
in the physical subspace. Then the chiral logs are physical and not enhanced,
but varying $\chi_V$ only allows a determination of $2L_8-L_5+2L_6-L_4$.
In fact, PQ simulations have been used by various groups to determine
$2L_8-L_5$, which is of particular interest as its value determines
whether it is possible for $m_u=0$ (which would solve the strong CP problem).
The answer is clearly negative\cite{MILCfpi}.

\subsubsection{Status of PQ\chpt\ calculations}\label{sec:statusPQChPT}

It is now standard to extend to PQ\chpt\ any \chpt\ calculation relevant for
extrapolating lattice results. 
Many quantities have been considered at one-loop,\footnote{%
I have chosen not to give references for the subsequent list---there is not space for
the $O(100)$ that would be needed.}
including the masses and form-factors of pions, baryons, vector mesons,
and heavy-light hadrons, 
structure functions of baryons,
the scalar two-point function,
weak matrix elements ($B_K$, $K\to\pi\pi$), 
the neutron electric dipole moment, and
pion and nucleon scattering amplitudes.
Similarly standard are partially quenched extensions of calculations including
discretization effects: tm\chpt, 
(rooted) staggered \chpt, and mixed-action \chpt.
In most cases this extension is straightforward.
The cases where it is not involve non-trivial generalizations of 
normal representations to those of graded groups.
Two notable examples are octect baryons\cite{LabrenzSharpe,Savage}
and four-fermion matrix elements\cite{GoltermanPallante}.
Particularly striking examples of the unphysical nature of PQQCD are the
negative contributions to correlators that are strictly positive in QCD,
e.g. the scalar two-point function\cite{negativescalar}, and the
appearance of non-unitary effects in two-pion 
correlators\cite{BGpipi,nonexponential}.
Another interesting application of PQQCD is to the calculation
of the spectrum of the continuum\cite{contspect} and lattice\cite{lattspect}
Dirac operators.

The MILC studies of pion and kaon properties show the potential power
of using partial quenching\cite{MILCfpi}, while at the same time exposing
the challenges of extrapolating using \chpt. With very precise numerical results,
and masses ranging from $m_s/10$ to $m_s$, an accurate description of their data
requires not only terms of NLO but also of NNLO and, at the higher masses, of NNNLO.
The loop contributions are not known beyond NLO in staggered \chpt, so only
analytic NNLO and NNNLO terms are kept. This is clearly a phenomenological approach,
but the key point to keep in mind is that these higher order terms have essentially
no impact on the resulting extrapolated results for physical quantities:
they are very small corrections for the physical up and down quark masses,
and the strange quark is anyway at, or close to, its physical value.
The MILC fits would not be possible without the large number of partially quenched
points, and their success provides {\em a posteriori} justification for
the assumptions needed to develop PQ\chpt.

The need to work beyond NLO in practice has spurred some heroic work from
continuum \chpt\ experts: there are now full NNLO (i.e. two-loop) calculations
for pion and kaon properties in PQ\chpt\cite{twoloopPQ}!

\subsubsection{A final example: $L_7$}
I close these lectures with a final example of which I am particularly fond 
and which nicely illustrates the power of PQQCD\cite{ShSh}. 
This concerns the LEC $L_7$, which multiplies a ``two (s)trace term'',
\begin{equation}
{\mathcal L}_{\chi,\rm PQ}^{(4)} = 
\dots - L_7 \str\;(\chi\Sigma^\dagger-\Sigma\chi^\dagger)^2 + \dots ,
\end{equation}
and which contributes to PGB masses only for non-degenerate quarks.
Its most significant contribution in QCD is to $m_{\eta}$,
and this leads to violations of the GMO relation:
\begin{equation}
4 m_K^2- m_\pi^2-3 m_\eta^2
=\frac{32(m_K^2-m_\pi^2)^2}{3f^2} (L_5-6L_8-12L_7)
+ \textrm{known chiral logs} \, .
\label{eq:L7GMO}
\end{equation}
Obtaining the physical result for $m_\eta$ would be a highly  non-trivial
check of the lattice methodology, as it involves quark-disconnected diagrams
with intermediate glue. For the same reason, this is a challenging calculation.
Partial quenching can help in the usual way by providing more data to fit,
but also by obviating the need to actually do the extrapolation to the physical
 $\eta'$. In particular, since $L_5$ and $L_8$ are known quite accurately,
as are the chiral logs, $\eta'$ physics is tested, using eq.~(\ref{eq:L7GMO}),
by any method that allows a calculation of $L_7$.
One such method is to calculate the residue of the double pole in
a disconnected valence-valence correlator:
\begin{equation}
\frac{\int d^3x \langle \Phi_{V1,V1}(t,\vec x) \Phi_{V2,V2}(0)\rangle}
  {\int d^3x \langle \Phi_{V1,V2}(t,\vec x) \Phi_{V2,V1}(0)
\rangle}\Bigg|_{m_{V1}=m_{V2}}
{\lower0.6ex\hbox{$\stackrel{t\to\infty}{\longrightarrow}$}}
\ \  \frac{{\mathcal D} t}{2 M_{VV}} \,.
\end{equation}
With $N=3$ degenerate sea quarks one finds
\begin{equation}
{\mathcal D} = \frac{\chi_V-\chi_S}{N} - 
\frac{16}{f^2}\left(L_7+ \frac{L_5}{2N}\right)(\chi_V-\chi_S)^2
+ \textrm{\rm known chiral logs} \,,
\end{equation}
so $L_7$ can be determined from the term quadratic (and thus even)
in the deviation from the unquenched theory. The generalization
of this formula to a $2+1$ theory has not been worked out,
but  should be straightforward.

I like this formula as it gets to the essence of partial quenching: using
unphysical phenomena (in this case the double pole) to obtain physical
results (here $L_7$). Of course, we have not removed the need to do the
challenging calculation of
quark-disconnected correlators, but rather have packaged the calculation
in a way that is more flexible because there are more knobs to turn.
The result has been very recently extended to NNLO\cite{L7NNLO}.

\section*{Acknowledgments}
I thank Oliver B\"ar for a helpful correspondence,
Will Detmold, Maarten Golterman and Andr\'e Walker-Loud for comments,
and Y.~Kuramashi and
co-organizers for a delightful school.
This work was supported in part by 
U.S. Department of Energy Grant No. DE-FG02-96ER40956.


\end{document}